\pdfoutput=1

\documentclass[12pt, hidelinks]{article}

\usepackage{import}
\usepackage[twoside=true, margin=1in]{geometry}
\usepackage{authblk}

\usepackage[english]{babel}
\usepackage{libertine}
\usepackage[subpreambles=false]{standalone}
\usepackage{amsmath}
\usepackage{amsthm}
\usepackage{amsfonts}
\usepackage{bbm}
\usepackage[normalem]{ulem}
\usepackage{aliascnt}
\usepackage[ruled,vlined]{algorithm2e}
\usepackage{graphicx}
\usepackage{url}            
\usepackage{subfiles}
\usepackage{enumitem}
\usepackage[numbers]{natbib}
\usepackage{amssymb}
\usepackage{float}
\usepackage{longtable}
\usepackage{array}
\newcolumntype{L}[1]{>{\raggedright\let\newline\\\arraybackslash\hspace{0pt}}m{#1}}

\newtheorem{theorem}{Theorem}[section]
\newtheorem{claim}[theorem]{Claim}
\newtheorem{lemma}[theorem]{Lemma}
\newtheorem{proposition}[theorem]{Proposition}
\newtheorem{corollary}[theorem]{Corollary}
\newtheorem{example}[theorem]{Example}

\newtheorem{definition}[theorem]{Definition}
\newtheorem{observation}[theorem]{Observation}

\newcommand\locallabel[1]{\label{\currentprefix:#1}}

\newcommand\localref[1]{\ref{\currentprefix:#1}}

\newif\ifcompact
\compactfalse

\newcommand{\toas}{\overset{a.s.}{\to}}

\newcommand{\tree}{\textsc{Tree}}
\newcommand{\chain}{\mathcal C}
\newcommand{\frontier}{\textsc{FRONTIER}}
\newcommand{\sm}{\textsc{SM}}
\newcommand{\nsm}{\textsc{NSM}}
\newcommand{\wait}{\text{Wait}}
\newcommand{\pow}{\ensuremath{\alpha^{\text{PoW}}}}
\newcommand{\pos}{\ensuremath{\alpha^{\text{PoS}}}}
\newcommand{\poss}{\ensuremath{\alpha^{\text{PoSNER}}}}
\newcommand{\cstring}{\gamma}
\newcommand{\half}{\textsc{Half}}

\newcommand{\unpublished}{\mathcal{U}}
\providecommand{\rev}{\textsc{Rev}}

\newcommand{\e}[1]{{\mathbb E}\left[ #1 \right]}
\newcommand{\pr}[1]{Pr\left[#1\right]}
\newcommand{\ind}[1]{\mathbbm{1}_{#1}}

\DeclareMathOperator{\Succ}{\textsc{Succ}}
\DeclareMathOperator{\Succs}{\textsc{Succ}_s}

\DeclareMathOperator{\potr}{\delta}

\DeclareMathOperator{\publishset}{\text{PublishSet}}
\DeclareMathOperator{\publishpath}{\text{PublishPath}}
\DeclareMathOperator{\publish}{\text{Publish}}
\DeclareMathOperator{\R}{r}
\DeclareMathOperator{\rew}{\R_\lambda}
\DeclareMathOperator{\ca}{Ca}
\DeclareMathOperator{\va}{\mathcal V}
\DeclareMathOperator{\vertex}{V}
\DeclareMathOperator{\edge}{E}

\author[1]{Matheus V. X. Ferreira\thanks{Contact: mvxf@cs.princeton.edu}}
\author[1]{S. Matthew Weinberg\thanks{Supported by NSF CAREER Award CCF-1942497}}
\affil[1]{Computer Science, Princeton University}

\begin{document}
\title{Proof-of-Stake Mining Games with Perfect Randomness}

\pagenumbering{gobble}
\maketitle
\begin{abstract}
Proof-of-Stake blockchains based on a longest-chain consensus protocol are an attractive energy-friendly alternative to the Proof-of-Work paradigm. However, formal barriers to ``getting the incentives right'' were recently discovered, driven by the desire to use the blockchain itself as a source of pseudorandomness~\cite{brown2019formal}.

We consider instead a longest-chain Proof-of-Stake protocol with perfect, trusted, external randomness (e.g.~a randomness beacon). We produce two main results.

First, we show that a strategic miner can strictly outperform an honest miner with just $32.5\%$ of the total stake. Note that a miner of this size \emph{cannot} outperform an honest miner in the Proof-of-Work model~\cite{sapirshtein2016optimal}. This establishes that even with access to a perfect randomness beacon, incentives in Proof-of-Work and Proof-of-Stake longest-chain protocols are fundamentally different.

Second, we prove that a strategic miner cannot outperform an honest miner with $30.8\%$ of the total stake. This means that, while not quite as secure as the Proof-of-Work regime, desirable incentive properties of Proof-of-Work longest-chain protocols can be approximately recovered via Proof-of-Stake with a perfect randomness beacon.

The space of possible strategies in a Proof-of-Stake mining game is \emph{significantly} richer than in a Proof-of-Work game. Our main technical contribution is a characterization of potentially optimal strategies for a strategic miner, and in particular, a proof that the corresponding infinite-state MDP admits an optimal strategy that is positive recurrent.
\end{abstract}
\newpage
\tableofcontents
\clearpage
\pagenumbering{arabic}

\section{Introduction}

Blockchains have been a resounding success as a disruptive technology. However, the most successful implementations (including Bitcoin~\cite{nakamoto2008bitcoin} and Ethereum~\cite{wood2014ethereum}) are built on a concept called proof-of-work. That is, participants in the protocol are selected to update the blockchain proportionally to their computational power. The consensus protocols underlying Bitcoin and Ethereum (and many other proof-of-work cryptocurrencies) have been secure in practice, and robust against strategic manipulation. There is even a theoretical foundation supporting this latter property: honestly following the Bitcoin protocol is a Nash equilibrium in a stylized model when no miner controls more than $\pow\approx 0.329$ of the total computational power in the network~\cite{sapirshtein2016optimal,kiayias2016blockchain} (we will use the notation $\alpha^{\text{model}}$ to denote the supremum $\alpha$ such that whenever no miner is selected to create the next block with probability bigger than $\alpha$, it is a Nash equilibrium for all miners to follow the longest-chain protocol in the referenced model).\footnote{\citet{kiayias2016blockchain} proves that $\pow \gtrapprox 0.308$, and~\cite{sapirshtein2016optimal} estimates $\pow$ to high precision as $\approx 0.329$.}

However, one major drawback of proof-of-work blockchains is their massive energy consumption. For example, Bitcoin currently consumes more electricity than all but 26 countries annually.\footnote{Source: \url{https://cbeci.org/}. Accessed 3/25/2021.} The need for specialized hardware and low-cost electricity/cooling/etc. also leads to concentration of the mining process among the few entities who have access to the necessary technology~\cite{ArnostiW19}. One popular emerging alternative is a paradigm termed proof-of-stake, where participants are selected proportionally to their stake in the currency itself. 

Proof-of-stake cryptocurrencies do not suffer from this drawback, but raise new technical challenges, especially from the incentives perspective. Indeed,~\citet{brown2019formal} identifies several formal barriers to designing incentive compatible longest-chain proof-of-stake cryptocurrencies (that is, proof-of-stake protocols ``like Bitcoin''). Their work highlights one key barrier: in existing proof-of-stake protocols, \emph{the blockchain itself serves as a source of pseudorandomness}, whereas in proof-of-work protocols \emph{the pseudorandom selection of participants is completely independent of the blockchain}. Specifically, they pose a stylized model with \emph{No External Randomness} and show that it is \emph{never} a Nash equilibrium for all miners to honestly follow the longest-chain protocol no matter how small they are (that is, $\poss = 0$).

In this work, we investigate the incentive compatibility of longest-chain proof-of-stake protocols with access to \emph{perfect external randomness, completely independent of the blockchain}, often termed a randomness beacon~\cite{rabin1983transaction} (for brevity of notation, we'll refer to this model simply as PoS). We provide two main results, which give a fairly complete picture:

\begin{itemize}
    \item We establish that $\pos < 0.325 < \pow$. That is, \emph{even with access to perfect external randomness, longest-chain proof-of-stake protocols admit richer strategic manipulation than their proof-of-work counterparts}. We do this by designing a new strategic deviation that we term nothing-at-stake selfish mining, and establish that it is strictly more profitable than honest behavior for any miner with $\gtrapprox 0.325$ of the total stake (Theorem~\ref{thm:enhanced-selfish-mining}).
    \item We prove that $\pos \gtrapprox 0.308$ (Theorem~\ref{thm:nash-equilibrium}). In particular, this means that access to a randomness beacon fundamentally changes longest-chain proof of stake protocols: without one $\poss = 0$, and any miner can profit by deviating. With a randomness beacon, the incentives are (quantitatively) almost as good as proof-of-work.
\end{itemize}

We now provide a high-level overview of our model (a significantly more detailed description of the model appears in Section~\ref{pos:sec:preliminaries}), a brief overview of the key technical highlights, and an overview of related work.

\subsection{Brief Overview of Model}
Seminal work of Eyal and Sirer poses an elegant abstraction of the Bitcoin protocol (that we call the PoW model)~\cite{eyal2014majority}. Specifically, the game proceeds in infinitely many discrete rounds. In each round, a single miner is chosen proportionally to their computational power, and creates a block. Immediately upon creating a block, the miner must choose its contents (including its predecessor in the blockchain). The strategic decisions a miner makes are: a) which predecessor to select when they create a block, and b) when to publish that block to the other miners. The fact that predecessors must be chosen \emph{upon creation} of the block captures that the contents of a block created via proof-of-work are fixed upon creation. 

A key concept in Bitcoin is the \emph{longest-chain protocol}. Specifically, the longest chain is the published block with the most ancestors.\footnote{An ancestor is any block that can be reached by following a path of predecessors. Observe that because each block has a single predecessor, there is a single path of predecessors out of any block.} Each miner's reward is equal to the fraction of blocks they produce in the longest chain (taking a limit as rounds go to $\infty$). A miner honestly follows the longest-chain protocol if: a) they always select the (current) longest chain as the predecessor of any created node, and b) they publish all created blocks during the round in which it's created.~\citet{eyal2014majority} establishes that $\pow \leq 1/3$ (previously, it was believed that $\pow = 1/2$ as proposed in~\cite{nakamoto2008bitcoin}), and follow-up work further nailed down $\pow \approx 0.329$~\cite{sapirshtein2016optimal}.

\citet{brown2019formal} modify this model to capture proof-of-stake with no external randomness. In their model, a random coin is selected in each round \emph{independently for each block}. That is, for each round, and each block $B$, an independent random miner is selected who is eligible to create a block with $B$ as a predecessor with probability equals to their fraction of all the coins in the system. This captures that the protocol must use the chain itself as a source of pseudorandomness. Their work establishes that $\poss = 0$: it is never a Nash equilibrium to honestly follow the longest-chain protocol in their model. This result is entirely driven by the fact that the protocol has no external randomness, and therefore, miners can make non-trivial predictions about future pseudorandomness.

Our model lies between these two, and captures proof-of-stake with perfect external randomness. Specifically, in each round a single coin is chosen to create a block. Thus a single miner is chose to create a block (just as in PoW) with probability proportional to their fraction of the coins in the system. The strategic decisions are now better phrased as: a) when to publish a created block, and b) which predecessor to select \emph{when publishing}. Our model captures the following: perfect external randomness allows the protocol to select a random miner independently of all previous selected miners and all previously published blocks. The distinction to proof-of-work is that it is now computationally tractable to set the contents of the block, including its predecessor, at any point before it is published.

Note that our model does stipulate that the winner of round $t$ can publish a single block with timestamp $t$. In a proof-of-stake protocol, there is no technical barrier to creating and publishing any number of blocks using the same timestamp (indeed, this is precisely because it is computationally efficient to produce blocks in proof-of-stake). However, it will be immediately obvious to the rest of the network that a miner has deviated from the longest-chain protocol in this specific way, and it will be immediately obvious which miner cheated.\footnote{Observe that deviations from the longest-chain protocol that select strategic predecessors or publish at strategic times cannot be definitively attributed to a cheating miner, as these deviations have an innocent explanation: latency. That is, perhaps the reason a miner chose the wrong predecessor is because news of the true longest chain had not yet reached them. Alternatively, perhaps the miner tried to publish their block during the correct round, but it only propagated through the network several rounds later due. Like all prior work, we do not rigorously model latency, and stick to the elegant model proposed in~\cite{eyal2014majority}.} A common solution to strongly disincentivize such behavior is a \emph{slashing protocol}: any miner can include pointers to two blocks created using the same timestamp and the cheating miner will be steeply fined. While we will not rigorously model the incentives induced by a slashing protocol, our model implicitly assumes a sufficiently strong disincentive for miners to publish multiple blocks (and capture this in our model by simply hard-coding that miners must publish at most a single block with each timestamp).

To get intuition for the types of protocols our stylized model aims to capture, below is a sample (simplified) protocol to have in mind:\footnote{We are not claiming that this protocol is secure in a rich model, nor will we reason formally about properties of the proposed slashing mechanism. We provide this just to give intuition for why our stylized model captures the salient features of a longest-chain protocol with trusted external randomness.}
\begin{itemize}
    \item In order to be eligible to mine, a coin must be frozen for (large) $T$ rounds, along with a deposit equal to (large) $L$ times its value (that is, the coin and its deposit must be owned by the same miner for $T$ rounds in a row).
    \item After being used for mining, a coin and its deposit must be frozen for $T$ rounds.
    \item During each round $t$, the randomness beacon outputs a random number. This is mapped to a random eligible coin, and its owner is the selected miner at round $t$. The selected miner can create blocks with timestamp $t$.
    \item If a miner ever publishes distinct blocks with the same timestamp, any other miner can include pointers to those two blocks in a block of their own. This will cause the deviant miner to lose their entire deposit (if desired, a $1-\varepsilon$ fraction of it can be destroyed, and an $\varepsilon$ fraction can be awarded to the altruistic miner).
\end{itemize}
Importantly, we are claiming neither that randomness beacons exist (either in theory or in practice), nor that slashing protocols that perfectly disincentivize detectable cheating (without affecting any other incentives) exist. Theorem~\ref{thm:enhanced-selfish-mining} shows that \emph{even if} these primitives existed, a longest-chain proof-of-stake protocol assuming them would still be (slightly) more vulnerable to strategic manipulation than a proof-of-work protocol. On the other hand, Theorem~\ref{thm:nash-equilibrium} establishes in some sense a reduction from proof-of-stake protocols that nearly match the incentive guarantees of proof-of-work protocols to the design of randomness beacons and slashing protocols. 
\subsection{Brief Technical Overview}
Theorem~\ref{thm:enhanced-selfish-mining} ($\pos \lessapprox 0.325$) follows by designing our nothing-at-stake selfish mining strategy, and analyzing its expected payoff. While the insights to design our strategy are novel, the analysis is similar to those used in prior work to analyze the payoff of the resulting Markov Decision Process (MDP). We defer to Section~\ref{sec:enhanced-selfish-mining} a description of our strategy and intuition for why it succeeds.

The proof of Theorem~\ref{thm:nash-equilibrium} is the bulk of our technical work. To start, we observe that our model admits an infinite-state MDP (just as in~\cite{sapirshtein2016optimal}). However, the space of strategies available to a miner in our setting is \emph{significantly} richer than in the PoW model. We provide several examples demonstrating why counterintuitive behavior (such as orphaning one's own blocks) could a priori be part of an optimal strategy. So our main technical results characterize possible optimal strategies for this infinite-state MDP, culminating in a strong enough characterization to lower bound the optimal payoff for Theorem~\ref{thm:nash-equilibrium} and concluding $\pos \geq 0.308$.

\subsection{Related Work}
The most related work is already overviewed above:~\citet{eyal2014majority} provide the PoW model, develop the selfish mining attack, and prove that $\pow \leq 1/3$.~\citet{sapirshtein2016optimal} estimates $\pow \approx 0.329$ by solving the associated MDP to high precision, and~\citet{kiayias2016blockchain} prove that $\pow \gtrapprox 0.308$.~\citet{brown2019formal} study a related proof-of-stake model with no external randomness, and show that $\poss = 0$. Other works also study similar questions in variants of this model (e.g.~\cite{CarlstenKWN16,Neuder2021Selfish}). 

There is a rapidly-growing body of work at the intersection of mechanism design and cryptocurrencies~\cite{LeshnoS20, ChenPR19, ArnostiW19, HubermanLM20, ferreira2021dynamic}. Some of these works further motiviate the consideration of proof-of-stake cryptocurrencies~\cite{ArnostiW19}, while others motivate the choice to restrict attention to Bitcoin's proportional reward scheme~\cite{ChenPR19, LeshnoS20}, but there is otherwise little overlap between our works.
 
In practice, implementing a random beacon is a complex task~\cite{boneh2018verifiable, bonneau2015bitcoin, clark2010use} and is outside the scope of this paper. As previously noted, our results can be viewed either as a reduction to designing a randomness beacon (Theorem~\ref{thm:nash-equilibrium}), or an impossibility result even under the assumption of a randomness beacon (Theorem~\ref{thm:enhanced-selfish-mining}).

Finally, it is worth noting that many existing proof-of-stake protocols fit the longest-chain paradigm~\cite{kiayias2017ouroboros, goodman2014tezos,DaianPS19}, while others are fundamentally different~\cite{gilad2017algorand}. Protocols based on Byzantine consensus are a growing alternative to the longest-chain paradigm, although both paradigms are well-represented in theory and in practice. Byzantine consensus protocols are outside the scope of our analysis.
 
\subsection{Roadmap}
Section~\ref{pos:sec:preliminaries} provides a very detailed description of our model, along with examples to help illustrate its distinction from proof-of-work. Section~\ref{sec:enhanced-selfish-mining} provides our nothing-at-stake selfish mining, and Theorem~\ref{thm:enhanced-selfish-mining}. Sections~\ref{sec:trimming-state} and~\ref{sec:trimming-strategy} narrow the space of optimal strategies through a series of reductions. Section~\ref{sec:nash-equilibrium} overviews Theorem~\ref{thm:nash-equilibrium}. Various helpful examples and all omitted proofs are in the appendix.

\section{Model}\label{pos:sec:preliminaries}

A mining protocol is a Nash equilibrium if no miner wishes to unilaterally change their strategy provided all miners are following the intended protocol. Thus it suffices to consider a two-player game between Miner 1 and Miner 2. Think of Miner 2 as the ``rest of the network'', which is honestly executing the longest-chain protocol, and think of Miner 1 as the ``potential attacker'' that optimizes his strategy provided Miner 2 is honest. Following~\cite{eyal2014majority, kiayias2016blockchain, sapirshtein2016optimal} and subsequent works, the game proceeds in discrete time steps (abstracting away the exponential rate at which blocks are found) which we call \emph{rounds}, and the rounds are indexed by $\mathbb{N}_+$. The state $B$ of the game is a tuple $(\tree(B), \unpublished_1(B), \unpublished_2(B), T_1(B), T_2(B))$ (each of these terms will be explained subsequently).

\vspace{1mm}\noindent\textbf{Rounds.} During every round $n$, a single miner creates a new block, and we denote that miner by $\cstring_n \in \{1,2\}$. We denote by $\cstring:= \langle \cstring_n \rangle_{n \in \mathbb{N}_{+}}$ the full ordered list of miners for each round. In an execution of the game, each $\cstring_n$ is drawn i.i.d., and equal to $1$ with probability $\alpha < 1/2$. We let $T_i:=\{n \ \ | \ \ \cstring_n = i\}$ as the rounds during which Miner $i$ creates a new block. $T_i(B)$ denotes all blocks created by Miner $i$ at state $B$. We abuse notation and might refer to $n$ as the state at round $n$ (after block $n$ is created and all actions are taken, but before round $n+1$ starts). That is, $(\tree(n), \unpublished_1(n), \unpublished_2(n), T_1(n), T_2(n))$ is the state at round $n$.

\vspace{1mm}\noindent\textbf{Blocks.} The second basic element is a \emph{block}. Each block has a label in $\mathbb N$. Blocks are totally ordered by their labels and we say block $s$ was created before block $v$ if $s < v$. We overload notation and also use $n$ to refer to the block produced in round $n$. All blocks are initially \emph{unpublished}, and can later become \emph{published} due to actions of the miners. Once a block $n$ is published, it has a pointer to exactly one predecessor $n'$ created earlier (that is $n' < n$) and we write $n \to n'$. 

\vspace{1mm}\noindent\textbf{Block Tree.} Because all published blocks have a pointer to an earlier block, this induces, at all rounds, a block tree $\tree$. We will also refer to $\vertex, \edge$ as the nodes and edges in $\tree = (\vertex, \edge)$. Here, the nodes are all blocks which have been published. Every node has exactly one outgoing (directed) edge towards its predecessor. Before the game begins, the block tree contains only block 0, which we refer to as the {\em genesis block} and not created by Miner 1 nor Miner 2. We let $\unpublished_i$ denote the set of blocks which have been created by Miner $i$, but are not yet published. We refer to
\begin{equation}\label{eq:b_0}
B_0 := ((\{0\}, \emptyset), \emptyset, \emptyset, \emptyset,\emptyset)
\end{equation}
as the initial state before any blocks are created and the block tree contains only the genesis block.

\vspace{1mm}\noindent\textbf{Ancestor Blocks.} We say that block $a$ is an \emph{ancestor} of block $b\in V$ if there is a directed path from $b$ to $a$ (so $b$ is an ancestor of itself). We write $A(b)$ to denote the set of all ancestors of $b$ (observe that a block can never gain new ancestors, so this is well-defined without referencing the particular state $S$, or the round, etc.). We use $h(b):=|A(b)|-1$ as short-hand for the height of block $b$ (the genesis block is the only block with height $0$).

\vspace{1mm}\noindent \textbf{Longest Chain.} The \emph{longest chain} $\chain:=\arg\max_{b \in V}\{h(b)\}$ is the leaf in $V$ with the longest path to the genesis block, breaking ties in favor of the first block published, and then in favor of the earliest-indexed block. We use $H_i$ to refer to the block in $v \in A(\chain)$ with height $i$ and we refer to $A(\chain)$ as the \emph{longest path}. We say a block $q$ is {\em forked} (or orphaned) when $q \in A(\chain)$, but a new block $\chain'$ becomes the longest chain and $q \not \in A(\chain')$.

\vspace{1mm}\noindent\textbf{Successor blocks.} We say block $a$ is a \emph{successor} of block $b \in V$ if $a \neq b$ is in the unique path from $\chain$ to $b$. We write $\Succ(b)$ to denote the set of successors of $b$.

\vspace{1mm}\noindent \textbf{Actions.} During round $n$, Miner $i$ knows $\cstring_\ell$ for all $\ell \leq n$, and can take the following actions:
\begin{enumerate}
    \item $\wait$: wait for the next round, and do nothing.
    \item $\publishset(V', E')$: publish a set of blocks $V'$ with pointers $E'$. This adds $V'$ to $V$, $E'$ to $E$, and changes all blocks in $V'$ from unpublished to published. To be valid, it must be that:
\begin{itemize}
\item $V' \subseteq \unpublished_i$ (Miner $i$ actually has blocks $V'$ to publish).
\item For all $v \to v' \in E'$, $v \in V'$, $v' \in V \cup V'$ (syntax check for edges in $E'$).
\item For all $v \to v' \in E'$, $v > v'$ (pointers are to earlier blocks).
\item For all $v \in V'$, there is exactly one outgoing edge in $E'$ (every block has exactly one pointer).
\end{itemize}
\end{enumerate}

\vspace{1mm}\noindent\textbf{Clarifying Order of Operations.} At the beginning of round $n$, there is a block tree $\tree = \tree(n-1)$, and each miner $i$ has a set of unpublished blocks $\unpublished_i = \unpublished_i(n-1)$. Then:
\begin{enumerate}
\item $\cstring_n$ is drawn, and equal to $1$ with probability $\alpha$, and $2$ with probability $1-\alpha$. This updates $\unpublished_{\cstring_n}:=\unpublished_{\cstring_n}(n-1)\cup \{n\}$. For the other miner, $\unpublished_{3-\cstring_n}:=\unpublished_{3-\cstring_n}(n-1)$.
\item Miner $2$ takes an action. If that action is $\publishset(V', E')$, add the nodes $V'$ and edges $E'$ to $\tree$, and update $\unpublished_2:=\unpublished_2\setminus V'$.
\item Miner $1$ takes an action. If that action is $\publishset(V', E')$, add the nodes $V'$ and edges $E'$ to $\tree$, and update $\unpublished_1:=\unpublished_1 \setminus V'$. 
\item At this point, round $n$ is over, so $\tree(n):=\tree$, $\unpublished_i(n):=\unpublished_i$, etc.
\end{enumerate}
\vspace{1mm}\noindent\textbf{Predecessor state.} For state $B$, we define $B^\half$ as the state prior to $B$ before Miner 1 took their most recent action and after Miner 2 took their most recent action. Similarly, we define $(\tree^\half(n), \unpublished_1^\half(n), \unpublished_2^\half(n), \chain^\half(n))$ as the subsequent state to $(\tree(n-1), \unpublished_1(n-1), \unpublished_2(n-1), \chain(n-1))$ after block $n$ was created, Miner 2 takes their action and before Miner 1 takes their action.

Recall that Miner 2 acts first in every round, so the second tie-breaker in deciding what is the longest chain is only used to distinguish between two blocks of Miner 1 published during the same round, and will never be invoked in a ``reasonable'' strategy for Miner 1 (see Observation~\ref{obs:timeserving}).\footnote{\citet{eyal2014majority} also considers the case where Miner 1 wins a tie-breaking with probability $\beta$. In principle, our model can easily accommodate any $\beta \in [0,1]$, but we focus on the case $\beta = 0$ since it is the most pessimistic for the attacker.} Like~\citet{kiayias2016blockchain}, we use $\frontier$ to refer to the honest strategy, which never forks and it will be Miner 2's strategy.

\begin{definition}[Frontier Strategy]\label{def:frontier}
During all rounds $n$, the $\frontier$ strategy for miner $i$ does the following:
\begin{itemize}
\item If $\cstring_n \neq i$, $\wait$.
\item If $\cstring_n = i$, $\publishset(\{n\}, \{n \to \chain\})$ (publishes the new block pointing to the longest chain).
\end{itemize}
\end{definition}
\begin{definition}[Rewards]\label{def:reward}
For any two states $B$ and $B'$, define \emph{Miner $k$'s reward} as the integer-valued function $r^k$ from state $B$ to $B'$ as the difference between the number of blocks created by Miner $k$ in the longest path at state $B'$ and $B$. That is,
\begin{equation}\label{eq:reward}
r^k(B, B') := |A(\chain(B')) \cap T_k(B')| - |A(\chain(B)) \cap T_k(B)|.
\end{equation}
\end{definition}
\vspace{1mm}\noindent\textbf{Payoffs.} As in~\cite{eyal2014majority} and follow-up work, miners receive steady-state revenue \emph{proportional to their fraction of blocks in the longest chain}. Recall that in our notation, $\chain(n)$ denotes the longest chain after the conclusion of round $n$, $A(\chain(n))$ denotes the ancestors of $\chain(n)$, and $T_i$ denotes all blocks created by Miner $i$. Therefore, we define the payoff to Miner $1$, when Miner $i$ uses strategy $\pi_i$ as:
$$\rev^{(n)}_\cstring(\pi_1,\pi_2):= \frac{|A(\chain(n)) \cap T_1|}{h(\chain(n))}, \qquad \text{and} \qquad \rev(\pi_1,\pi_2) := \mathbb{E}_{\cstring}\bigg[\liminf_{n \to \infty}\rev^{(n)}_\cstring(\pi_1,\pi_2)\bigg],$$
where the expectation is taken over $\gamma$, recalling that each $\gamma_n$ is i.i.d. and equals to $1$ with probability $\alpha$, and $2$ with probability $1-\alpha$. 

We will study in particular the payoff of strategies $\pi_1$ against $\frontier$. For simplicity of notation later, we will refer to $\unpublished := \unpublished_1$ (because $\frontier$ has no unpublished blocks, so $\unpublished_2$ is unnecessary), $\pi := \pi_1$ (because $\pi_2:=\frontier$). We will also list all states only as $(\tree(n),\unpublished(n),T_1(n))$ (because the other variables can be inferred from these, conditioned on $\pi_2 :=\frontier$). We further define:
\begin{equation}\label{eq:payoff}
\rev^{(n)}_\cstring(\pi_1):= \rev^{(n)}_\cstring(\pi_1,\frontier), \quad \rev(\pi_1):=\rev(\pi_1,\frontier).
\end{equation}
A strategy $\pi^*$ is optimal if $\rev(\pi^*) = \max_{\pi} \rev(\pi)$. Thus $\frontier$ is a {\em Nash equilibrium} if $\frontier$ is an optimal strategy for Miner 1.

\vspace{1mm}\noindent\textbf{Proof-of-Work vs. Proof-of-Stake.} Our model, as stated, captures Proof-of-Stake protocols with perfect/trusted external randomness. Importantly, this means that once a miner knows they created a block during round $n$, they do not need to decide precisely the contents of that block until they publish it (because there is no computational difficulty to produce a block). In Proof-of-Work and the model of~\citet{eyal2014majority}, the miner of block $n$ must decide \emph{in round $n$} the contents of that block (because the contents are locked in as soon as the miner succeeds in proving work). Crucially, the miner \emph{must} decides the ancestor of block $n$ during time $t = n$ while in our model, the miner decides the ancestor of block $n$ any time $t \geq n$ before block $n$ is published. This is the only difference between the two models. Observe that any Proof-of-Work strategy is also valid in our model: this would just be a strategy which chooses the pointer for block $n$ in round $n$, and does not change it when publishing later. Example~\ref{example:mining-game} helps illustrate this distinction.
\begin{figure}[ht]
	\centering
    \begin{minipage}[c]{0.5\textwidth}
    \includegraphics[width=\textwidth]{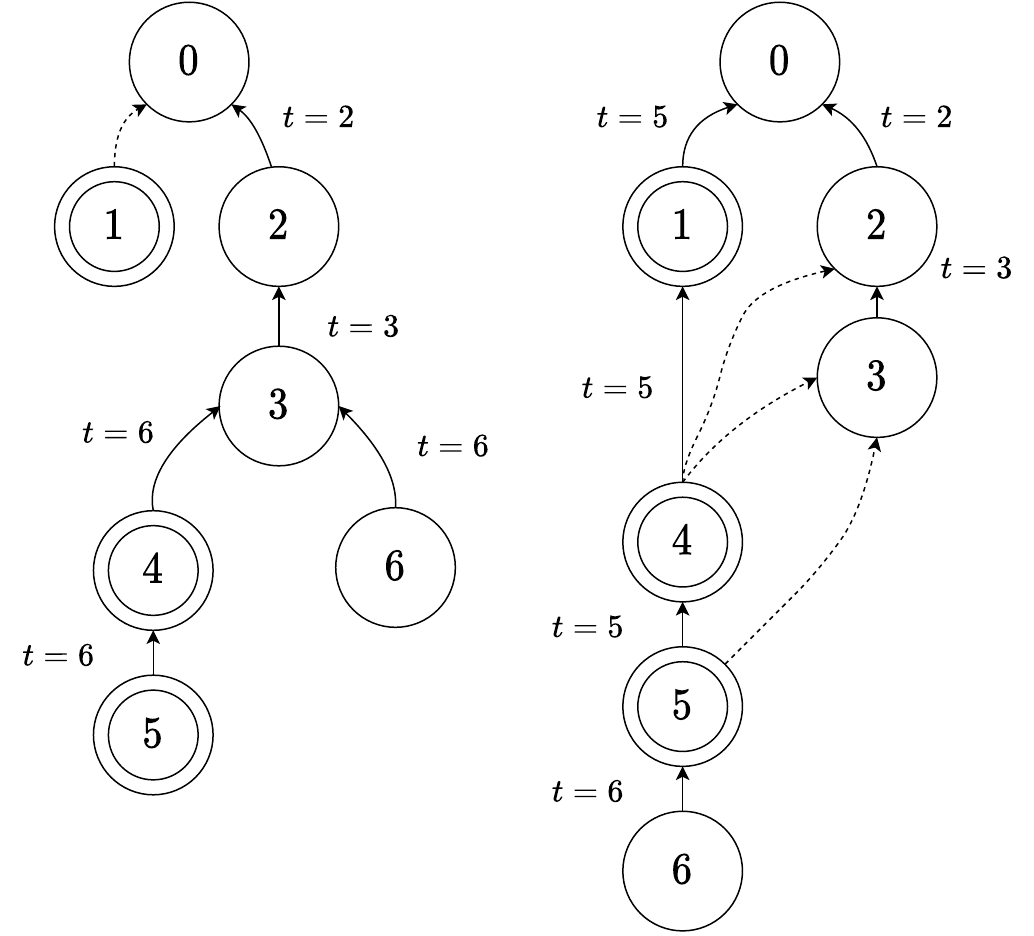}
    \end{minipage}
    \begin{minipage}[c]{0.45\textwidth}
    \caption{Diagram representing the mining game in Example~\ref{example:mining-game}. We use double circles for blocks owned by Miner 1 and single circles for blocks owned by Miner 2. The genesis block---block $0$---is not owned by neither Miner 1 nor Miner 2. The time stamps near the solid edges represent the round where the edge was published, and the number inside the circle represents the round when the block was created. The dashed edges represent some edges that could have been created (edges from any nodes in $\{4, 5\}$ to any nodes in $\{0, 1, 2, 3\}$ are also feasible --- i.e., any edge from a later node to an earlier node is feasible).}
    \label{fig:mining-game}
    \end{minipage}
\end{figure}
\begin{example}[Proof-of-Work vs Proof-of-Stake]\label{example:mining-game} 
Consider the following 6-round example where Miner 1 creates blocks 1, 4 and 5, and Miner 2 create blocks 2, 3 and 6 as depicted in Figure~\ref{fig:mining-game}. If Miner 1 was following the Selfish Mining strategy~\cite{eyal2014majority} (Definition~\ref{def:sm}), they would decide to create and withhold $1 \to 0$ (at time $t = 1$). Miner 2 would then publish $3 \to 2 \to 0$. At this point, Miner 1 would never attempt to publish $1 \to 0$ and we say block 1 becomes permanently orphan. Next, Miner 1 creates and withhold $4 \to 3$ (at $t = 4$) and $5 \to 4$ (at $t = 5$). When Miner 2 publishes $6 \to 3$, Miner 1 publishes $5 \to 4 \to 3$ (at $t = 6$), {\em forking} block $6$ from the longest path. This nets 2 blocks in the longest path for Miner 1, and 2 for Miner 2.

Here is a viable strategy in the Proof-of-Stake mining game: Miner 1 still withholds block 1 (\emph{but does not yet decide where it will point}), and it is still initially orphaned when Miner 2 publishes blocks 2 and 3. When Miner 1 creates block 4, they withhold it (\emph{but does not yet decide where it will point}). When they create block 5, they decide to publish block 4 (\emph{deciding only now to point to block 1}) and block 5 (\emph{deciding only now to point to block 4}). This creates a new longest path. Miner 2 then publishes $6 \to 5$. This nets 3 blocks in the longest path for Miner 1, and 1 for Miner 2. 

Importantly, observe that in the proof-of-work model, it would be exceptionally risky for Miner 1 to pre-emptively decide to point block 4 to block 1 at time $t=4$ without knowing that they will create block 5 (because maybe Miner 2 creates block 5, and then they would be an additional block behind). But in the proof-of-stake model, Miner 1 can wait to gather more information before deciding where to point. In particular, if they happened to instead create block 6 but not 5, they could have published $6 \to 4 \to 3$. In proof-of-stake, Miner 1 has the flexibility to make this decision later. In proof-of-work, they have to decide immediately whether to have block 4 pointing to 1 or 3.
\end{example}
\vspace{1mm}\noindent\textbf{Reminder of Notation.} Table~\ref{tab:notation} in Appendix~\ref{sec:table} is a reminder of our notation.
\subsection{Payoff as Fractional of Blocks in the Longest Path}\label{sec:steady-payoff}
\citet{eyal2014majority} motivates the use of the fraction of blocks as a miner's utility due to the difficulty adjustment in Bitcoin's PoW protocol: Bitcoin adjusts PoW difficult so that, on expectation, miners create one block every 10 minutes and the creator of each block receives new Bitcoins as block reward. Thus a miner maximizes their expected number of blocks in the longest path up to time $T$ by maximizing their expected fraction of blocks in the longest path up to time $T$.

For PoS, a random beacon outputs a random string at a fixed rate, independent of the blockchain state.\footnote{The National Institute of Standards and Technology (NIST) random beacon outputs 512 bits every 60 seconds~\cite{kelsey2019reference}.} Although difficult adjustment is absent in proof-of-stake, the probability of a miner creating the next block is proportional to $\alpha$, their fraction of coins in the system. Although $\alpha$ is approximately constant over short time horizons, over long time horizons, $\alpha$ will depend on the fraction of block rewards Miner 1 collects. Thus, in the {\em long-term}, Miner 1 will maximize block rewards by maximizing their fraction of blocks in the longest path.

\begin{figure}[H]
    \centering
    \includegraphics[width=0.6\textwidth]{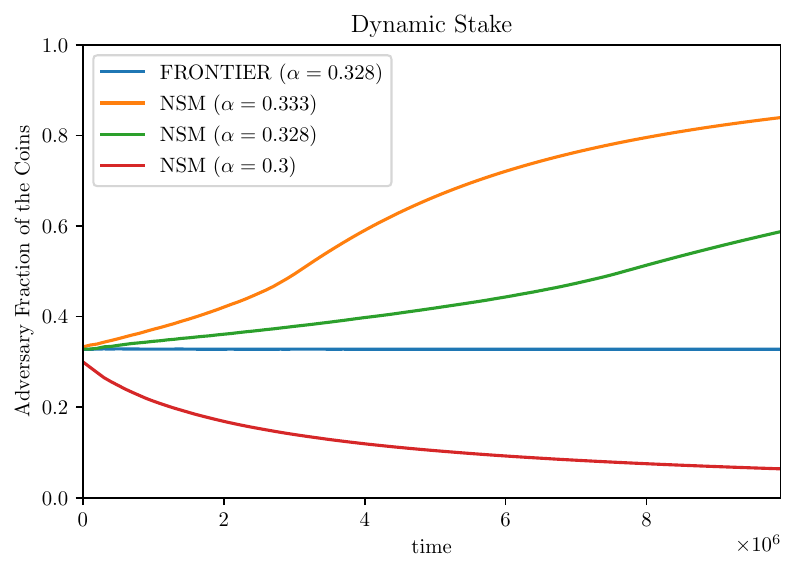}
    \caption{Simulation of Miner 1's dynamic stake over one million rounds when Miner 1 either follow $\frontier$ or the Nothing-at-Stake Selfish Mining ($\nsm$) strategy for distinct initial values of $\alpha$. As initial condition, the system has 100 thousand coins with Miner 1 initialing owning $\alpha$ fraction of the coins. We observe when following $\frontier$, Miner 1 cannot increase their fraction of the stake. We also observe $\nsm$ allows Miner 1 to increase their fraction of the stake when they initially own $32.8\%$ of the coins, but $\nsm$ is not profitable when Miner 1 owns less than $32.4\%$, as we find in Theorem~\ref{thm:enhanced-selfish-mining}.}
    \label{fig:dynamic-stake}
\end{figure}

In Figure~\ref{fig:dynamic-stake}, we simulate the fraction of coins owned by Miner 1 overtime when Miner 1 follows $\frontier$ or the Nothing-at-Stake Selfish Mining ($\nsm$ in Definition~\ref{def:nsm}) and Miner 2 follows $\frontier$. From the simulation, we observe $\nsm$ allows Miner 1 to add a higher fraction of blocks in the longest path when compared with $\frontier$ as long as Miner 1 owns more than $32.8\%$ of the coins. We confirm this empirical result in Theorem~\ref{thm:enhanced-selfish-mining}. This allows {\em Miner 1 to eventually own an arbitrarily large fraction of the coins}. Thus the security of a longest chain proof-of-stake protocol depend on a formal guarantees that {\em no strategy is more profitable than $\frontier$} when a profit maximizing miner is maximizing their fraction of blocks in the longest path. We accomplish this task in Theorem~\ref{thm:nash-equilibrium}.
\subsection{Capitulating a State}
For some states of the game, Miner 1 might follow a strategy that will never fork some blocks from the block tree. Then, it is safe to say that Miner 1 deletes those blocks from the state (or treats the one with highest height as the new genesis block) and consider a trimmed version of the state variable. As example, define $B_{0, 1}$ where Miner 2 creates and publishes block 1. Thus
\begin{equation}\label{eq:b_0_1}
B_{0, 1} := ((\{0, 1\}, \{1 \to 0\}), \emptyset, \emptyset).
\end{equation}
If Miner 1 never forks block 1, then it is safe to say that in the view of Miner 1 state $B_{0, 1}$ is equivalent to state $B_0$ (after treating block 1 as the new genesis block). Then, we say Miner 1 capitulates from state $B_{0, 1}$ to $B_0$. Since Miner 1 can induce the mining game to return to prior states, it is convenient to think of Miner 1 optimizing an underlying Markov Decision Process. Next, we provide a definition and formalize the payoff of the MDP in Appendix~\ref{sec:mdp-appendix}.

\vspace{1mm}\noindent\textbf{Markov Decision Process.} A {\em Markov Decision Process} (MDP) for the mining game where Miner 1 follows strategy $\pi$ and Miner 2 follows $\frontier$ is a sequence $(X_t)_{t \geq 0}$ where $X_t$ is a random variable representing the state by the end of round $t$ and before any actions have been take in round $t+1$. Unless otherwise stated, we initialize $X_0 = B_0$ (Equation~\ref{eq:b_0}). The game transitions from state $X_t$ to $X_{t+1}$ once the next block is created followed by Miner 2 taking their action followed by Miner 1 taking their action.

For a mining game $(X_t)_{t \geq 0}$ that starts at state $X_0 = B_0$, let
\begin{equation}\label{eq:tau}
\tau := \min\{t \geq 1: \text{State $X_t$ is equivalent to state $B_0$ in the view of Miner 1}\}
\end{equation}
be the first time step Miner 1 capitulates to state $B_0$. Similary, let $\tau'$ be the second time step Miner 1 capitulates to state $B_0$. Then, the sequences of rewards
$$r^k(X_0, X_1), r^k(X_1, X_2), \ldots, r^k(X_{\tau-1}, X_\tau),$$
$$r^k(X_{\tau}, X_{\tau + 1}), r^k(X_{\tau+1}, X_{\tau+2}), \ldots, X_{\tau'-1}, X_{\tau'})$$ 
are independent and identically distributed for $k = 1, 2$. One fundamental question is to understand if $\mathbb E[\tau] < \infty$ when Miner 1 is following an optimal strategy (that is, does Miner 1 capitulate to state $B_0$ at some point with probability $1$?). In the proof-of-stake setting, this is not obviously true, and Example~\ref{example:mining-game} gives some intuition why: while a proof-of-work Selfish Miner capitulates to state $B_0$ at round 3 (allowing Miner 2 to keep block 2 in the longest path), a Proof-of-Stake miner might prefer to wait for an opportunity to use block 1, and it is not a priori clear at what point it is safe to conclude that any optimal strategy would have given up on block 1 by now.
\subsection{Recurrence}
\begin{definition}[Recurrence]
Consider a mining game starting at state $X_0 = B_0$ where Miner 1 follows strategy $\pi$. Let $E$ be the event Miner 1 capitulates to state $B_0$ at some time $\tau < \infty$. We say $\pi$ is
\begin{itemize}
\item Transient if $\pr{E} < 1$.
\item Recurrent if $\pr{E} = 1$.
\item Null recurrent if it is recurrent and $\e{\tau} = \infty$.
\item Positive recurrent if it is recurrent and $\e{\tau} < \infty$.
\end{itemize}
\end{definition}

Observe Miner 1 never forks the longest chain when using $\frontier$. Thus Miner 1 capitulates to state $B_0$ at every time step and $\tau = 1$.
\begin{observation}\label{obs:frontier-positive-recurrent}
$\frontier$ is positive recurrent.
\end{observation}

For Proof-of-Work mining games, \citet{kiayias2016blockchain} and \citet{sapirshtein2016optimal} assumes Miner 1 will always follow a positive recurrent strategy. To motivate this technical assumption, they assume Miner 1 will never fork a block published by himself, which is sensible because Miner 1 can only mine on a single branch of the blockchain. This is not the case for Proof-of-Stake blockchains, and it is not a priori clear that Miner 1 will never fork a block that they created themselves. To see why this may occur, consider the following example.

First, define $B_{k, 0}$, for $k \geq 0$, as the state where Miner 1 creates and withhold blocks $\{1, 2, \ldots, k\} = [k]$. Thus
\begin{equation}\label{eq:b_k_0}
    B_{k,0} := ((\{0\}, \emptyset), [k], [k]).
\end{equation}
Then define $B_{1, 1}$ as the state after $B_{1, 0}$ where Miner 2 creates block 2 and publishes $2 \to 0$. Thus
\begin{equation}\label{eq:b_1_1}
B_{1, 1} := ((\{0, 2\}, \{2 \to 0\}), \{1\}, \{1\}).
\end{equation}
\begin{example}\label{example:fork-own-block}
Consider a game at state $B_{1, 1}$. After round 2, Miner 2 creates blocks $3, 4, \ldots, 9, 10, 12$ and publishes $12 \to 10 \to 9 \to \ldots \to 4 \to 3 \to 2$ and Miner 1 creates and withholds blocks $11, 13, 14, \ldots, 24$. At time step $13$, Miner 1 publishes $13 \to 11 \to 10$ (this follows the classical selfish mining strategy: it gives up on block 1, but publishes $11 \to 10$ and $13 \to 11$ to fork block 12). This is reasonable, because it is unlikely that Miner 1 can add block 1 to longest path and Miner 1 risks losing blocks 11 and 13 if Miner 2 creates and publishes block 14. However, in the event Miner 1 is lucky and creates blocks $14, 15, \ldots, 24$, Miner 1 can fork all blocks from the current longest path (including his own blocks $11$ and $13$), resulting in a new longest path with blocks $1, 14,  15, \ldots, 24$ (consisting entirely of Miner 1's blocks). Indeed, upon creating block 14, Miner 1 need not immediately decide whether to make its predecessor 13, or whether to wait and see if they get an extremely lucky run to override the entire chain.

Note that we are not claiming that this is the optimal decision for Miner 1 from this state, or even that an optimal strategy may ever find itself in this state.\footnote{For example, if this were the optimal decision from this state, it would likely be because $\alpha$ is small, and Miner 1 should just take whatever opportunities they have to publish blocks. However, if $\alpha$ is small, that may mean it is better for Miner 1 to just be honest, and they would never find themselves in this situation. The point is that some quantitative comparison is necessary in order to determine whether the optimal strategy for Miner 1 would ever take this action.} However, this example helps demonstrate that a significantly richer space of strategies are potentially optimal in our model, as compared to proof-of-work.
\end{example}

This example shows that we must be careful to not exclude optimal strategies when claiming any restrictions on strategies considered by Miner 1. We will eventually address this by introducing the notion of a \emph{checkpoint}, Section~\ref{sec:checkpoint}, and prove that there are \emph{some} conditions that allow us to claim that any optimal strategy for Miner 1 will not fork a checkpoint (however, we do not prove that Miner 1 will never consider forking their own blocks).

\section{Enhancing Selfish Mining with Nothing-at-Stake}\label{sec:enhanced-selfish-mining}

In this section, we show explicitly an strategy that outperforms $\frontier$ even when $\alpha = 0.325$. From \citet{sapirshtein2016optimal}, $\frontier$ is optimal for Proof-of-Work mining games when Miner 1 has mining power $\alpha \leq 0.329$ (i.e., $\pow \approx 0.329$). Thus our strategy witnesses that Proof-of-Stake mining games admits strategies that are more profitable than any strategy in a Proof-of-Work mining game -- that is, will establish $\pos < \pow$.

Our strategy will be a subtle modification from the Selfish Mining strategy of \citet{eyal2014majority} that leverages the Nothing-at-Stake vulnerability in Proof-of-Stake blockchains.\footnote{The nothing-at-stake vulnerability refers to the fact the algorithm for which a miner can verify a block validity is computationally efficient. Thus miners have no cost to choosing the block content (including its ancestor) at the moment the block is about to be published
.}

\begin{figure}[H]
	\centering
    \includegraphics[width=\textwidth]{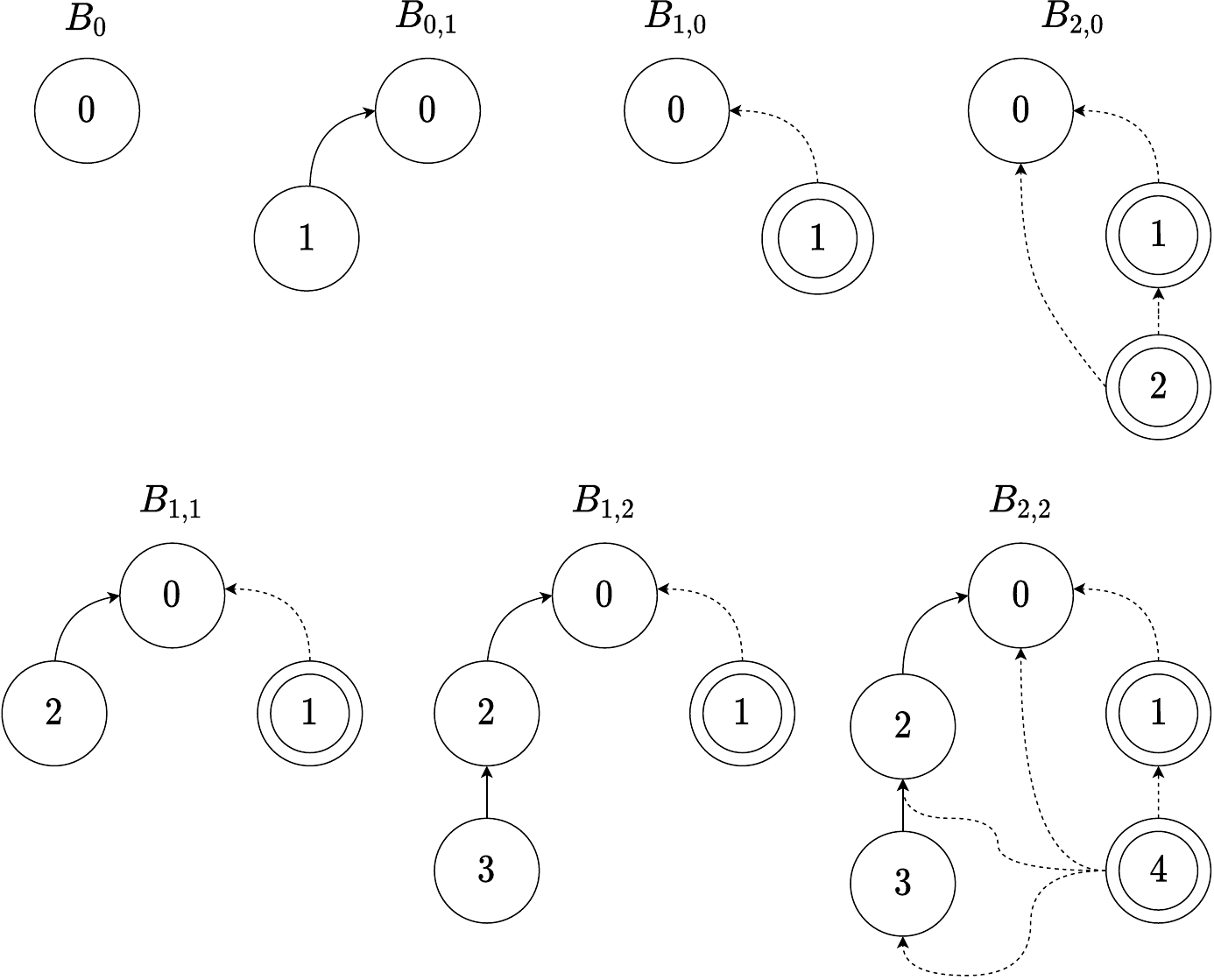}
    \caption{Diagram representing states $B_0$, $B_{0, 1}$, $B_{0, 2}$, $B_{1, 0}$, $B_{1, 1}$, $B_{1, 2}$ and $B_{2, 2}$. Block 0 has no owner. All double circles are Miner 1's hidden blocks. All circles are Miner 2's published blocks. Dashed lines are edges Miner 1 can publish.}
    \label{fig:states}
\end{figure}

Let's first define the states of interest for our strategy. Recall $B_0$ is the state where the block tree contains only the genesis block; $B_{1, 0}$ is the state after Miner 1 creates and withholds block 1 (Equation~\ref{eq:b_k_0}); $B_{2, 0}$ is the state after Miner 1 creates and withholds blocks 1 and 2 (Equation~\ref{eq:b_k_0}); $B_{0, 1}$ is the state after Miner 2 publishes $1 \to 0$ (Equation~\ref{eq:b_0_1}); $B_{1, 1}$ is the state after $B_{1,0}$ if Miner 2 publishes $2 \to 0$ (Equation~\ref{eq:b_1_1}). Additionally, define the following states.
\begin{itemize}
\item $B_{1,2}$ is the state after $B_{1, 1}$ if Miner 2 creates block $3$ and publishes $3 \to 2$:
\begin{equation}\label{eq:b_1_2}
B_{1, 2} := ((\{0, 2, 3\}, \{3 \to 2 \to 0\} ), \{1\}, \{1\}).
\end{equation}
\item $B_{2, 2}$ is the state after $B_{1, 2}$ if Miner 1 creates and withhold block 4:
\begin{equation}\label{eq:b_2_2}
    B_{2, 2} := ((\{0, 2, 3\}, \{3 \to 2 \to 0\} ), \{1, 4\}, \{1, 4\}).
\end{equation}
\end{itemize}
These and other relevant states are depicted in Figure~\ref{fig:states}.

\begin{figure}[H]
	\centering
    \includegraphics[width=0.9\textwidth]{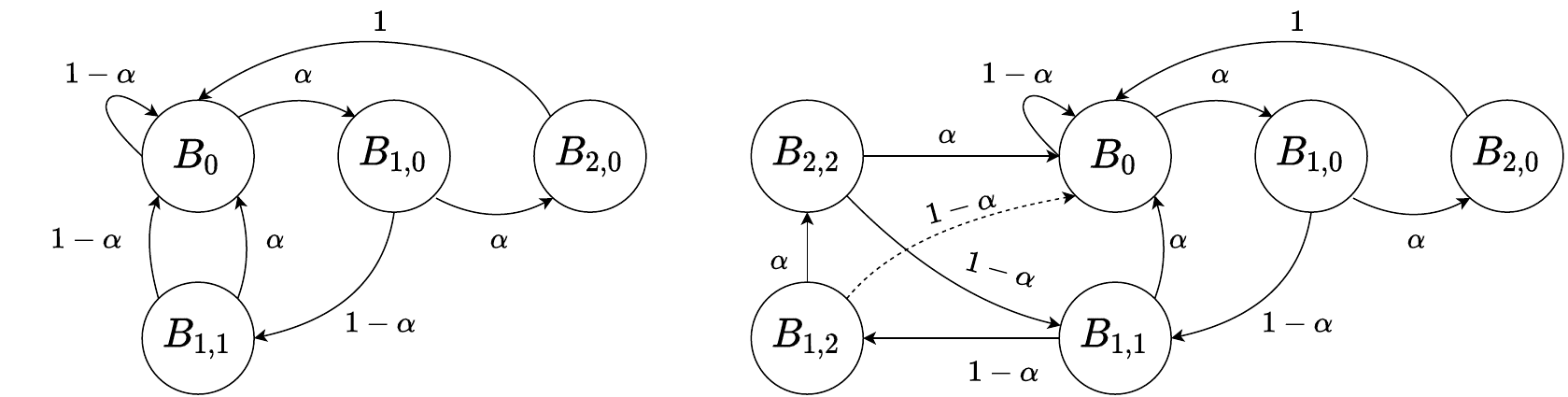}
	\caption{Markov chain representing the Selfish Mining strategy (left) and the Nothing-at-Stake Selfish Mining strategy (right).}
    \label{fig:sm}
\end{figure}

\begin{definition}[Selfish Mining \cite{eyal2014majority}]\label{def:sm}
Let $(X_t)_{t \geq 0}$ be a mining game starting at state $X_ 0 = B_0$. Miner 1 uses the {\em Selfish Mining ($\sm$)} strategy, Figure~\ref{fig:sm}, which takes the following actions:
\begin{itemize}
    \item Wait at states $B_0$ and $B_{1, 0}$.
    \item At state $B_{0, 1}$, capitulate to state $B_0$.
    \item If $X_2 = B_{1, 1}$ and Miner 1 creates block 3, publishes $3 \to 1 \to 0$, then capitulates to state $B_0$.
    \item If $X_2 = B_{1, 1}$ and Miner 2 creates block 3 and publishes $3 \to 2$, then Miner 1 capitulates to state $B_0$.
    \item If $X_2 = B_{2, 0}$, Miner 1 plays $\wait$ until the first time step $\tau \geq 3$ where $|T_2(X_\tau)| = |T_1(X_\tau)| - 1$. At time step $\tau$, Miner 1 publishes all of $\unpublished(X_\tau) = T_1(X_\tau)$ pointing to $0$ and forking $T_2(X_\tau)$, then capitulates to state $B_0$.
\end{itemize}
\end{definition}

\begin{theorem}[Equation 8 in \cite{eyal2014majority}]\label{thm:selfish-mining}
For the selfish mining strategy
$$\rev(\sm) = \frac{\alpha^2(4\alpha^2-9\alpha+4)}{\alpha^3-2\alpha^2-\alpha+1}.$$
Moreover, $\rev(\sm) > \alpha = \rev(\frontier)$ for $\alpha > 1/3$.
\end{theorem}
Our Nothing-at-Stake Selfish Mining is similar, with one key difference: In Selfish Mining, Miner 1 capitulates immediately after a loss (specifically, if Miner 2 creates block 3 and publishes $3 \to 2$ from state $B_{1,1}$, Miner 1 immediately accepts that block 1 is now permanently orphaned and capitulates to $B_{0}$). Nothing-at-Stake Selfish Mining instead remembers this orphaned block, and considers bringing it back later. Importantly, Nothing-at-Stake Selfish Mining can wait to see whether it finds many blocks (in which case it will try to publish the block 1) or not (in which case it will let block 1 remain orphaned) before deciding what to do. Below is a formal description.
\begin{definition}[Nothing-at-Stake Selfish Mining]\label{def:nsm}
Let $(X_t)_{t \geq 0}$ be a mining game starting at state $X_0 = B_0$. Miner 1 uses the {\em Nothing-at-Stake Selfish Mining ($\nsm$)} strategy , right of Figure~\ref{fig:sm}, which takes the following actions:

\begin{itemize}
    \item Wait at states $B_0$, $B_{1, 0}$ and $B_{1, 2}$.
    \item At state $B_{0, 1}$, capitulate to state $B_0$.
    \item If $X_t = B_{1, 1}$ and Miner 1 creates block 3, publishes $3 \to 1 \to 0$, then capitulates to state $B_0$.
    \item If $X_t = B_{1, 2}$ and Miner 2 creates block 4 and publishes $4 \to 3$, then Miner 1 capitulates to state $B_0$.
    \item If $X_t = B_{2, 2}$ and Miner 1 creates block 5, publishes $5 \to 4 \to 1 \to 0$, then Miner 1 capitulates to state $B_0$.
    \item If $X_t = B_{2, 2}$ and Miner 2 creates block 5 and publishes $5 \to 3$, Miner 1 capitulates to state $B_{1, 1}$. That is, Miner 1 allows Miner 2 to walk away with blocks 2 and 3 and forgets about unpublished block 1, but remembers unpublished block 4 in the hope of forking block 5 in the future. The resulting state is equivalent to $B_{1, 1}$ since we can relabel block 3 as 0, 4 as 1 and 5 as 2.
    \item If $X_2 = B_{2, 0}$, Miner 1 plays $\wait$ until the first time step $\tau \geq 3$ where $|T_2(X_\tau)| = |T_1(X_\tau)| - 1$. At time step $\tau$, Miner 1 publishes all of $\unpublished(X_\tau) = T_1(X_\tau)$ pointing to $0$ forking blocks $T_2(X_\tau)$, then Miner 1 capitulates to state $B_0$.
\end{itemize}
\end{definition}
Let's quickly understand why this strategy is not possible in the proof-of-work model. Zero in on Miner 1's behavior at $B_{2,2}$. If Miner 1 creates block 5, Miner 1 publishes blocks $1,4,5$, and in particular has their block $4$ point to block $1$. However, if Miner 2 creates block 5, Miner 1 capitulates to state $B_{1, 1}$. From here, if Miner 1 creates block 6, they immediately publish $4$ and $6$, \emph{having block $4$ point to block $3$}. 

That is, while using this strategy, Miner 1 does not decide where block 4 will point upon mining it, but only upon publishing it. In a Proof-of-Work blockchain, Miner 1 must commit to the ancestor of block 4 at time step 4, so this strategy cannot be used. Intuitively, a nothing-at-stake selfish miner remembers an orphaned block to see if they might get lucky in the future. Importantly, in the PoS model they can \emph{still} wait to decide whether to try and bring this block into the longest chain \emph{even after finding their next block} (but before deciding where to publish it). This extra power enables not only a slight improvement over standard selfish mining but also a strategy the is strictly better than any other valid strategies for the Proof-of-Work mining game.
\begin{theorem}\label{thm:enhanced-selfish-mining}
For the nothing-at-stake selfish mining strategy,
$$\rev(\nsm) = \frac{4 \alpha^2 - 8 \alpha^3 - \alpha^4 + 7 \alpha^5 - 3 \alpha^6}{1 - \alpha - 2 \alpha^2 + 3 \alpha^4 - 3 \alpha^5 + \alpha^6}.$$
Moreover $\rev(\nsm) > \alpha$ for $\alpha > 0.324718$.
\end{theorem}
Recall \citet{sapirshtein2016optimal} estimates $\alpha^{PoW} \approx 0.329$. Thus Theorem~\ref{thm:enhanced-selfish-mining} implies
$$\alpha^{PoS} < 0.325 < 0.329 \approx \alpha^{PoW}.$$

\begin{figure}[H]
    \centering
    \begin{minipage}[t]{0.475\textwidth}
    \includegraphics[width=\textwidth]{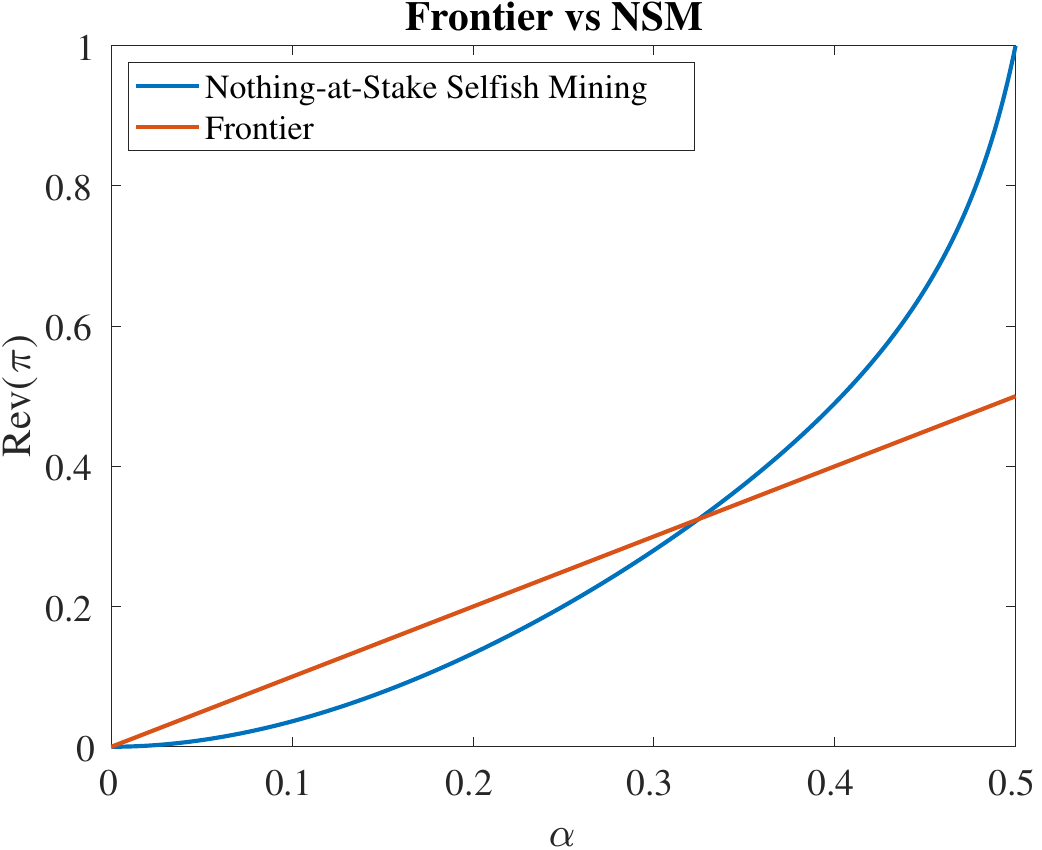}
    \caption{Payoff comparison between $\frontier$ and Nothing-at-Stake Selfish Mining.}
    \label{fig:payoff-nas-selfish}
    \end{minipage}
    \begin{minipage}{0.025\textwidth}\ 
    \end{minipage}
    \begin{minipage}[t]{0.475\textwidth}
    \includegraphics[width=\textwidth]{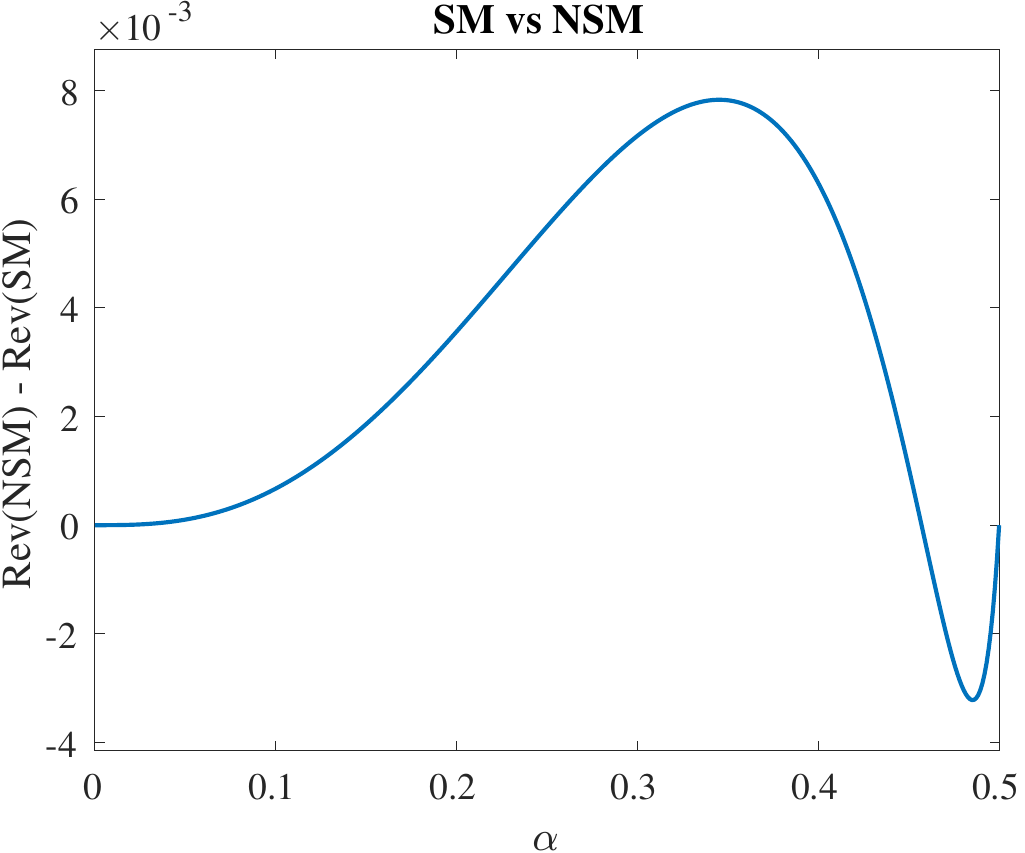}
    \caption{Payoff comparison between Selfish Mining and Nothing-at-Stake Selfish Mining. Observe $\nsm$ is slightly better than $\sm$ for $\alpha$ close to $1/3$, but $\sm$ outperforms $\nsm$ for $\alpha > 0.458$.}
    \label{fig:payoff-comparison}
    \end{minipage}
\end{figure}

Interestingly, Nothing-at-Stake Selfish Mining is not better than Selfish Mining for all $\alpha$. In Figure~\ref{fig:payoff-comparison}, by plotting the difference $\rev(\nsm) - \rev(\sm)$ as a function of $\alpha$, we observe Selfish Mining is better than Nothing-at-Stake Selfish Mining for $\alpha > 0.458$.

To get intuition why this happens, consider how $\sm$ and $\nsm$ differs in the event Miner 1 creates blocks 1, 4, 5 and Miner 2 creates blocks 2 and 3. By the end of the 5-th round, $\sm$ is at state $B_{2, 0}$ while $\nsm$ just moved from state $B_{2, 2}$ to $B_0$. The main intuition is that being at state $B_{2, 0}$ is a highly profitable for Miner 1 when $\alpha$ is large. In the proof of Theorem~\ref{thm:selfish-mining}, Appendix~\ref{sec:selfish-appendix}, we show Miner 1 creates, on expectation, $\frac{\alpha}{1-2\alpha}$ blocks from the moment the game reaches state $B_{2, 0}$ until the moment the game first returns to state $B_0$. Moreover, $\sm$ has a bigger probability of being at state $B_{2, 0}$ than $\nsm$ because $\sm$ capitulates to state $B_0$ once it reaches state $B_{1, 2}$ but $\nsm$ does not.

\section{Trimming the Strategy Space}\label{sec:trimming-strategy}

Analyzing the revenue of \emph{all} possible strategies for Miner 1 is quite unwieldy. Therefore, our first goal is to reduce the space of possible strategies to ones which are simpler to analyze \emph{while guaranteeing that this simpler space still contains an optimal strategy}. We accomplish this through a series of reductions. This section provides a series of three ``elementary'' reductions which build upon each other. That is, the conclusions in each section should not be surprising, although it is challenging to rigorously prove this (examples throughout Appendix~\ref{sec:trimming-proof} are used to highlight the challenges). Sections~\ref{sec:timeserving} through~\ref{sec:lcm} provide our three reductions. Section~\ref{sec:wrapup} provides the main theorem statement of this section: there is an optimal \emph{trimmed} strategy. Many proofs are omitted, and can be found in Appendix~\ref{sec:trimming-proof}.

\subsection{Step 1: Timeserving}\label{sec:timeserving}

We first show that, w.l.o.g., every strategy only publishes blocks which will be ancestors of the longest chain at the end of that round. 

\begin{definition}[Timeserving]
The action $\publishset(V', E')$ is \emph{Timeserving} if all blocks in $V'$ immediately enter the longest chain (formally: if the action is taken during round $n$, then $V' \subseteq A(\chain(n))$). A strategy is Timeserving if when played against $\frontier$, with probability 1, all $\publishset(V', E')$ actions it takes are Timeserving.
\end{definition}

It is easy to see that $\frontier$ is itself Timeserving: it publishes at most a single block at a time, and that block is the new unique longest chain. We first argue that there exists an optimal strategy against $\frontier$ which is also Timeserving.

\begin{theorem}[Timeserving]\label{thm:timeserving}
For any strategy $\pi$, there is a strategy $\tilde{\pi}$ that is Timeserving, takes a valid action at every step, and satisfies $\rev_{\cstring}^{(n)}(\tilde{\pi}) = \rev_{\cstring}^{(n)}(\pi)$ for all $\cstring$ and $n \in \mathbb N$.
\end{theorem}

We now state three basic properties of Timeserving strategies.

\begin{observation}\label{obs:timeserving}
If $\pi$ is Timeserving, then
\begin{enumerate}[label=(\roman*)]
    \item \label{label:timeserving-3} Whenever $\pi$ publishes blocks, it publishes a single path. Formally, whenever $\pi$ takes action $\publishset(V', E')$ in round $n$, with $V'= \{b_1,\ldots, b_k\}$ ($b_i < b_{i+1}$ for all $i$), then $E'$ contains an edge $b_{i+1} \to b_i$ for all $i \in [k-1]$, and an edge $b_1 \to b$ for some $b \in V$. 
    \item \label{label:timeserving-1} There are never two leaves of the same height. Formally, for all leaves $q\neq \tilde{q} \in V$, $h(q) \neq h(\tilde{q})$.
    \item \label{label:timeserving-2} Whenever $\pi$ forks, it publishes at least two blocks. Formally, whenever $\pi$ takes the action $\publishset(V', E')$ which removes the old longest chain from the longest path, then $|V'| \geq 2$.
\end{enumerate}
\end{observation}

Observation~\ref{obs:timeserving} gives us some nice structure about Timeserving strategies (and Theorem~\ref{thm:timeserving} asserts that it is w.l.o.g. to study such strategies). In particular, we only need to consider strategies which publish a single path at a time. Formally, we may w.l.o.g. replace the action $\publishset(V', E')$ with the action:

\begin{definition}[PublishPath] Taking action $\publishpath(V', u)$ with $u \in V$  and $V' \subseteq \unpublished$ is equivalent to taking action $\publishset(V', E')$, where $E'$ contains an edge from the minimum element of $V'$ to $u$, and an edge from $v$ to the largest element of $V'$ strictly less than $v$, for all other $v \in V'$. 
\end{definition}

\subsection{Step 2: Orderly}\label{sec:orderly}

Section~\ref{sec:timeserving} provides structure on \emph{when} we may assume blocks are announced, but it does not yet provide structure on \emph{which blocks} are announced. Specifically, for all we know right now it could still be that when a strategy chooses to take action $\publishpath(V', u)$, and $|V'|=k$, the precise $k$ blocks it chooses to publish matter (e.g. in state $B_{2, 0}$ it could choose to publish $2 \to 0$ versus $1 \to 0$). Our next reduction shows that it is without loss to consider only strategies which are \emph{Orderly}, and always publish the earliest legal blocks. Intuitively, this gives the strategy more flexibility later on. For simplicity of notation, we introduce the terms $\min^{(k)}\{S\} \subseteq S$ to refer to the $\min\{k, |S|\}$ smallest elements in $S$ and $\max^{(k)}\{S\} \subseteq S$ to refer to the $\min\{k, |S|\}$ largest elements in $S$.

\begin{definition}[Orderly]
The action $\publishpath(V', u)$ is \emph{Orderly} if $V' = \min^{(|V'|)} (\unpublished \cap (u, \infty))$. That is, an action is Orderly if it publishes the smallest $|V'|$ blocks it could have possibly published on top of $u$. A strategy is Orderly if when played against $\frontier$, with probability $1$, all $\publishpath(\cdot, \cdot)$ actions it takes are Orderly.
\end{definition}

\begin{theorem}[Orderly]\label{thm:orderly}
Let $\pi$ be any Timeserving strategy. Then there is a valid, Timeserving, Orderly strategy $\tilde \pi$ that satisfies $\rev_\gamma^{(n)}(\tilde{\pi}) = \rev_\gamma^{(n)}(\pi)$ for all $\gamma$ and $n \in \mathbb N$.
\end{theorem}

We conclude this section by noting that, after restricting attention to Orderly strategies, we can further replace the action $\publishpath(V', u)$ with the action:

\begin{definition}[Publish] Taking action $\publish(k, u)$ with $k \in \mathbb N_+$ and $u \in V$ is equivalent to taking the action $\publishpath(\min^{(k)} (\unpublished \cap (u, \infty)), u)$.
\end{definition}

\subsection{Step 3: Longest Chain Mining}\label{sec:lcm}

We now have structure on \emph{when} blocks are published, and \emph{which} blocks are published, but not yet on \emph{where} those blocks are published. Specifically, an \emph{orphaned chain} is a path in $\tree$ that used to be part of the longest path $A(\chain)$ but was overtaken by another path. Intuitively, a chain can only be orphaned by Miner 1 and if Miner 1 is playing according to an optimal strategy, publishing blocks which build on top of orphaned chains should be sub-optimal. We define a strategy as \emph{Longest Chain Mining} if it never publishes on top of a block in an orphaned chain.
\begin{definition}[Longest Chain Mining]
Action $\publish(k,u)$ is \emph{Longest Chain Mining} (LCM) if $u \in A(\chain)$ is a block in the longest path. That is, an action is LCM if it builds on top of some block within the longest path (not necessarily the leaf). A strategy is LCM if, with probability $1$, every $\publish$ action it takes against $\frontier$ is LCM.
\end{definition}
Previous work on Proof-of-Work mining games \cite{kiayias2016blockchain, sapirshtein2016optimal} assume all strategies are LCM. For Proof-of-Stake mining games, Theorem~\ref{thm:lcm} proves that it is w.l.o.g. to assume an LCM strategy.
\begin{theorem}[LCM]\label{thm:lcm}
Let $\pi$ be any Timeserving, Orderly strategy. Then there is a $\tilde{\pi}$ that is Timeserving, Orderly, LCM, takes a valid action at every step, and satisfies $\rev^{(n)}_{\cstring}(\tilde{\pi}) \geq \rev^{(n)}_{\cstring}(\pi)$ for all $\cstring$ and $n \in \mathbb N$.
\end{theorem}

\subsection{Step 4: Trimmed}\label{sec:wrapup}

With Theorems~\ref{thm:timeserving},~\ref{thm:orderly} and~\ref{thm:lcm}, we can immediately conclude that there exists an optimal strategy satisfying several structural properties. We wrap up by showing one final property, and will show that there exists an optimal strategy which is \emph{Trimmed}.

\begin{definition}[Trimmed Action] Action $\publish(k, v)$ is \emph{Trimmed} if $v \in A(\chain)$ and whenever $v$ is not \emph{the} longest chain (that is, $v \neq \chain$), and $u$ is the unique node in $A(\chain)$ with an edge to $v$, then $u$ was created by Miner $2$ (that is, $u \in T_2$).
\end{definition}

Put another way, every $\publish(k, v)$ either builds on top of the longest chain (in which case it is vacuously Trimmed), or kicks out the successors of $v$. In the latter case, an action is Trimmed if and only if the \emph{minimum} successor of $v$ was created by Miner $2$.

\begin{definition}[Trimmed Strategy] A strategy is \emph{Trimmed} if every action it takes is either $\wait$ or Trimmed.
\end{definition}

We now conclude our main theorem of this section.

\begin{theorem}[Trimming]\label{thm:trimmed}
For all strategies $\pi$, there is a Trimmed strategy $\tilde{\pi}$ that take valid actions in every step, and $\rev^{(n)}_{\cstring}(\tilde{\pi}) \geq \rev^{(n)}_{\cstring}(\pi)$ for all $\cstring$ and $n \in \mathbb N$. 
\end{theorem}

\section{Trimming the State Space}\label{sec:trimming-state}

So far, we have greatly simplified strategies which we need to consider. However, we still have not even established that there exists an optimal strategy which is \emph{recurrent}. That is, for all we know so far, the optimal strategy might need to store not only the entire longest chain, but also all blocks which have ever been published, and all unpublished blocks which they ever created. The goal in this section is to establish that an optimal strategy exists which is recurrent: it will eventually (with probability $1$) reach a ``checkpoint'' which the strategy treats as a new genesis block that will never be overridden.

\subsection{Checkpoints and Weak Recurrence}\label{sec:checkpoint}

We iteratively define a sequence of blocks $P_0, P_1,\ldots$ in the longest path $A(\chain)$ to be checkpoints as follows.
\begin{definition}[Checkpoints]\label{def:checkpoint} Based on the current state, checkpoints are iteratively defined as follows.
\begin{itemize}
    \item The first checkpoint, $P_0$, is the genesis block.
    \item If $P_{i-1}$ is undefined, then $P_{i}$ is undefined as well. 
\item If $P_{i-1}$ is defined, then $v$ is a \emph{potential $i^{th}$ checkpoint} if:
\begin{itemize}
\item $v > P_{i-1}$.
\item $v \in A(\chain)$. 
\item Among blocks that Miner $1$ created between $P_{i-1}$ and $v$ (including $v$, not including $P_{i-1}$), more are in the longest chain than unpublished. That is, $|A(\chain) \cap (P_{i-1},v] \cap T_1| \geq |\unpublished_1 \cap (P_{i-1},v]|$.
\end{itemize}
\item If there are no potential $i^{th}$ checkpoints, then $P_i$ is undefined.
\item Else, then $P_{i}$ is defined to be the minimum potential $i^{th}$ checkpoint.
\end{itemize}
\end{definition}
Note that each $P_i$ is again a random variable, meaning that a priori $P_i$ might change over time, including from undefined to defined. For example, $P_i(n)$ would denote the $i^{th}$ checkpoint, as defined by the state after the conclusion of the $n^{th}$ round (we will later prove that there exists an optimal strategy which never changes or undefines $P_i$ once it is defined. But this will be a result, and not a definition).
\begin{example}\label{example:checkpoint}
\begin{figure}[H]
\begin{minipage}{0.5\textwidth}
\includegraphics[width=\textwidth]{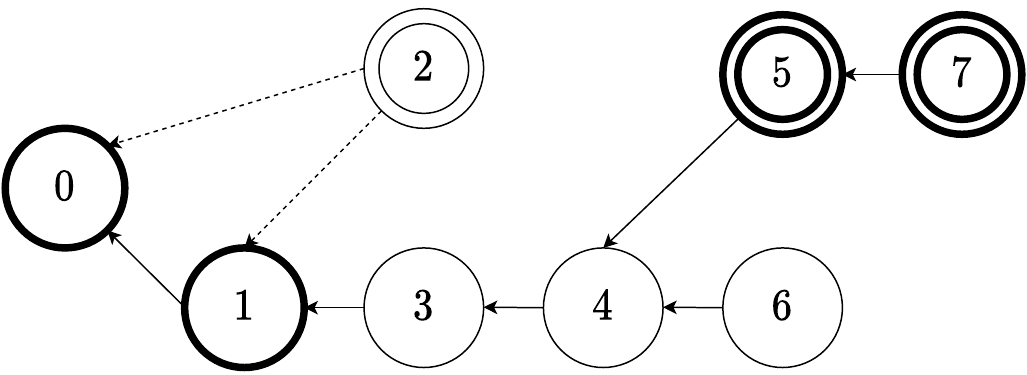}
\end{minipage}
\begin{minipage}{0.05\textwidth}\ 
\end{minipage}
\begin{minipage}{0.40\textwidth}
\caption{Example of a state and its checkpoints. Blocks with thicker lines denote blocks that are checkpoints and blocks with thinner lines denote blocks that are not checkpoints. From Definition~\ref{def:checkpoint}, blocks 0, 1, 5, and 7 are checkpoints.}
\label{fig:checkpoint}
\end{minipage}
\end{figure}
Consider the state in Figure~\ref{fig:checkpoint} where blocks 0, 1, 5, and 7 are the checkpoints. By definition, block 0 is the base-case and is always a checkpoint. Block 1 is a checkpoint because Miner 1 has no unpublished blocks in the interval $(0, 1]$. Block 2 is unpublished and thus not a checkpoint. Block 3 is not a checkpoint because Miner 1 has one unpublished block in the interval $(1, 3]$ and zero published block in the path from $3 \to 1$ (not counting block 1). From a similar reasoning, blocks 4 and 6 are not checkpoints. Block 5 is a checkpoint because Miner 1 has one unpublished block in the interval $(1, 5]$ and one block in the path $5 \to 4 \to 3 \to 1$ (not counting block 1). Block 7 is a checkpoint because Miner 1 has no unpublished blocks in the interval $(5, 7]$.
\end{example}
The main result of this section is stated below, and claims that there exists an optimal strategy which treats checkpoints like the genesis block.
\begin{definition}[Checkpoint Recurrent] A strategy $\pi$ is \emph{Checkpoint Recurrent} if when $\pi$ is played against $\frontier$:
\begin{itemize}
\item For all $i \in \mathbb N$, if $P_i$ changes from undefined to defined, $P_i$ never changes again (in particular, this implies that once $P_i$ is defined, it remains in $A(\chain)$ forever).
\item Immediately when $P_i$ becomes defined, neither player has any unpublished blocks $>P_i$.
\end{itemize}
\end{definition}

When bullet one is satisfied, checkpoints are never overridden. Given that bullet one holds, bullet two implies that immediately when $P_i$ is defined, it is essentially a genesis block (because bullet one holds, no unpublished blocks $< P_i$ can ever enter the longest chain. If bullet two also holds, then there are no unpublished blocks $> P_i$, so there are no relevant unpublished blocks, and $P_i$ is in the longest chain forever, just like the genesis block in round $0$). This implies that when optimizing over Checkpoint Recurrent strategies, it suffices to consider only strategies that reset its state space whenever a new checkpoint is defined. That is, whenever a new checkpoint is defined Miner 1 capitulates to state $B_0$.

\begin{theorem}[Weak-Recurrence]\label{thm:weak-recurrence}
There exists an optimal strategy which is checkpoint recurrent.
\end{theorem}

The weak-recurrence theorem provides a useful tool to reduce the state space of optimal strategies; however, it does not say how often (if ever) the block tree reaches a new checkpoint. Fortunately, each new checkpoint give us important information about the payoff of a strategic miner: \emph{if the block tree never reaches a checkpoint, then at all times Miner 1 has at least half of their blocks unpublished}. Next, we check such strategies are not better than $\frontier$ before diving into the proof of the weak-recurrence theorem.
\begin{proposition}\label{prop:checkpoint-recurrent}
If $\pi$ is checkpoint recurrent and $P_1$ is never defined, then $\rev(\pi)\leq \rev(\frontier)$.
\end{proposition}

As a first step toward proving upper bounds in the revenue, we will require a simply but useful fact about rate of growth of the block tree.
\begin{lemma}[Minimum Growth Rate]\label{lemma:tree-growth}
For any mining game starting at state $X_0 = B_0$,
\begin{equation}
\liminf_{n \to \infty} \frac{h(\chain(X_n))}{n} \geq 1-\alpha, \quad \text{with probability 1}.
\end{equation}
\end{lemma}
\begin{corollary}\label{cor:opt-revenue}
For any optimal strategy $\pi$, $\rev(\frontier) = \alpha \leq \rev(\pi) \leq \frac{\alpha}{1-\alpha}$.
\end{corollary}
Both Lemma~\ref{lemma:tree-growth} and Corollary~\ref{cor:opt-revenue} will be useful to prove Proposition~\ref{prop:checkpoint-recurrent}. We need to understand one property of checkpoints, and then we can complete the proof. Intuitively, Proposition~\ref{prop:checkpoint-inequality} and Corollary~\ref{cor:checkpoint-reward-bound} just apply the definition of checkpoints to relate the number of blocks that Miner 1 has unpublished vs.~published in the longest path.
\begin{proposition}\label{prop:checkpoint-inequality}
For all $v \in A(\chain)$,
\begin{enumerate}[label=(\roman*)]
\item \label{prop:checkpoint-inequality-1} If $v$ is a checkpoint, then for all checkpoints $P_i > v$, $|A(\chain) \cap (v, P_i] \cap T_1| \geq |\unpublished \cap (v, P_i)|$.
\item \label{prop:checkpoint-inequality-2} If $v$ is not a checkpoint and $P_i = \max\{ P_j : P_j < v\}$, then $|A(\chain) \cap (P_i,  v] \cap T_1| < |\unpublished \cap (P_i,  v]|$.
\item \label{prop:checkpoint-inequality-3} If $v$ is not a checkpoint, then for all checkpoints $P_i > v$, $|A(\chain) \cap (v,  P_i] \cap T_1| > |\unpublished \cap (v,  P_i]|$.
\end{enumerate}
\end{proposition}
\begin{corollary}\label{cor:checkpoint-reward-bound}
Suppose $v \in A(\chain)$ is not a checkpoint and let $P_i$ be the highest checkpoint below $v$. Then Miner 1 publishes in the longest path less than half of all blocks they created from time $P_i+1$ to $v$. That is, $|A(\chain) \cap (P_i, v] \cap T_1| < \frac{|T_1 \cap (P_i, v]|}{2}$.
\end{corollary}
\begin{proof}
Bullet \ref{prop:checkpoint-inequality-2}, Proposition~\ref{prop:checkpoint-inequality}, implies
$$|A(\chain) \cap (P_i, v] \cap T_1| +  |\unpublished \cap (P_i, v]| > 2|A(\chain) \cap (P_i, v] \cap T_1|.$$
Suppose for contradiction $|A(\chain) \cap (P_i, v] \cap T_1| \geq \frac{|T_1 \cap (P_i, v)|}{2}$. Then, the number of unpublished blocks plus blocks in the longest path would be strictly bigger than the number of blocks Miner 1 created, a contradiction.
\end{proof}
\begin{proof}[Proof of Proposition~\ref{prop:checkpoint-recurrent}]
From the strong law of large numbers, Corollary~\ref{cor:checkpoint-reward-bound} and Lemma~\ref{lemma:tree-growth},
\begin{align*}
\limsup_{n \to \infty} \frac{|A(\chain(X_n)) \cap T_1|}{|A(\chain(X_n))|} \leq  \limsup_{n \to \infty} \frac{\frac{|T_1 \cap (0, n]|}{2n}}{\frac{|A(\chain(X_n))|}{n}} \leq \frac{\limsup_{n \to \infty}\frac{|T_1 \cap (0, n]|}{2n}}{\liminf_{n \to \infty} \frac{|A(\chain(X_n))|}{n}}\leq \frac{\alpha/2}{(1-\alpha)} \leq \alpha.
\end{align*}
where the last inequality uses the fact $\alpha \leq 1/2$. Intuitively, if Miner 1 never reaches a checkpoint, then they are publishing at most $\alpha n/2$ blocks in expectation by round $n$ (and clearly at most $\alpha n/2$ of these can be in the longest path). But Lemma~\ref{lemma:tree-growth} asserts that there are at least $(1-\alpha)n$ blocks in expectation in the longest path by round $n$. Therefore, the fraction produced by Miner 1 cannot be too high (and in particular, honesty would have been better in expectation).
\end{proof}

We prove Theorem~\ref{thm:weak-recurrence} in two steps, Section~\ref{sec:checkpoint-preserving} and Section~\ref{sec:opportunistic}.

\subsection{Step 1: Checkpoint Preserving}\label{sec:checkpoint-preserving}

The first step is to show the existence of an optimal strategy that never forks a checkpoint. For that, we will give an explicitly procedure $f$ to transform any strategy $\pi$ that could fork checkpoints into another strategy $f(\pi)$ that does not fork checkpoints satisfying $\rev(f(\pi)) \geq \rev(\pi)$.
\begin{definition}[Finality]\label{def:finality}
A block $q \in A(\chain)$ reaches {\em finality} with respect to strategy $\pi$ if, with probability 1, $\pi$ takes no action that removes $q$ from longest path.
\end{definition}
\begin{definition}[Checkpoint Preserving]
A strategy $\pi$ is {\em checkpoint preserving} if whenever a new checkpoint $P_i$ is defined, $P_i$ reaches finality with respect to $\pi$.
\end{definition}
\begin{theorem}\label{thm:checkpoint-preserving}
For every strategy $\pi$, there is a trimmed, checkpoint preserving strategy $f(\pi)$ with $\rev(f(\pi)) \geq \rev(\pi)$.
\end{theorem}

\subsection{Step 2: Opportunistic}\label{sec:opportunistic}

We have shown the existence of an optimal strategy that would never fork the longest chain $\chain$ when it becomes a checkpoint. However, to be checkpoint recurrent, we must also show Miner 1 has no unpublished blocks bigger than $\chain$ when the longest chain is a checkpoint. The converse can only happen when Miner 1 is about to take action $\publishpath(Q, v)$ and $\max\ Q$ will reach finality with respect to Miner 1's strategy, but Miner 1 would leave an unpublished block $q > \max\ Q$. The intuition is that Miner 1 can wait instead of publishing $Q$ pointing to $v$ in current round. If Miner 1 creates the next block, Miner 1 can publish $Q$ pointing to $v$ as before. If Miner 2 creates the next block, Miner 1 can still take action $\publishpath(Q \cup \{q\}, v)$ adding $Q \cup \{q\}$ to the longest path.
\begin{definition}[Opportunistic]\label{def:opportunistic}
Let $\pi$ be a strategy and let $B$ be a state. Action $\publishpath(Q, v)$ is {\em opportunistic} with respect to $B$ and $\pi$ if
\begin{itemize}
\item $\publishpath(Q, v)$ is a valid action at state $B$.
\item If $\pi$ takes action $\publishpath(Q, v)$ where $\max\ Q$ reaches finality with respect to $\pi$ (Definition~\ref{def:finality}), then $Q = \unpublished(B) \cap (v, \infty)$.
\end{itemize}
Strategy $\pi$ is {\em opportunistic} if at all states $B$, $\pi$ waits or takes an opportunistic action with respect to $B$ and $\pi$.
\end{definition}
\begin{theorem}\label{thm:opportunistic}
For any strategy $\pi$, there is a valid, trimmed, checkpoint preserving and opportunistic strategy $f(\pi)$ with $\rev(f(\pi)) \geq \rev(\pi)$.
\end{theorem}
\begin{proof}[Proof of Theorem~\ref{thm:weak-recurrence}]
Theorem~\ref{thm:opportunistic} directly implies the weak-recurrence theorem since a checkpoint preserving and opportunistic strategy is also checkpoint recurrent.
\end{proof}

\subsection{Step 3: Strong Recurrence}\label{sec:positive-recurrence}

So far, we have shown that there exist an optimal strategy that is checkpoint recurrent. That is, once we reach a state $X_t$ where $\chain(X_t)$ is a checkpoint, Miner 1 capitulates to state $B_0$. Next, we will aim for a stronger result.
\begin{theorem}[Strong recurrence]\label{thm:strong-recurrence}
There exists an optimal checkpoint recurrent and positive recurrent strategy.
\end{theorem}
For a proof sketch, observe the Weak Recurrence Theorem implies there exists an optimal strategy $\pi$ that is checkpoint recurrent. We will assume $\pi$ is not positive-recurrent (i.e., the expected time $\e{\tau}$ to define a new checkpoints is infinite) and derive that $\rev(\pi) \leq \alpha = \rev(\frontier)$. The case where Miner 1 never defines checkpoint $P_1$ -- i.e., $\tau = \infty$ with probability 1 in Proposition~\ref{prop:checkpoint-recurrent} -- give us intuition why the claim should hold to the more general case where $\e{\tau} = \infty$. Once we proof $\rev(\pi) \leq \rev(\frontier)$, we just observe $\frontier$ is checkpoint and positive recurrent. Thus there exists an optimal checkpoint and positive recurrent strategy.

\section{Nash Equilibrium}\label{sec:nash-equilibrium}

We briefly give intuition behind our second main result, which leverages Theorem~\ref{thm:strong-recurrence} to lower bound $\pos$.
\begin{theorem}\label{thm:nash-equilibrium}
For $\alpha \leq 0.308$, $\frontier$ is an optimal strategy for Miner 1 when Miner 2 follows $\frontier$.
\end{theorem}
We defer the proof to Appendix~\ref{sec:nash-equilibrium-proof}. The main idea behind the proof is to show that Nothing-at-Stake Selfish Mining is almost optimal when $\alpha < 1/3$. The following proof-sketch highlights the main insights of the proof.

\vspace{1mm}\noindent\textit{Selfish Mining is optimal when Miner 2 creates the first block (when $\alpha < 1/3$).} We know that there is an optimal checkpoint recurrent strategy, Theorem~\ref{thm:strong-recurrence}. Therefore, it is optimal for Miner 1 to capitulate to state $B_0$ if Miner 2 creates and publishes $1 \to 0$ (which is exactly what selfish mining does).

\vspace{1mm}\noindent\textit{Selfish Mining is optimal after Miner 1 creates and withholds blocks 1 and 2.} Starting from state $B_{2, 0}$, selfish mining will wait until the first time step $\tau$ when the lead decrease to a single block to fork all of Miner 2 blocks. We show that waiting until time $\tau$ is indeed optimal for any value of $\alpha$ (which is not surprising). Less obvious is why Miner 1 must publishes all his blocks at time $\tau$ when they still have a lead of a single block. Indeed we should not expect this to be optimal for all values of $\alpha$. If Miner 1 waits at time $\tau$ and creates the next block, they will again have a lead of two blocks and can resume to ``selfish mine''. Here, we shown that Miner 1 creates $\frac{\alpha}{1-2\alpha}$ blocks on expectation from the time they have a lead of two blocks to the moment the lead decreases to a single block. This quantity can be arbitrarily large but it is at most $1$ when $\alpha < 1/3$ so it is a risky action for Miner 1. That is because if (instead) Miner 2 creates next block, there is a tie (Miner 1 does not have enough blocks to fork the longest chain) and we can show that the probability Miner 1 will ever publish any blocks created before time $\tau$ is at most $\frac{\alpha}{1-\alpha}$. Formally, we will prove Miner 1 maximizes rewards by waiting until time $\tau$ and immediately publishing all unpublished blocks (which is what selfish mining does).

\vspace{1mm}\noindent\textit{There is little window to improve Selfish Mining when Miner 1 creates and withhold block 1 and Miner 2 publishes $2 \to 0$.} Winning the tie-breaking at state $B_{1, 1}$ is another source of revenue for Miner 1. In fact, the only improvement that Nothing-at-Stake Selfish Mining provides over standard Selfish Mining is increasing the probability that Miner 1 wins the tie-breaking between blocks $1$ and $2$. From a similar argument from previous bullet, we can show the probability of adding block 1 to the longest path is at most $\frac{\alpha}{1-\alpha}$. Next, we observe Miner 1 has no more advantage of being the creator of the block at height $\ell \geq 2$ at state $B_{1, 1}$ than being the creator of the block at height $\ell - 1$ at state $B_0$. We formalize this intuition by showing that, by ignoring blocks 1 and 2, any action taken on a state reachable from $B_{1, 1}$ can be converted into an action for an state reachable from $B_0$. The only advantage state $B_{1, 1}$ provides over state $B_0$ is that Miner 1 has a probability (of at most $\frac{\alpha}{1-\alpha}$) of adding block 1 to the longest path.

\vspace{1mm}\noindent\textit{Wrapping up.} From the discussion above, state $B_{1, 1}$ is the only state where we could possible search for a better strategy than Nothing-at-Stake Selfish Mining when $\alpha < 1/3$, but there is little window to improve Miner 1's action at state $B_{1, 1}$. As a result, we will obtain $\frontier$ is optimal when $\alpha \leq 0.308$ as desired.

\section{Conclusion}\label{sec:conclusions}

We study miner incentives in longest-chain proof-of-stake protocols with perfect external randomness. We show both that such protocols are strictly more vulnerable to manipulation than those based on proof-of-work (Theorem~\ref{thm:enhanced-selfish-mining}), but also that it is a Nash equilibrium for all miners to follow the longest-chain protocol as long as no miner has more than $\approx 0.308$ of the total stake (Theorem~\ref{thm:nash-equilibrium}). Our main technical results characterize potentially optimal strategies in a complex, infinite-state MDP (Theorem~\ref{thm:strong-recurrence}). Our work motivates several natural open problems: 
\begin{itemize}
    \item Theorem~\ref{thm:strong-recurrence}, combined with the analysis in Theorem~\ref{thm:nash-equilibrium}, provides strong structure on optimal strategies. It is therefore conceivable that a simulation-based approach with MDP solvers (as in~\cite{sapirshtein2016optimal}) could estimate $\pos$ to high precision. 
    \item Our Theorem~\ref{thm:nash-equilibrium} provides a reduction from incentive-compatible longest-chain proof-of-stake protocols to designing a randomness beacon and a slashing protocol. Clearly, it is important for future work to construct these primitives, although these are well-known and ambitious open problems. In our setting, it is further important to understand what are the \emph{minimal} assumptions on a randomness beacon or slashing protocol necessary to leverage Theorem~\ref{thm:nash-equilibrium}.
\end{itemize}

\bibliographystyle{plainnat}
\bibliography{masterbib}

\newpage
\appendix

In Appendix~\ref{sec:real-analysis}, we give relevant real analysis background. In Appendix~\ref{sec:probability-theory}, we give relevant probability theory background. In Appendix~\ref{sec:mdp-appendix}, we introduce the Markov Decision Process formulation. In Appendix~\ref{sec:table}, we provide a table of notation. In the remaining sections, we provide omitted proves.

\section{Real Analysis Background}\label{sec:real-analysis}

Let $a_1, a_2, \ldots$ be a sequence of real numbers. The limit of a sequence $a_1, a_2, \ldots$ exists if the sequence converges to $a \in \mathbb R$. We write
$$a_n \to a, \quad \text{or} \quad \lim_{n \to \infty} a_n = a.$$
The limit inferior and limit superior of a sequence $a_1, a_2, \ldots$ is defined as
$$\liminf_{n \to \infty} a_n := \lim_{n \to \infty} \inf_{k \geq n} a_k := \sup_{n \geq 1} \inf_{k \geq n} a_k,$$
$$\limsup_{n \to \infty} a_n := \lim_{n\to \infty} \sup_{k \geq n} a_k := \inf_{n \geq 1} \sup_{k \geq n} a_k,$$
respectively. The limit of $a_1, a_2, \ldots$ exists and is equals to $a$ if and only if
$$\liminf_{n \to \infty} a_n = \limsup_{n \to \infty} a_n = a.$$
\begin{lemma}[Properties of $\liminf$ and $\limsup$]\label{lemma:limit-algebra} Whenever the right hand side is well-defined (not of the form $\pm\infty + \mp\infty$ or $0\cdot\pm\infty$).
\begin{enumerate}[label=(\roman*)] 
\item Supperadditivity. \label{liminf-addition} $\liminf_{n \to \infty} (a_n + b_n) \geq \liminf_{n \to \infty} a_n + \liminf_{n \to \infty} b_n$.
\item Subadditivity. \label{limsup-addition} $\limsup_{n \to \infty} (a_n + b_n) \leq \limsup_{n \to \infty} a_n + \limsup_{n \to \infty} b_n$.
\item Supermultiplicativity.\label{liminf-multiplication} $\liminf_{n \to \infty} (a_n \cdot b_n) \geq \liminf_{n \to \infty} a_n \cdot \liminf_{n \to \infty} b_n$.
\item Submultiplicativity.\label{limsup-multiplication} $\limsup_{n \to \infty} (a_n \cdot b_n) \leq \limsup_{n \to \infty} a_n \cdot \limsup_{n \to \infty} b_n$.
\item \label{liminf-1} If $a_n \to a$, $\liminf_{n \to \infty} a_n + b_n = a + \liminf_{n \to \infty} b_n$.
\item \label{limsup-1} If $a_n \to a$, $\limsup_{n \to \infty} a_n + b_n = a + \limsup_{n \to \infty} b_n$.
\item \label{liminf-2} If $a_n \to a$ and $b_n$ is bounded, $\liminf_{n \to \infty} a_n b_n = a \liminf_{n \to \infty} b_n$.
\item \label{limsup-2} If $a_n \to a$ and $b_n$ is bounded, $\limsup_{n \to \infty} a_n b_n = a \limsup_{n \to \infty} b_n$.
\end{enumerate}
\end{lemma}

\section{Probability Theory Background}\label{sec:probability-theory}

\subsection{Convergence of Random Variables}
\begin{definition}[Almost sure convergence]
We say a sequence of random variables $X_1, X_2, \ldots$ converges almost surely to random variable $X$ (and we write $X_n \toas X$) if
$$\pr{\lim_{n \to \infty} X_n = X} = 1.$$
\end{definition}
\begin{proposition}
The following are equivalent:
\begin{itemize}
    \item $X_n \toas X$.
    \item $\lim_{n \to \infty} \pr{\cup_{i = n}^\infty \left\{|X_i - X| \geq 1/k\right\}} = 0$ for any $k \geq 1$.
    \item $\lim_{n \to \infty} \pr{\cap_{i = n}^\infty \left\{|X_i - X| < 1/k\right\}} = 1$ for any $k \geq 1$.
\end{itemize}
\end{proposition}
\subsection{Laws of Large Numbers}
\begin{definition}[Absolutely Integrable]
A random variable $X$ is {\em absolutely integrable} if $\e{|X|} < \infty$.
\end{definition}
\begin{lemma}[Strong Law of Large Numbers (SSLN), Theorem 2.4.1 in \cite{durrett2019probability}]\label{lemma:slln}
Let $X_1, X_2, ...$ be an i.i.d. sequence of copies of absolutely integrable random variable $X$. Then
$$\frac{1}{n} \sum_{i = 1}^n X_i \overset{a.s.}{\to} \e{X}.$$
\textit{Remark:} The strong law can be generalized for the case where $X$ is non-negative and {\em not} absolutely integrable. That is, if $\e{X} = \infty$ and $\pr{X \geq 0} = 1$, then $\frac{1}{n}\sum_{i = 1}^n X_i \toas \infty$.
\end{lemma}
\begin{definition}[Counting Process]\label{def:counting-process}
The {\em interarrival times} $\tau_1, \tau_2, ...$ is an i.i.d. sequence of positive and absolutely integrable random variables. Let $X_n = \sum_{i = 1}^n \tau_i$ with $X_0 = 0$. The random variable $N_n = \sum_{i = 1}^\infty \mathbbm 1_{X_i \leq n}$ denotes the {\em counting proces}  associated with interarrival times $\tau_1, \tau_2, \ldots$.
\end{definition}
\begin{observation}\label{obs:counting-process}
If $N_n = \sum_{i = 1}^\infty \ind{X_i \leq n}$ is a counting process, for all $n \in \mathbb N$, $X_{N_n} \leq n < X_{N_n+1}$.
\end{observation}
\begin{lemma}[SLLN for Counting Processes, Theorem 2.4.7 in \cite{durrett2019probability}]\label{lemma:slln-counting-process}
If $N_n = \sum_{i = 1}^\infty \ind{\tau_i \leq n}$ is a counting process, then
$$\frac{N_n}{n} \overset{a.s.}{\to} \frac{1}{\mathbb E[\tau_i]}.$$
\end{lemma}

\section{Markov Decision Process}\label{sec:mdp-appendix}

In this section, we show we can obtain a closed form for the revenue of positive recurrent strategies by using the strong law of large numbers. See Appendix~\ref{sec:probability-theory} for basic background on probability theory. A sequence of random variables $X_1, X_2, \ldots$ \emph{converges almost surely} (or with probability 1) to random variable $X$ (and we write $X_n \toas X$) if $Pr[\lim_{n \to \infty} X_n = X] = 1$. Our first result, is a generalization of Corollary 9 in \cite{sapirshtein2016optimal} to Proof-of-Stake mining games with positive recurrent strategies.
\begin{theorem}\label{thm:slln-mining-games}
Let $X_0, X_1, \ldots$ be a mining game starting at state $X_0 = B_0$ where Miner 1 follows a positive recurrent strategy. Let $R^k = \sum_{t = 1}^\tau r^k(X_{t-1}, X_t)$ be Miner k's total reward before the first time step $\tau \geq 1$ where Miner 1 capitulates to state $B_0$. Then
\begin{enumerate}[label=(\roman*)]
\item For $k = 1, 2$, $\frac{|A(\chain(X_n)) \cap T_k|}{n} \overset{a.s.}{\to} \frac{\mathbb E[R^k]}{\mathbb E[\tau]}$.
\item \label{eq:recurrent-revenue} $\rev(\pi) = \frac{\mathbb E[R^1]}{\mathbb E[R^1 + R^2]}$.
\end{enumerate} 
\end{theorem}
\begin{proof}
Without loss of generality assume $X_0 = B_0$ and let $n_0 = 0$. Let $X_{n_1}, X_{n_2}, \ldots$ be the sequence of states where Miner 1 capitulates to state $B_0$. Let $R_i^k$ be Miner $k$'s reward from time step $n_{i-1}$ to time $n_i$. That is,
$$R_i^k = |A(\chain(X_{n_i})) \cap T_k| - |A(\chain(X_{n_{i-1}})) \cap T_k|.$$
Observe $R_1^k, R_2^k, \ldots$ and $n_1-n_0, n_2-n_1, \ldots$ are i.i.d. sequences of random variables (because Miner 1 capitulates to state $B_0$ at time step $n_i$). Additionally $\e{n_{i+1}-n_i} < \infty$ because Miner 1's strategy is positive recurrent. Thus define the counting process $N_n = \sum_{i = 1}^\infty \mathbbm 1_{n_i \leq n}$. From Observation~\ref{obs:counting-process}, $n_{N_n} \leq n < n_{N_n+1}$, and since Miner 2 never forks Miner 1's blocks, the number of Miner 1's blocks in the longest path is an increasing function of time. Thus
\begin{align*}
    \frac{|A(\chain(X_{n_{N_n}})) \cap T_1|}{n} \leq \frac{|A(\chain(X_n))\cap T_1|}{n} \leq  \frac{|A(\chain(X_{n_{N_n+1}})) \cap T_1|}{n}.
\end{align*}
Observe $\sum_{i = 1}^{N_n} R_i^k = |A(\chain(X_{N_n})) \cap T_k|$ is a telescoping sum. Thus
$$\frac{N_n}{n}\frac{\sum_{i = 1}^{N_n} R_i^1}{N_n} \leq \frac{|A(\chain(X_n))\cap T_1|}{n} \leq\frac{N_n+1}{n} \frac{\sum_{i = 1}^{N_n+1} R_i^1}{N_n+1}.$$
Observe $N_n \to \infty$ as $n \to \infty$ (because Miner 1's strategy is recurrent) and $\mathbb E[R^k] \leq \mathbb E[\tau] < \infty$ (because Miner 1's strategy is positive recurrent). From the SLLN for Counting Processes (Lemma~\ref{lemma:slln-counting-process}) and SLLN (Lemma~\ref{lemma:slln}),
$$\frac{N_n}{n}\frac{1}{N_n}\sum_{i = 1}^{N_n} R_i^k \overset{a.s.}{\to} \frac{\mathbb E[R^k]}{\mathbb E[\tau]} \quad \text{and} \quad \frac{N_n}{n}\frac{1}{N_n+1}\sum_{i = 1}^{N_n+1} R_i^k \overset{a.s.}{\to} \frac{\mathbb E[R^k]}{\mathbb E[\tau]}.$$
From the Sandwich Theorem,
$$\frac{|A(\chain(X_n)) \cap T_1|}{n} \overset{a.s.}{\to} \frac{\mathbb E[R^1]}{\mathbb E[\tau]}.$$
Observe the height of the block tree is also a increasing function of time. Thus with an similar argument from above, we sandwich $h(\chain(X_n))$ between $h(\chain(X_{n_{N_n}}))$ and $h(\chain(X_{n_{N_n+1}}))$ to get
$$\frac{N_n}{n}\frac{h(\chain(X_n))}{n} \overset{a.s.}{\to} \frac{\mathbb E[R^1 + R^2]}{\mathbb E[\tau]}.$$
From linearity of expectation, we get $\frac{|A(\chain(X_n)) \cap T_2|}{n} \overset{a.s.}{\to} \frac{\mathbb E[R^2]}{\mathbb E[\tau]}$. This proves the first bullet. From the first bullet,
$$\rev(\pi) = \e{\liminf_{n \to \infty}\frac{\frac{1}{n}|A(\chain(X_n) \cap T_1|}{\frac{1}{n}|A(\chain(X_n) \cap T_1| + \frac{1}{n}|A(\chain(X_n)) \cap T_2|}} = \frac{\mathbb E[R^1]}{\mathbb E[R^1] + \mathbb E[R^2]}.$$
This proves the second bullet.
\end{proof}
Theorem~\ref{thm:slln-mining-games} will be useful to study the revenue of positive recurrent strategies such as the nothing-at-stake selfish mining strategy---which will be positive recurrent by design. It also come in hand when studying optimal strategies; however, it require us to assume the existence an optimal positive recurrent strategy. In fact, \citet{sapirshtein2016optimal} and \citet{kiayias2016blockchain} relies on the structure of proof-of-work mining games to conjecture the existence an optimal positive recurrent strategy. In proof-of-stake, there is no obvious incentives for a miner to forget unpublished blocks. Thus it is not clear if an optimal strategy would be (positive) recurrent (recall Example~\ref{example:fork-own-block} where Miner 1 might be motivated to fork their own blocks). As one of our technical contributions, Theorem~\ref{thm:strong-recurrence} proves proof-of-stake mining games admits optimal strategies that are positive recurrent.

Additionally, \citet{sapirshtein2016optimal} uses Theorem~\ref{thm:slln-mining-games} to compute an optimal positive recurrent strategy by optimizing a Markov Decision Process (MDP). Instead of explicitly computing optimal strategies, we will use their MDP to give sufficient conditions on $\alpha$ for which $\frontier$ is an optimal strategy (Theorem~\ref{thm:nash-equilibrium}). Given a parameter $\lambda$, their MDP defines the real-valued function $r_\lambda$ as the reward from transition from state $B$ to $B'$:
\begin{definition}[Mining Game Reward]\label{def:game-reward}
For $\lambda \in \mathbb R$, the \emph{mining game reward} is the real-valued function $r_\lambda$ from states $B$ and $B'$ to
\begin{equation}\label{eq:game-reward}
\R_{\lambda}(B, B') := (1-\lambda) r^1(B, B') - \lambda r^2(B, B').
\end{equation}
\textit{Remark:} For any states $X_t$, $X_{t+1}$, and $X_{t+2}$, the mining game reward function satisfy the identity
\begin{equation}\label{eq:reward-identity}
\R_{\lambda}(X_t, X_{t+1}) + \R_{\lambda}(X_{t+1}, X_{t+2}) = \R_{\lambda}(X_t, X_{t+2}).
\end{equation}
\end{definition}
\begin{definition}[Value Function]\label{def:value-function}
The \emph{objective function} for mining game $(X_t)_{t \geq 0}$ is a real-value function $\va_\pi^\lambda$ from state $B$ to
\begin{equation}\label{eq:objective-function}
\va_\pi^\lambda(B) := \e{r_{\lambda}(X_0, X_\tau)\big | X_0 = B} = \e{r_{\lambda}(X_0, X_1) + \va_\pi^{\lambda}(X_1) \cdot \ind{\tau \neq 1} | X_0 = B},
\end{equation}
the expected game reward from a mining game starting at state $X_0 = B$ and stopping at state $X_\tau$ where $\tau \geq 1$ is the first time step Miner 1 capitulates to state $B_0$. Define the \emph{value function} $\va$ as the real-valued function from state $B$ to
\begin{equation}\label{eq:value-function}
\va(B) := \va_{\pi^*}^{\lambda^*}(B)
\end{equation}
where $\lambda^* = \max_\pi \rev(\pi)$ and $\pi^*$ is an optimal positive recurrent strategy.
\end{definition}
First, they observed that by setting $\lambda^* = \max_\pi \rev(\pi)$, any positive recurrent strategy $\pi$ that maximizes $\va_\pi^{\lambda^*}(B_0)$ is an optimal positive recurrent strategy.
\begin{lemma}\label{lemma:b-0-optimization}
Let $\pi^*$ be an optimal positive recurrent strategy and let $\lambda^* = \rev(\pi^*)$. Then $\pi^* \in \arg\max_\pi \va_\pi^{\lambda^*}(B_0)$ and $\va(B_0) = 0$.
\end{lemma}
\begin{proof}
Given two positive recurrent strategies $\pi$ and $\pi'$, the following claim allow us to compare their revenue.
\begin{claim}\label{claim:optimality-test}
For positive recurrent strategies $\pi$ and $\tilde \pi$,
\begin{enumerate}[label=(\roman*)]
\item \label{optimality-test-1} $\va_\pi^\lambda(B_0) = 0$ if and only if $\lambda = \rev(\pi)$.
\item \label{optimality-test-2}$\rev(\pi) > \rev(\tilde \pi)$ if and only if $\va_\pi^{\rev(\tilde \pi)}(B_0) > 0$.
\item \label{optimality-test-3}$\rev(\pi) < \rev(\tilde \pi)$ if and only if $\va_\pi^{\rev(\tilde \pi)}{\pi}(B_0) < 0$.
\end{enumerate}
\end{claim}
\begin{proof}
Recall $\tau$ denotes the first time step Miner 1 capitulates to state $B_0$, Equation~\ref{eq:tau}, and let $R^k = \sum_{t = 1}^\tau r^k(X_{t-1}, X_t)$. Observe
$$\va_\pi^\lambda(B_0) = \e{(1-\lambda)R^1 - \lambda R^2 | X_0 = B_0}.$$
From Theorem~\ref{thm:slln-mining-games},
$$\frac{\e{R^1}\e{R^2}}{\e{R^1} + \e{R^2}} = \rev(\pi) \e{R^2} \quad \text{and} \quad \frac{\e{R^1}\e{R^2}}{\e{R^1} + \e{R^2}} = (1-\rev(\pi))\e{R^1}.$$
Taking the difference and setting $\lambda = \rev(\pi)$,
\begin{align*}
\va_\pi^{\rev(\pi)}(B_0) &= \e{(1-\rev(\pi))R^1 - (1-\rev(\pi))R^2}\\
&= \frac{\e{R^1}\e{R^2}}{\e{R^1} + \e{R^2}} - \frac{\e{R^1}\e{R^2}}{\e{R^1} + \e{R^2}} = 0.
\end{align*}
This proves \ref{optimality-test-1}. Rewrite $\va_\pi^\lambda(B_0)$ as
$$\va_\pi^\lambda(B_0) = \mathbb E\big[R^1] - \lambda \mathbb E [R^1 + R^2 ].$$
Note that $\mathbb E[R^1 + R^2]$ is the expected height of the block tree at time $\tau$. Since $\tau \geq 1$ and Miner 2 mines a block with probability $1-\alpha > 0$, the expected height of the block tree at time $\tau$ is at least $1-\alpha > 1/2$. Thus $\va_\pi^\lambda(B_0)$ is strictly decreasing function of $\lambda$. This proves \ref{optimality-test-2}, and \ref{optimality-test-3}.
\end{proof}
Thus if $\pi^*$ is an optimal positive recurrent strategy with $\lambda^* = \rev(\pi^*)$, we directly obtain $\va(B_0) = \va_{\pi^*}^{\lambda^*} (B_0) = 0$. For any positive recurrent strategy $\pi$, $\va_{\pi}^{\lambda^*}(B_0) \leq \va(B_0) = 0$. Therefore, $\pi^* \in \arg\max_\pi \va_\pi^{\lambda^*}(B_0)$. 
\end{proof}
Thus if $\pi$ maximizes $\va_\pi^{\lambda^*}(B_0)$ (Equation~\ref{eq:objective-function}), then $\pi$ must maximize $\va_\pi^{\lambda^*}(X_t)$ for all subsequent states $X_t$. As a corollary, we recover the well known Bellman's principle of optimality.
\begin{lemma}[Bellman's Principle of Optimality]\label{lemma:bellman}
For all states $B$, for all positive recurrent strategies $\pi$, $\va(B) \geq \va_\pi^{\max_\pi \rev(\pi)}(B)$.
\end{lemma}
\begin{proof}
Let $\pi^*$ be an optimal positive recurrent strategy with $\lambda^* = \rev(\pi^*)$. Let $(X_t^\pi)_{t \geq 0}$ be a mining game starting from state $X_0 = B_0$ when Miner 1 follows strategy $\pi$. From Definition~\ref{def:value-function},
$$\va(B_0) = \va_{\pi^*}^{\lambda^*}(B_0) = \e{r_{\lambda^*}(B_0, X_1^{\pi^*}) + V(X_1^{\pi^*}) | X_0^{\pi^*} = B_0},$$
and from Lemma~\ref{lemma:b-0-optimization}, $\va(B_0) = \max_\pi \va_\pi^{\lambda^*}(B_0) = 0$. Let $\pi(X_t)$ be the action $\pi$ takes at state $X_t^\half$ and let $\pi : \pi(X_t)$ be all strategies conditioned on $\pi$ taking action $\pi(X_t)$ at state $X_t^\half$. Note $X_t^\pi$ depends only on actions taken up to time $t$. Thus
\begin{align*}
\va(B_0) &= \max_\pi \e{r_{\lambda^*}(B_0, X_1^\pi) + \va_\pi^{\lambda*}(X_1^\pi) | X_0^\pi = B_0}\\
&= \max_{\pi(X_0)} \e{r_{\lambda^*}(B_0, X_1^{\pi(X_0)}) + \max_{\pi : \pi(X_0)} \va_\pi^{\lambda*}(X_1^{\pi(X_0)}) | X_0^{\pi(X_0)} = B_0}\\
&= \max_{\pi(X_0)} \e{r_{\lambda^*}(B_0, X_1^{\pi(X_0)}) + \va(X_1^{\pi^*}) | X_0^{\pi(X_0)} = B_0}.
\end{align*}
The last line observes the expected value takes its maximum by taking action $\pi(X_0) = \pi^*(X_0)$. Thus $\va(B) = \max_\pi \va_\pi^{\lambda^*}(B) \geq \va_\pi^{\lambda^*}(B)$ for all positive recurrent $\pi$ and states $B$.
\end{proof}
We will use Lemma~\ref{lemma:bellman} to witnesses when a particular action is optimal for a particular state $B$. That is, we can guess a particular strategy $\pi$ takes an optimal action at state $B$ and assume the optimal strategy $\pi^*$ takes a distinct action. By deriving $\va_\pi^{\lambda^*}(B) \geq \va(B)$, we conclude that $\pi$'s action at state $B$ is indeed optimal.

\section{Omitted Proofs from Section~\ref{sec:enhanced-selfish-mining}}\label{sec:selfish-appendix}

First, we compute the expected number of blocks Miner 1 creates from the moment it reaches state $X_2 = B_{2, 0}$ until they capitulate to state $B_0$. For that, we will define a coupling between the mining game $X_2, X_3, \ldots, X_\tau$ with a biased one-dimensional random walk.
\begin{lemma}\label{lemma:random-walk}
Let $(N_t)_{t \geq 0}$ be a biased one-dimensional random walk with initial state $N_0 \in \mathbb Z$ and for $t \geq 1$, $N_t = N_{t-1} + 1$ with probability $\alpha$; otherwise, $N_t = N_{t-1} - 1$. Let $\tau = \min\{t \geq 1 : N_t = N_0 - 1\}$. We say the state $N_t$ increments if $N_{t+1} = N_t +1$ and decrements if $N_{t+1} = N_t -1$. Let $X = \sum_{t = 1}^\tau \ind{N_t = N_{t-1} + 1}$ be the number of state increments up to time $\tau$. Similarly let $Y = \sum_{t = 1}^\tau \ind{N_t = N_{t-1}-1}$ be the number of state decrements up to time $\tau$. Then
$$\e{X} = \frac{\alpha}{1-2\alpha} \qquad \e{Y} = \frac{1-\alpha}{1-2\alpha} \qquad \e{\tau} = \frac{1}{1-2\alpha}.$$
\end{lemma}
\begin{proof}
For the case $N_1 = N_0 - 1$ and $X = 0$, $Y = 1$ (with probability $1-\alpha$). For the case $N_1 = N_0 + 1$ (with probability $\alpha$), the expected number of state increments until the Markov process first returns to state $N_0$ is equals to $\e{X}$. Once the Markov process first returns to state $N_0$, the expected number of time steps until the Markov process first reaches state $N_0 - 1$ is also $\e{X}$. Thus
\begin{align*}
   	\e{X} &= \alpha (1 + \e{X} + \e{X})\\
    &= \alpha + 2\alpha \e{X}.
\end{align*}
Solving for $\e{X}$ proves $\e{X} = \frac{\alpha}{1-2\alpha}$. A similar argument proves $\e{Y} = \frac{1-\alpha}{1-2\alpha}$. Since $\tau = X + Y$, linearity of expectation proves $\e{\tau} = \frac{1}{1-2\alpha}$ as desired.
\end{proof}
\begin{lemma}\label{lemma:selfish-mining-reward}
Let $X_0 = B_{2, 0}$ and let $\tau = \min\{t \geq 1 : |T_1(X_\tau)| = |T_2(X_\tau)| + 1\}$. Then the expected number of blocks Miner 1 creates from time 1 to $\tau$ is $\e{|T_1(X_\tau)|-1} = \frac{\alpha}{1-2\alpha}$. Moreover, the expected number of blocks Miner 2 creates from time 1 to $\tau$ is $\frac{1-\alpha}{1-2\alpha}$.
\end{lemma}
\begin{proof}
Define the biased one-dimensional random walk $Y_t = |T_1(X_t)| - |T_2(X_t)| - 1$ for $t \geq 0$. Then $\tau$ is the first time step where $Y_\tau = 0$ (observe $Y_0 = 1$). The random variable $|T_1(X_t)|-2$ counts the number of time steps where $Y_t = Y_{t-1} + 1$ for $t \leq \tau$. Thus from Lemma~\ref{lemma:random-walk}, the expected number of blocks Miner 1 creates from time 1 to $\tau$ is $\e{|T_1(X_t)| - 2} = \frac{\alpha}{1-2\alpha}$ and the expected number of blocks Miner 2 creates from time 1 to $\tau$ is $\frac{1-\alpha}{1-2\alpha}$ as desired.
\end{proof}
\begin{proof}[Proof of Theorem~\ref{thm:selfish-mining}]
Let $(X_t)_{t \geq 0}$ be a mining game starting at state $X_0 = B_0$ where Miner 1 uses the $\sm$ strategy (Definition~\ref{def:sm}). Let $\tau \geq 1$ be the first time step where Miner 1 capitulates to state $B_0$. If Miner 1 did not capitulate by time step $t = 3$, it is because $X_2 = B_{2, 0}$. In this case, $\tau = |T_1(X_\tau)| + |T_2(X_\tau)| = 2|T_1(X_\tau)| - 1$. From Lemma~\ref{lemma:selfish-mining-reward}, $\mathbb E[\tau] < 3 + \frac{2\alpha}{1-2\alpha}$. Thus $\sm$ is positive recurrent and from Lemma~\ref{lemma:b-0-optimization},
$$0 = \va_\sm^\lambda(B_0) = \e{\rew(B_0, X_\tau) | X_0 = B_0}$$
where $\lambda = \rev(\sm)$. To compute this quantity, we consider the following events:
\begin{itemize}
    \item When the game reaches state $B_{0, 1}$, Miner 1 capitulates to state $B_0$. Miner 2 owns one block in the longest path. Thus the reward is $\rew(B_0, X_\tau) = -\lambda$.
    \item When the game reaches state $B_{2, 0}$, Miner 1 wait until the first time step $\tau$ where $|T_2(X_\tau)| = |T_1(X_\tau)| - 1$, then Miner 1 publishes blocks $T_1(X_\tau)$ and capitulates to state $B_0$. Miner 1 owns $|T_1(X_\tau)|$ blocks in the longest path. From Lemma~\ref{lemma:selfish-mining-reward}, the expected reward is
    $$\e{\rew(B_0, X_\tau) | X_2 = B_{2,0}} = \e{T_1(X_\tau) | X_2 = B_{2, 0}}(1-\lambda) = \left(2 + \frac{\alpha}{1-2\alpha}\right)(1-\lambda).$$
    \item When the game reaches state $B_{1, 1}$, we consider two cases:
    \begin{itemize}
    \item Miner 1 creates block 3, publishes $3 \to 1 \to 0$ and capitulates to state $B_0$. Miner 1 owns two blocks in the longest path. Thus the reward is $2(1-\lambda)$.
    \item Miner 2 publishes $3 \to 2$, then Miner 1 capitulates to state $B_0$. Miner 2 owns two blocks in the longest path. Thus the reward is $-2\lambda$.
    \end{itemize}
\end{itemize}
Substituting in the equation above,
$$0 = \va_\sm^\lambda(B_0) = \alpha^2\bigg(2+\frac{\alpha}{1-2\alpha}\bigg)(1-\lambda) + 2\alpha^2(1-\alpha)(1-\lambda)-2\alpha(1-\alpha)^2\lambda - (1-\alpha)\lambda.$$
Solving for $\lambda$ and observing $\lambda = \rev(\sm)$ proves the theorem.
\end{proof}

\begin{proof}[Proof of Theorem~\ref{thm:enhanced-selfish-mining}]
Consider the mining game starting at state $X_0 = B_0$. Let $\tau \geq 1$ be the first time step Miner 1 capitulates to state $B_0$. Let's first check the Markov chain induced by the $\nsm$ strategy (Definition~\ref{def:nsm}) is positive recurrent. If Miner 1 did not capitulate by time step 2, $X_2 \in \{B_{2, 0}, B_{1, 1}\}$. For the case $X_2 = B_{2, 2}$, $\tau = |T_1(X_\tau)| + |T_2(X_\tau)| = 2|T_1(X_\tau)| - 1$. From Lemma~\ref{lemma:selfish-mining-reward}, $\e{\tau | X_2 = B_{2, 2}} = 3 + \frac{2\alpha}{1-2\alpha} < \infty$. For the case $X_2 = B_{1, 1}$, the probability that the Markov process returns to state $B_0$ before returning to state $B_{1, 1}$ is $\alpha + (1-\alpha)^2 + \alpha^2(1-\alpha)$. Thus $\e{\tau |X_2 = B_{1, 1}} = 2 + \frac{1}{\alpha + (1-\alpha)^2 + \alpha^2(1-\alpha)} < \infty$. Thus the Markov chain is positive recurrent and from Lemma~\ref{lemma:b-0-optimization},
\begin{align*}
    0 = \va_\nsm^\lambda(B_0) = \e{\rew(B_0, X_\tau) | X_0 = B_0}
\end{align*}
where $\lambda = \rev(\nsm)$. To compute the expected value above, we consider the following events:
\begin{itemize}
    \item When $X_1 = B_{0, 1}$, Miner 1 capitulates to state $B_0$. Miner 2 owns one block in the longest path. Thus the reward is $-\lambda$.
    \item When $X_2 = B_{2, 0}$, Miner 1 waits until the first time step $\tau \geq 3$ where $|T_2(X_\tau)| = |T_1(X_\tau)| - 1$, then Miner 1 publishes blocks $T_1(X_\tau)$ and capitulates to state $B_0$. Miner 1 owns $|T_1(X_\tau)|$ blocks in the longest path at time $\tau$. From Lemma~\ref{lemma:selfish-mining-reward}, the expected reward is 
    $$\e{\rew(B_0, X_\tau) | X_2 = B_{2, 0}} = \e{T_1(X_\tau) | X_2 = B_{2, 0}}(1-\lambda) = \bigg(2 + \frac{\alpha}{1-2\alpha}\bigg)(1-\lambda).$$
    \item When the game reaches state $X_t$ equivalent to state $B_{1, 1}$, there is four scenarios:
\begin{itemize}
    \item Miner 1 creates block 3, then publishes $3 \to 1 \to 0$ and  capitulates to state $B_0$. Miner 1 owns two blocks in the longest path. Thus the reward is $2(1-\lambda)$.
    \item Miner 2 publishes $4 \to 3 \to 2$, then Miner 1 capitulates to state $B_0$. Miner 2 owns three blocks in the longest path. Thus the reward is $-3\lambda$.
    \item Miner 2 publishes $3 \to 2$ while Miner 1 creates block 4 and 5. Then Miner 1 publishes $5 \to 4 \to 1 \to 0$ and capitulates to state $B_0$. Miner 1 owns three blocks in the longest path. Thus the reward is $3(1-\lambda)$.
    \item Miner 2 publishes $5 \to 3 \to 2$ while Miner 1 creates and withholds block 4. Then Miner 1 capitulates to state $B_{1, 1}$ allowing Miner 2 to stay with blocks 2 and 3. Thus the reward is $-2\lambda$ (repeat until Miner 1 capitulates to state $B_0$).
\end{itemize}
\end{itemize}
From the work above, the reward when $X_2 = B_{1, 1}$ is
\begin{align*}
    &\e{\rew(X_0, X_\tau) | X_2 = B_{1, 1}} = \\
    &\qquad \sum_{i = 0}^\infty (\alpha(1-\alpha)^2)^i(1-\alpha(1-\alpha)^2) \left(\e{\rew(X_0, X_\tau) | X_2 = B_{1, 1}, X_\tau = B_0} - 2i\lambda\right)
\end{align*}
where $\e{\rew(X_0, X_\tau) | X_2 = B_{1, 1}, X_\tau = B_0}$ is the expected reward conditioned on Miner 1 capitulating to $X_\tau = B_0$ given $X_2 = B_{1, 1}$. Thus
$$\e{\rew(X_0, X_\tau) | X_2 = B_{1, 1}, X_\tau = B_0} = \frac{2\alpha(1-\lambda) -3(1-\alpha)^2\lambda + 3\alpha^2(1-\alpha)(1-\lambda)}{1 - \alpha(1-\alpha)^2}.$$
To get a closed form, recall the geometric series (for $x < 1$)
$$\sum_{i = 0}^\infty x^i = \frac{1}{1-x} \quad \text{and} \quad \sum_{i = 1}^\infty i x^{i} = \frac{x}{(1-x)^2}.$$
Substituting in the equation above, we get
\begin{align*}
0 &= \va_\nsm^\lambda(B_0) = \alpha^2\bigg(2+\frac{\alpha}{1-2\alpha}\bigg)(1-\lambda) - (1-\alpha)\lambda\\
&\quad \quad + \frac{\alpha(1-\alpha)}{1-\alpha(1-\alpha)^2}\bigg(2\alpha(1-\lambda)-3(1-\alpha)^2\lambda + 3\alpha^2(1-\alpha)(1-\lambda)-2\alpha(1-\alpha)^2\lambda\bigg).
\end{align*}
Solving for $\lambda$ and observing $\lambda = \rev(\nsm)$ proves
$$\rev(\nsm) = \frac{4 \alpha^2 - 8 \alpha^3 - \alpha^4 + 7 \alpha^5 - 3 \alpha^6}{1 - \alpha - 2 \alpha^2 + 3 \alpha^4 - 3 \alpha^5 + \alpha^6}$$
as desired.
\end{proof}

\section{Omitted Proofs from Section~\ref{sec:trimming-strategy}}\label{sec:trimming-proof}

\subsection{Timeserving Reduction}

We will now define a reduction that takes as input a strategy $\pi$, and produces a new strategy $\tilde{\pi}$. The strategy $\tilde{\pi}$ will \emph{simulate} $\pi$, which we clarify below.

\begin{definition}[Simulating a strategy] A strategy $\tilde{\pi}$ may wish to \emph{simulate} $\pi$, and take actions based on this simulation. Specifically, we will add a subscript of $_\pi$ to random variables like $\tree, V, \chain$, etc. to denote the state of this simulation. Formally, $X_\pi$ denotes ``what the state of variable $X$ would have been, had Miner $1$ been using strategy $\pi$ all along (with the same $\cstring$)''.

If there is no subscript of $\pi$, then these variables retain their original meaning (and refer to the state when Miner $1$ uses their actual strategy $\tilde{\pi}$).
\end{definition}

\begin{definition}[Timeserving Reduction]\label{def:timeserving-reduction} For any strategy $\pi$, define the \emph{Timeserving Reduction} of $\pi$ to take the following action during round $n$:
\begin{itemize}
\item If $A(\chain_\pi(n)) \cap \unpublished(n) = \emptyset$, then \wait. That is, if the longest chain from simulating $\pi$ is already published, don't publish further.
\item Else, $\publishset(A(\chain_\pi(n)) \cap \unpublished(n), E')$, where $E'$ contains all pointers leaving $A(\chain_\pi(n)) \cap \unpublished(n)$ in $E_\pi(n)$. That is, if the longest chain from simulating $\pi$ is not yet published, publish whatever part of the chain is not yet published.
\end{itemize}
\end{definition}

Essentially, the Timeserving Reduction of $\pi$ simply holds off on publishing blocks which are not yet in the longest chain, and only publishes them when necessary to support a longest chain. Below, note that the conclusion of Theorem~\ref{thm:timeserving} does not necessarily hold against \emph{arbitrary} strategies for Miner $2$, but it does hold against $\frontier$ (so we must necessarily use properties of $\frontier$ in the proof).

\begin{proof}[Proof of Theorem~\ref{thm:timeserving}]
Let $\tilde{\pi}$ denote the Timeserving Reduction of $\pi$. Here is the main idea: all actions taken by the $\frontier$ strategy \emph{only depend on the longest chain, and not on what other nodes are published}. This is formally stated in the following observation.

\begin{observation}\label{obs:identical}
Let $\pi, \tilde{\pi}$ be two strategies such that when used against $\frontier$, with probability $1$, for all $n$, $\chain_\pi(n) = \chain_{\tilde{\pi}}(n)$. Then in every round $\frontier$ takes identical actions against $\pi$ and $\tilde{\pi}$.
\end{observation}

It is clear that $\pi$ and $\tilde{\pi}$ have the same longest chain at the end of each round, by definition of $\tilde{\pi}$. Specifically, $\tilde{\pi}$ will never publish a block that has not yet been published by $\pi$, and it makes sure the longest chain under $\pi$ is always published. Therefore, $\frontier$ will take the same actions against both. Because $\frontier$ is taking the same actions, and the longest chain is the same at the end of each round, this guarantees that $\rev(\pi) = \rev(\tilde{\pi})$.

Finally, it is clear that $\tilde{\pi}$ is Timeserving. The longest chain at the end of round $n$ is indeed $\chain_\pi(n)$, and $\tilde{\pi}$ only publishes ancestors of $\chain_\pi(n)$ (if it publishes at all).
\end{proof}

\begin{proof}[Proof of Observation~\ref{obs:timeserving}]
\vspace{1mm}\noindent \textit{Proof of~\ref{label:timeserving-3}.} Ancestors of the longest chain form a single path. If $\pi$ does not publish a single path, then it is not possible for all of these nodes to immediately be ancestors of the longest chain.

\vspace{1mm}\noindent \textit{Proof of~\ref{label:timeserving-1}.} Suppose for contradiction there is some state where $\tree$ has two distinct leaves $q \neq \tilde{q}$ with the same height, and consider when this first happens. First, $q$ and $\tilde{q}$ can't have been published by the same action by Property~\ref{label:timeserving-3}. If they are published during distinct actions, then w.l.o.g. let $q$ be published before $\tilde{q}$ (and therefore the most recent action to update $\tree$ must have published $\tilde{q}$). 

Because $q$ has the same height as $\tilde{q}$ and was published first, $\tilde{q}$ is not in $\chain$. Also, because $\tilde{q}$ is a leaf, it cannot be in $A(\chain)$ unless it is $\chain$ itself. Therefore, no Timeserving strategy could have published $\tilde{q}$ (because Miner 1 is Timeserving, and $\frontier$ only publishes a longest chain no matter what), a contradiction.

\vspace{1mm}\noindent \textit{Proof of~\ref{label:timeserving-2}.} Suppose for contradiction Miner 1 forks by publishing a single block $q$ on top of $r$. If Miner 1 can fork the longest chain with a single block, then there must have previously been two leaves of the same height, a contradiction to~\ref{label:timeserving-1}.
\end{proof}

\subsection{Orderly Reduction}

\begin{example}\label{ex:orderly}
Consider the game in Figure~\ref{fig:orderly} where Miner 1 creates blocks 1, 2, 3, 4 and 5. Looking at strategy $\pi$ on the left, observe that the action $\publishpath(\{4,5\}, 0)$ taken in round $5$ is not Orderly. Instead, if an Orderly action is to publish two blocks pointing to $0$ in round $5$, it must publish $2 \to 1 \to 0$. This initially suggests a local fix: simply swap the roles of $1$ and $4$, and also $2$ and $5$. This switch is shown in the center.

Unfortunately, there is a problem: while it is certainly safe to publish $1 \to 0$ instead of $4 \to 0$ (this makes it only easier to build later blocks, as $1 < 4$), it is \emph{not safe} to publish $4$ where $1$ used to be. Indeed, $\pi$ previously used the action $\publishpath(\{1\}, 0)$ in round $6$, and then the action $\publishpath(\{3,6\}, 1)$ in round $7$. Simply swapping all the $1$s and $4$s results in an infeasible action in round $7$ because the edge $3 \to 4$ is invalid. 

Instead, what we want is to apply this reduction ad infinitum. Indeed, the key observation is that the action $\publishpath(\{4\}, 0)$ is itself not Orderly, and should instead be changed to $\publishpath(\{3\}, 0)$, as is done in $\tilde{\pi}$ on the right of Figure~\ref{fig:orderly}. The same reduction should again be applied in round $7$.

This example illustrates our reduction (defined formally in Definition~\ref{def:orderlyreduction}), but also establishes the importance of doing a reduction ``all at once'' rather than a sequence of local changes.
\begin{figure}[H]
    \centering
    \includegraphics[width=0.90\textwidth]{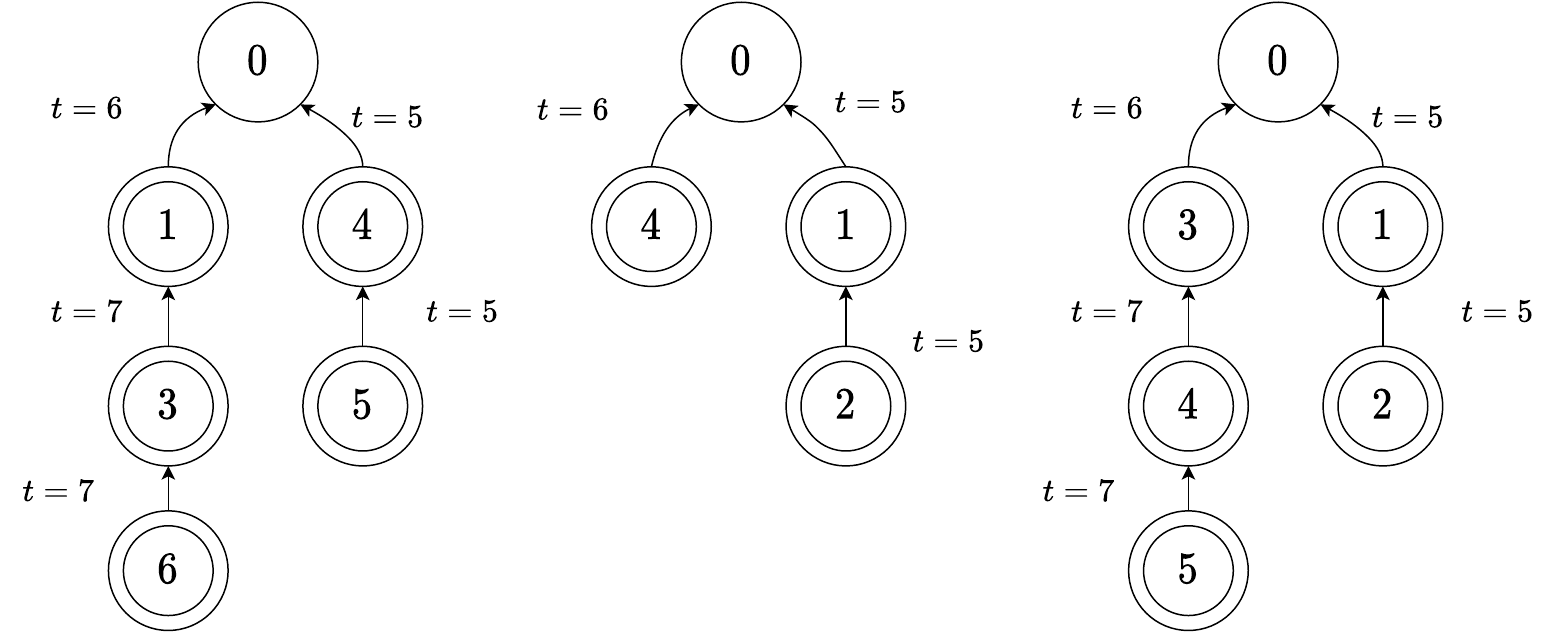}
    \caption{Mining game with a strategy $\pi$ (left) where the action in round $5$ is not Orderly. An intermediate ``reduction'' which is infeasible (center), and a transformed strategy $\tilde{\pi}$ (right) where all actions are feasible and Orderly.}
    \label{fig:orderly}
\end{figure}
\end{example}

We now provide our Orderly Reduction, which generalizes the ideas in Example~\ref{ex:orderly}. 

\begin{definition}[Orderly Reduction]\label{def:orderlyreduction} Let $\pi$ be any timeserving strategy. The \emph{Orderly Reduction} of $\pi$ is the strategy $\tilde{\pi}$ which does the following:
\begin{itemize}
\item Maintain a mapping $\sigma:\mathbb{N}_+\rightarrow \mathbb{N}_+$. Initialize $\sigma(b) = b$ for all $b \in \mathbb{N}_+$.
\item During round $n+1$, if $\pi$ takes action $\wait$ from state $\tree^\half_\pi(n)$, then \wait. That is, if $\pi$ waits in the simulation, then also wait.
\item Else, if $\pi$ takes the action $\publishpath(V', u)$ from state $\tree^\half_\pi(n)$, let $V'':= \min^{|V'|}(\unpublished \cap (\sigma(u),\infty))$. Take the action $\publishpath(V'', \sigma(u))$, and make the following updates to $\sigma$:
\begin{itemize}
\item For all $v \in V'$, if $v$ is the $\ell^{th}$ smallest element of $V'$, update $\sigma(v)$ to be the $\ell^{th}$ smallest element of $V''$. That is, update $\sigma$ so that it maps newly-published blocks under $\pi$ to newly-published blocks under $\tilde{\pi}$, and is monotone increasing within this domain.
\item For all $v \in (\unpublished)_\pi$, if $v$ is the $\ell^{th}$ smallest element of $(\unpublished)_\pi$, update $\sigma(v)$ to be the $\ell^{th}$ smallest element of $\unpublished$. That is, update $\sigma$ so that it maps still-unpublished blocks under $\pi$ to still-unpublished blocks under $\tilde{\pi}$, and is monotone increasing within this domain.
\end{itemize}
\end{itemize}
\end{definition}

Intuitively, this reduction hard-codes that every $\publishpath(\cdot, \cdot)$ action is Orderly, and uses $\sigma(\cdot)$ to remember which ``new blocks'' after the reduction correspond to the ``old blocks'' before the reduction. Specifically, the intent of $\sigma$ is to be an isomorphism between $\tree_\pi$ and $\tree$, and to also map the $\ell^{th}$ smallest element of $(\unpublished)_\pi$ to the $\ell^{th}$ smallest element of $\unpublished$ at all times. Note that the second bullet updating $\sigma(\cdot)$ is simply to maintain $\sigma(\cdot)$ as a useful helper for proofs, and not necessary to actually execute the reduction.

Observe that, by definition, whenever $\tilde{\pi}$ takes a $\publishpath(\cdot, \cdot)$ action, that action is Orderly (as long as it is valid). So our only task is to confirm that every action is valid, and also that the payoffs are identical against \frontier. Observe further that the only way the $\publishpath(\cdot, \cdot)$ action could be invalid is if there simply aren't $|V'|$ elements of $\unpublished_i \cap (\sigma(u),\infty)$: all elements of $(\sigma(u),\infty)$ clearly come after $\sigma(u)$, so the action would otherwise be valid.

We first prove a helper lemma, which describes how $\sigma$ evolves over time. In this lemma (and subsequent proofs), it will be helpful to introduce the following notation. Observe that every time a PublishPath action is taken by Miner $1$, that the function $\sigma(\cdot)$ changes. In order to cleanly reference ``the state of $\sigma(\cdot)$, before the action was taken'', we use the notation $\sigma_{\text{old}}(\cdot)$. In order to cleanly reference ``the state of $\sigma(\cdot)$, after the action was taken'', we use the notation $\sigma_{\text{new}}(\cdot)$. More generally, we will add a subscript of $_\text{old}$ to reference the state of a variable before a referenced action is taken, and $_\text{new}$ to reference the state of a variable after a referenced action is taken.

\begin{lemma}\label{lem:sigma}
Consider any action $\publishpath(V', u)$ action taken by Miner $1$ using $\pi$. If it further holds that for all $v \in V'$, $\sigma_{\text{old}}(v) > \sigma_{\text{old}}(u)$, then for all $v$, the following hold:
\begin{itemize}
\item If $v$ is already published, then $\sigma_{\text{new}}(v) = \sigma_{\text{old}}(v)$.
\item If $v\in V'$, then $\sigma_{\text{new}}(v) \leq \sigma_{\text{old}}(v)$.
\item If $v\notin V'$, then $\sigma_{\text{new}}(v) \geq \sigma_{\text{old}}(v)$.
\end{itemize}
\end{lemma}

\begin{proof}
Observe that bullet one clearly holds, as we don't change $\sigma(v)$ for any $v$ which is already published. The two interesting bullets will follow from the following observation: \emph{If for all $v \in V'$, $\sigma_{\text{old}}(v) > \sigma_{\text{old}}(u)$, then the set $\sigma_{\text{old}}(V')$ is valid to publish on top of $\sigma_{\text{old}}(u)$}. Because $V''$ is the smallest $|V'|$ elements which are valid to publish on top of $\sigma_{\text{old}}(u)$, this immediately implies: For all $v \in V'$, if $v$ is the $\ell^{th}$ smallest element of $V'$, then the $\ell^{th}$ smallest element of $V''$ is at most $\sigma_{\text{old}}(v)$. As $\sigma_{\text{new}}(v)$ is exactly the $\ell^{th}$ smallest element of $V''$, we conclude bullet two.

Now that we have bullet two, bullet three readily follows. Consider when $v$ was previously the $\ell^{th}$ smallest element of $((\unpublished)_\pi)_\text{old}$. If there are $k$ elements smaller than $v$ in $V'$, then $v$ is now the $(\ell-k)^{th}$ smallest element of $((\unpublished)_\pi)_\text{new}$. If there are $\geq k$ elements smaller than $\sigma_{\text{old}}(v)$ in $V''$, then the $(\ell-k)^{th}$ smallest element of $(\unpublished)_\text{new}$ is \emph{at least as large} as the $\ell^{th}$ smallest element of $(\unpublished)_{\text{old}}$ (which is exactly $\sigma_{\text{old}}(v)$). Observe, however, that there are exactly $k$ elements smaller than $\sigma_{\text{old}}(v)$ in $\sigma_{\text{old}}(V')$. By bullet two, this means that there are indeed at least $k$ elements smaller than $\sigma_{\text{old}}(v)$ in $V''$, and we conclude bullet three.

\end{proof}

We use this technical lemma to conclude the following, which describes how $\tree_\pi$ and $\tree$ evolve isomorphically over time. 
\begin{corollary}\label{cor:orderly}
At all points in time, the following hold:
\begin{itemize}
\item The graphs $\tree_\pi$ and $\tree$ are isomorphic. Specifically:
\begin{itemize}
\item For all $v \in V_\pi$, $\sigma(v)$ is published by $\tilde{\pi}$ during the same round that $v$ is publishedby $\pi$, and these are the only nodes in $V$.
\item For all $(u,v) \in E_\pi$, $(\sigma(u),\sigma(v)) \in E$, and these are the only edges in $E$.
\end{itemize}
\item $\sigma(\cdot)$ is injective and surjective (from $\mathbb{N}_+$ onto $\mathbb{N}_+$).
\item For all $v$ created by Miner $2$, $\sigma(v) = v$.
\item For any $v\notin V_\pi$, and any $u \in \mathbb{N}_+$, $v > u \Rightarrow \sigma(v) > \sigma(u)$.
\end{itemize}
\end{corollary}
\begin{proof}
We'll prove the claim by induction. Clearly at initialization, all claims hold because both $\tree_\pi$ and $\tree$ contain just the genesis block and $\sigma(b) = b$ for all $b$.

Now assume that all four claims hold prior to an action being taken (in the execution with $\pi$) for inductive hypothesis, and we will establish that all four claims continue to hold after the action is taken. Clearly, if that action is $\wait$, then the execution with $\tilde{\pi}$ is also wait, and neither the graphs nor $\sigma$ change, so the claim still holds.

If that action is a PublishPath action taken by Miner $2$, it will publish a single block $r$ on top of the longest chain $\chain_\pi$. Observe that by inductive hypothesis, specifically bullet one, that $\chain = \sigma(\chain_\pi)$. Therefore, Miner $2$ will take the action $\publishpath(\{r\}, \chain) :=\publishpath(\{r\}, \sigma(\chain_\pi))$. This maintains that $\sigma$ maps $\tree_\pi$ to $\tree$. Because no changes are made to $\sigma$, bullets two, three, and four continue to hold.

The interesting case to consider is if the action taken is $\publishpath{V'}{u}$ by Miner $1$ using $\pi$. Then the action taken by $\tilde{\pi}$ is $\publishpath(V'', \sigma(u))$. Because we update $\sigma(V'):=V''$ upon publishing (while being monotone increasing), this would guarantee that the first bullet continues to hold, \emph{if we can guarantee that the action $\publishpath(V'', \sigma(u))$ is valid}. However, bullet four readily lets us claim this. Indeed, because $\publishpath(V', u)$ is valid for $\pi$, we have both: (a) all $v \in V'$ are not in $(V _\pi)_{\text{old}}$ and (b) $v > u$ for all $v \in V'$. Bullet four immediately yields that $\sigma_{\text{old}}(v) > \sigma_{\text{old}}(u)$ for all $v \in V'$, meaning that there are at least $|V'|$ nodes which can be published on top of $\sigma_{\text{old}}(u)$, and therefore $\publishpath(V'', \sigma_{\text{old}}(u))$ is valid. This establishes that bullet one continues to hold.

Now we just need to confirm that the three bullets regarding $\sigma(\cdot)$ continue to hold. It is easy to see that $\sigma$ remains injective and surjective. It is also easy to see that $\sigma$ remains unchanged for any blocks created by Miner $2$, so bullet three continues to hold. It is also easy to verify that bullet four continues to hold \emph{when both $v$ and $u$ are unpublished}. This is simply because $\sigma(\cdot)$ maps the $\ell^{th}$ smallest unpublished element under $\pi$ to the $\ell^{th}$ smallest unpublished element under $\tilde{\pi}$. If $u$ is published, then we will use Lemma~\ref{lem:sigma}.

Indeed, the fact that bullet four previously held lets us conclude that $v > u \Rightarrow \sigma_{\text{old}}(v) > \sigma_{\text{old}}(u)$ prior to the action. By Lemma~\ref{lem:sigma}, $\sigma_{\text{new}}(v) \geq \sigma_{\text{old}}(v)$ (if $v$ was not published, and therefore $\notin (V_\pi)_{\text{new}}$). By the same Lemma~\ref{lem:sigma}, $\sigma_{\text{new}}(u) \leq \sigma_{\text{old}}(u)$ if $u$ is published by the action (or was already published). Therefore, we get the following chain of inequalities (the outer two are by Lemma~\ref{lem:sigma}, the middle inequality is by inductive hypothesis).

$$v \notin (V_\pi)_{\text{new}} \text{ AND } u \in (V_\pi)_{\text{new}}\text{ AND } v > u \Rightarrow \sigma_{\text{new}}(v) \geq \sigma_{\text{old}}(v) > \sigma_{\text{old}}(u) \geq \sigma_{\text{new}}(u),$$

as desired. We have now completed the inductive step, and shown that if all four properties hold prior to an action, they all hold after that action as well. Because they all hold at initialization, they hold at all times.
\end{proof}

\begin{proof}[Proof of Orderly Theorem~\ref{thm:orderly}]
The proof readily follows from bullet one of Corollary~\ref{cor:orderly}. Indeed, because $\tree_\pi$ and $\tree$ are isomorphic, and because $\sigma(b)$ is published by $\tilde{\pi}$ during the same action that $\pi$ publishes $b$ for all $b$, it's always the case that $\chain = \sigma(\chain_\pi)$. The fact that they are isomorphic also implies that every action taken by $\tilde{\pi}$ is valid), and also that for all $n, \gamma$: $\rev_\gamma^{(n)}(\pi) = \rev_\gamma^{(n)}(\tilde{\pi})$. 
\end{proof}

\subsection{Longest Chain Mining Reduction}

As in the previous sections, we devise a reduction from any strategy which is Timeserving, and Orderly, but not LCM, to one which is Timeserving, Orderly, and LCM.

\begin{definition}[LCM Reduction]\label{def:lcmreduction}
Let $\pi$ be any Timeserving, Orderly strategy. The LCM Reduction of $\pi$ for Player $1$ is the strategy $\tilde{\pi}$ which does the following:
\begin{itemize}
\item During round $n+1$, if $\pi$ takes action $\wait$ from state $\tree^\half_\pi(n)$, then $\wait$. That is, if $\pi$ waits in the simulation, then also wait.
\item Else, if $\pi$ takes the action $\publish(k,u)$ from state $\tree_\pi^\half(n)$, let $v$ be the unique block in $A(\chain)$ with $h_\pi(u) = h(v)$. Take the action $\publish(k,v)$.
\end{itemize}
\end{definition}

In bullet two above, it should be clear that the desired block $v$ is unique \emph{if it exists}. It is not immediately obvious that $v$ exists (but we will prove that it does).

\begin{figure}[H]
    \centering
    \includegraphics[scale=0.5]{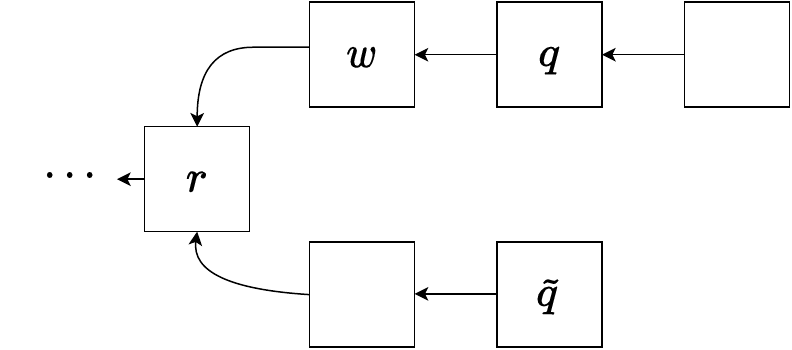}
    \caption{Instance example for the Fork Ownership Lemma. Empty squares represent blocks that are irrelevant to the claim. Lemma~\ref{lem:fork-ownership} states that if Miner 1 plays a strategy which is Timeserving and LCM, and $q$ is a block in the longest chain, and there is a block $\tilde{q}$ with the same height as $q$, then Miner 1 must have created $q$ and $w$ (but perhaps not $r$, and perhaps not $\tilde{q}$).}
    \label{fig:fork-ownership}
\end{figure}

The following observation will be useful, both in proving Theorem~\ref{thm:lcm}, and in drawing conclusions from it.

\begin{observation}\label{obs:lcm} Let $\pi$ be any LCM strategy. Then if $v$ is ever published, but not in the longest chain, $v$ will never again enter the longest chain (formally, if $v \in V$ has $v \notin A(\chain)$, then $v \notin A(\chain)$ for the rest of the game).
\end{observation}
\begin{proof}
If $v \notin A(\chain)$, then all of $v$'s descendants are also not in $A(\chain)$. Therefore, no LCM action can point to $v$ or any of its descendants, and the longest chain will never contain $v$.
\end{proof}

Next, we prove a helper lemma stating that Miner $1$ must have created all blocks in the longest chain \emph{since the most recent fork}. Figure~\ref{fig:fork-ownership} provides an example illustrating the claims of the lemma. The intuition is as follows: if Miner $1$ is using a Timeserving, LCM strategy, and $\tilde{q}$ was published, then $\tilde{q}$ was at some point in the longest chain. Therefore, Miner $2$ (using $\frontier$) would not publish elsewhere (so if a new longest chain is created, it must have come entirely from Miner $1$).

\begin{lemma}[Fork Ownership Lemma]\label{lem:fork-ownership}
Let $\pi$ be any Timeserving, LCM strategy. Let $q \in A(\chain)$ be a block in the longest chain, and let $\tilde{q} \in V$ be another block of the same height (formally: $\tilde{q} \neq q, h(\tilde{q}) = h(q)$). If $r \in A(\chain)$ is the least common ancestor of $\tilde{q}$ and $q$, then Miner $1$ created all blocks between $r$ and $q$ (including $q$, not necessarily including $r$). Formally, $A(\chain) \cap (r, q] \subseteq T_1$.
\end{lemma}
\begin{proof}
Observe that if we can prove the lemma when $\tilde{q}$ is a leaf, then we will have in fact proved the lemma for all $\tilde{q}$ (for arbitrary $\tilde{q}$, simply pick any leaf $\tilde{q}_2$ which is a descendent of $\tilde{q}$, and apply the lemma statement with $q_2$, the node in $A(\chain)$ with $h(q_2) = h(\tilde{q}_2)$. The lemma statement for $q_2,\tilde{q}_2$ implies the lemma statement for $q,\tilde{q}$). So we proceed assuming that $\tilde{q}$ is a leaf.

Let now $\tilde{Q}:=A(\tilde{q}) \cap (r,\tilde{q}]$ denote the ancestors of $\tilde{q}$ (including $\tilde{q}$, up to but not including $r$). Let $Q:=A(q)\cap (r,q]$ denote the same for $q$. Recall that as $r$ is the least common ancestor of $q, \tilde{q}$, these two sets are disjoint. 

Observe before continuing that because Miner $1$ is Timeserving (and Miner $2$ is using $\frontier$, which is also Timeserving), and that $\tilde{q}$ is a leaf, that $\tilde{q}$ is the longest chain immediately following the action in which it is published. In particular, this means that $\tilde{q}$ is published before $q$. 

Our goal is to prove that Miner $1$ created every block in $Q$, so assume for contradiction that there exists a $w \in Q$ which was created by Miner $2$. First, observe that $w$ clearly was not published during the same action as $\tilde{q}$, as $\frontier$ publishes only a single block at a time (and $w \in Q$, while $\tilde{q} \notin Q$, so $w \neq \tilde{q}$). 

Observe also that $w$ cannot have been published \emph{before} $\tilde{q}$. Indeed, if $w$ were published before $\tilde{q}$, then immediately after $\tilde{q}$ is published, $w$ is both published and not in the longest chain. By Observation~\ref{obs:lcm}, $w$ would never again be in the longest chain, contradicting that $q$ becomes the longest chain.

Finally, we show that $w$ cannot be published by an action \emph{after} $\tilde{q}$. Indeed, once $\tilde{q}$ is published, no element of $Q \cup \{r\}$ can possibly be the longest chain (because $\tilde{q}$ is longer), and therefore $\frontier$ would publish on top of $\tilde{q}$ rather than any block in $Q \cup \{r\}$. Therefore, Miner $2$ could not possibly have published a block $w$ in $Q$ (which necessarily would point to a block in $Q \cup \{r\} \setminus \{q\}$) after $\tilde{q}$ was already published.

In summary, we have shown that no block $w \in Q$ could have been created by Miner $2$ during the same round that $\tilde{q}$ was published (because Miner $2$ is using $\frontier$), after $\tilde{q}$ was published (also because Miner $2$ is using $\frontier$), or before $\tilde{q}$ was published (because Miner $1$ and Miner $2$ are both LCM). Therefore, Miner $2$ cannot have created any blocks $w \in Q$, and Miner $1$ must have created them all, completing the proof.
\end{proof}

From here, a complete proof of Theorem~\ref{thm:lcm} becomes unwieldy to do entirely in one shot (like Theorem~\ref{thm:orderly}). Instead, we will complete the proof by changing $\pi$ one action at a time, and interleaving these changes with the Orderly Reduction. Specifically, we will need the following definitions:

\begin{definition}[$N$-LCM and $N$-Orderly] A strategy is $N$-LCM if, when played against $\frontier$, every action it takes up to and including round $N$ is LCM (with probability $1$). A strategy is $N$-Orderly if, when played against $\frontier$, every action it takes up to and including round $N$ is Orderly.

Observe that a strategy is LCM/Orderly if and only if it is $N$-LCM/$N$-Orderly for all $N$.
\end{definition}

Towards Theorem~\ref{thm:lcm}, we will now define a simpler one-step reduction. Importantly, note that we return to the language PublishPath instead of simply Publish.

\begin{definition}[Step-$N$ LCM Reduction] 
Let $\pi$ be any Timeserving, $N$-Orderly, $N$-LCM strategy. The Step-$N$ LCM Reduction of $\pi$ for Player $1$ is the strategy $\tilde{\pi}$ which does the following:
\begin{itemize}
\item During round any round $\neq N+1$, if $\pi$ takes the action $\wait$, then also $\wait$. If $\pi$ takes the action $\publishpath(V', u)$, take action $\publishpath(V', u)$.
\item During round $N+1$, if $\pi$ takes action $\wait$ from state $\tree^\half_\pi(N)$, then $\wait$. That is, if $\pi$ would have waited, then also wait.
\item Else, if $\pi$ takes the action $\publishpath(V', u)$ from state $\tree_\pi^\half(n)$, let $v$ be the unique block in $A(\chain)$ with $h_\pi(u) = h(v)$. That is, let $v$ be the block of height $h_\pi(u)$ within the longest chain of $\tree^\half(n)$. Take the action $\publishpath(V', u)$.
\end{itemize}
\end{definition}

\begin{figure}[H]
    \centering
    \includegraphics[width=0.75\textwidth]{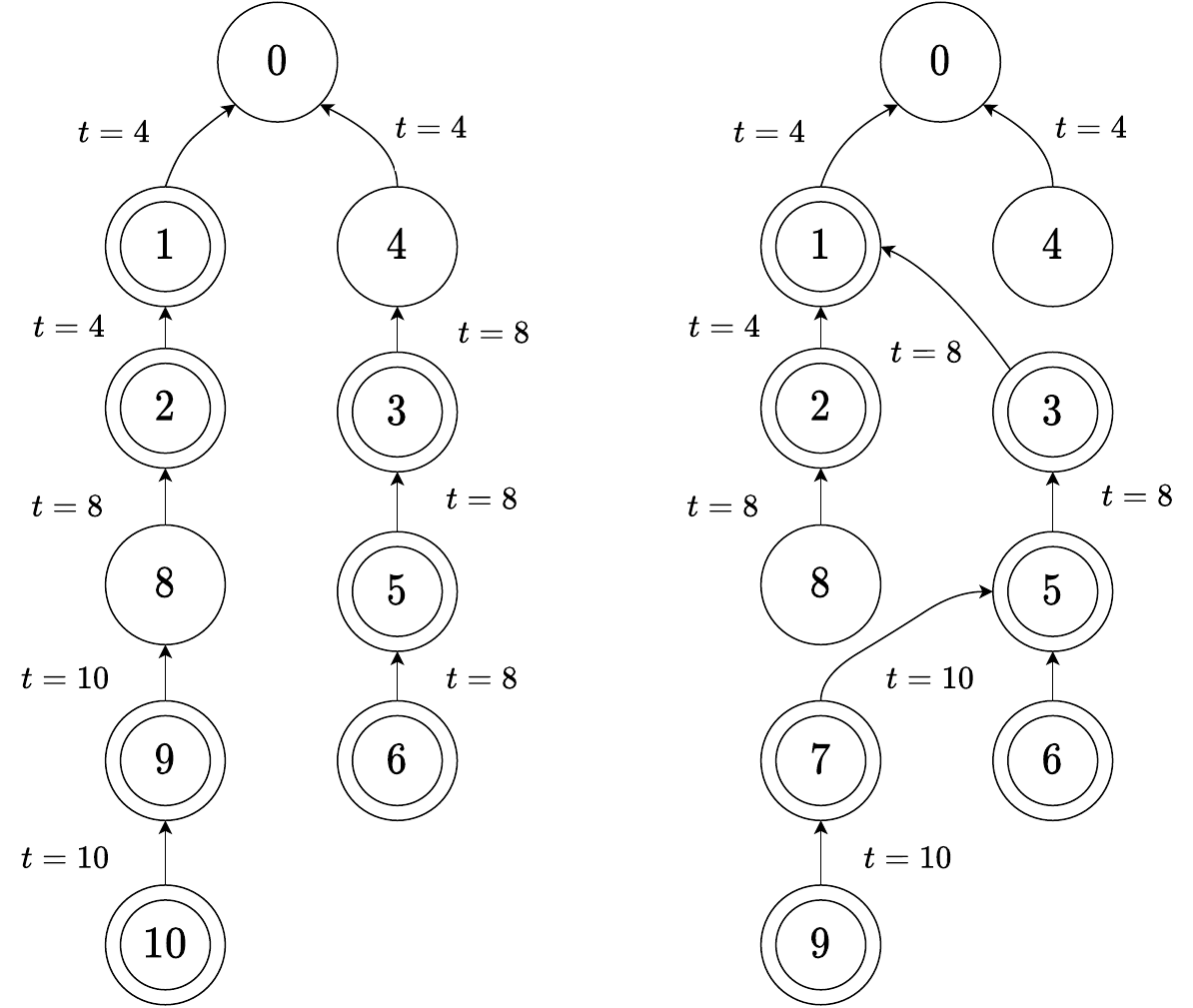}
    \caption{On the left is the history for a Timeserving and Orderly strategy $\pi$. In the middle is the Step-$9$ LCM Reduction of $\pi$, which is not $10$-Orderly. The history on the right further applies the Orderly Reduction.}
    \label{fig:orderly-lcm}
\end{figure}

Importantly, observe that the Step-$N$ Reduction only changes the action taken in step $N+1$, and in a very simple way: it tries to publish exactly the same blocks, just on top of a different node. This makes it significantly simpler to prove claims about validity. See Figure~\ref{fig:orderly-lcm} for an example.

\begin{lemma}\label{lem:nlcm} Let $\pi$ be any Timeserving, $N$-Orderly, $N$-LCM strategy, and let $\tilde{\pi}$ be its Step-$N$ LCM Reduction. Then $\tilde{\pi}$ is Timeserving, $N$-Orderly, $(N+1)$-LCM, and takes a valid action during every step. Moreover, for all $\gamma, n$: $\rev_\gamma^{(n)}(\tilde{\pi}) \geq \rev_\gamma^{(n)}(\pi)$.
\end{lemma}

\begin{proof}
Observe, importantly, that $\tilde{\pi}$ takes identical actions to $\pi$ \emph{except during round $N+1$}, and moreover that during round $N+1$ $\pi$ \emph{attempts to publish the exact same path at the exact same height} (just perhaps at a different location). This means that all the desired properties immediately follow if we can prove that the action taken in round $N+1$ is valid. Indeed, as long as $\tilde{\pi}$ can publish the same path (just perhaps in a different location), this implies that the longest chain will be the same (just perhaps with different ancestors), and therefore that Miner $2$ using $\frontier$ will use the same action in every future round, and therefore that all actions taken by Miner $1$ will be Timeserving and valid in future rounds. It is also clear that if the action taken by $\tilde{\pi}$ is valid, then it is LCM. Also note that we have \emph{not} claimed that $\tilde{\pi}$ is $(N+1)$-Orderly.

So, to confirm that the action taken during round $N+1$ is valid, we just need to confirm that whenever $\pi$ tries to publish a set of nodes $V'$ on top of $u$ (and this is valid), that it is also valid to publish the same $V'$ on top of $v$ (where $v \in A(\chain)$ has the same height as $u$). Note that if we knew that $v \leq u$, the claim would be trivial, but we might have $v > u$.

To this end, we use Lemma~\ref{lem:fork-ownership} and the fact that $\pi$ is $N$-Orderly. Let $r$ denote the least common ancestor of $v$ and $u$. Then whenever $v \neq u$, Lemma~\ref{lem:fork-ownership} asserts that Miner $1$ published not only $v$, but the entire path from $v$ to $r$ (not necessarily including $r$). Moreover, because $\pi$ is $N$-Orderly, there must not be any blocks in $\unpublished(N) \cap (r,v]$ (otherwise, $\pi$ must have previously taken a non-Orderly action). But now, as $u > r$, this immediately implies that any $V'\subseteq \unpublished(N)$ which is valid to publish on top of $u$ is also valid to publish on top of $v$. Specifically, because $\unpublished(N) \cap (r,v] = \emptyset$, and $V' \subseteq \unpublished(N)$, and $w > u > r$ for all $w \in  V'$, $w > v$ for all $w \in V'$ as well. 

This concludes the first half of Lemma~\ref{lem:nlcm}: we have just shown in the previous path that the action taken in round $N+1$ is valid. The previous paragraphs establish that this implies that all actions taken in all rounds are valid, Timeserving, and LCM. Because no actions are changed during the first $N$ rounds, $\tilde{\pi}$ is still $N$-Orderly. The remaining step is to confirm that for all $\gamma, n$: $\rev_\gamma^{(n)}(\tilde{\pi}) \geq \rev_\gamma^{(n)}(\pi)$.

To see this, observe we have just argued that $\chain_\pi(n) = \chain(n)$ for all $n$. The only difference is that perhaps $A(\chain_\pi(n)) \neq A(\chain(n))$. In fact, the only potential difference is that perhaps $A(\chain_\pi(n))$ contains the path $(r,u]$, while $A(\chain(n))$ contains instead the path $(r,v]$, but they must otherwise be the same. Fortunately, Lemma~\ref{lem:nlcm} also asserts that Miner $1$ created \emph{every node} in $(r,u]$. Therefore, we have both that $|A(\chain(n))| = |A(\chain_\pi(n))|$ for all $n$, and also that $|A(\chain(n)) \cap T_1| \geq |A(\chain_\pi(n)) \cap T_1|$ for all $n$. This directly implies that $\rev_\gamma^{(n)}(\tilde{\pi}) \geq \rev_\gamma^{(n)}(\pi)$ for all $\gamma, n$.
\end{proof}

Now, we can complete the proof of Theorem~\ref{thm:lcm}.

\begin{proof}[Proof of Longest Chain Mining Theorem~\ref{thm:lcm}] 
The proof will boil down to showing that we can alternate between the Step-$N$ LCM Reduction and the Orderly Reduction to implement the LCM reduction. Specifically, recursively define $\pi_0:=\pi$, and $\pi_{N+1}$ to be the Orderly Reduction applied to the Step-$N$ LCM Reduction applied to $\pi_{N}$. We make a series of observations.

\begin{observation} For all $N$, $\pi_N$ is Orderly, $N$-LCM, and Timeserving.
\end{observation}
\begin{proof}
We prove by induction. $\pi_0$ is clearly Orderly, $0$-LCM, and Timeserving. Assume then that $\pi_{N}$ is Orderly, $N$-LCM, and Timeserving. Then the Step-$N$ LCM Reduction applied to $\pi_N$ results in a strategy which is $(N+1)$-LCM, and Timeserving. The Orderly Reduction applied to this strategy does not change where blocks are published (up to the isomorphism), but only which blocks are published. Therefore, the Orderly reduction preserves Timeserving and $(N+1)$-LCM, but makes the strategy Orderly, as desired.
\end{proof}

\begin{observation} For all $N$, and all $n \leq N+1$, $\tilde{\pi}$ and $\pi_N$ take the same action during round $n$.
\end{observation}
\begin{proof}
We also prove this by induction. The claim clearly holds for $N = 0$. Assume now that it holds for some $N$ and we will prove it for $N+1$. $\pi_N$ and $\pi_{N+1}$ take the same actions for the first $N+1$ rounds, so the inductive hypothesis immediately implies the desired conclusion for all $n \leq N+1$. We just need to prove that the action taken in round $N+2$ by $\tilde{\pi}$ and $\pi_{N+1}$ are the same. Indeed, they will publish the same number of blocks, and at the same location. Because both are Orderly, they will publish the same set of blocks as well.
\end{proof}

By the two observations above, we immediately conclude that for all $N$, $\tilde{\pi}$ takes a valid action during round $N$, and is also $N$-LCM. Therefore, $\tilde{\pi}$ takes a valid action during every step, and is LCM. Also, it is clear that as long as $\tilde{\pi}$ takes a valid action during each timestep that $\tilde{\pi}$ is Timeserving and Orderly.
\end{proof}

\subsection{Trimmed Strategies}

\begin{proof}[Proof of Theorem~\ref{thm:trimmed}]
We can immediately apply Theorems~\ref{thm:timeserving},~\ref{thm:orderly} and~\ref{thm:lcm} to conclude that there exists a strategy $\hat{\pi}$ such that every non-$\wait$ action it takes satisfies bullet (i) of being Trimmed. We just need to show that $\hat{\pi}$ can be modified to also satisfy bullet (ii).

So consider any action $\hat{\pi}$ takes which does not satisfy bullet (ii). This means that the action is of the form $\publish(k,v)$, $u$ is the unique node in $A(\chain)$ with an edge to $v$, but $u$ was created by Miner $1$. Let us further consider the entire set of descendents of $v$ in $A(\chain)$ created by Miner $1$. That is, let $U$ denote the maximal path in $A(\chain)$, which terminates at $v$, where every node is created by Miner $1$. By hypothesis that the taken action violates bullet (ii), we have that $|U| \geq 1$. Let $w$ denote the maximal node in $U$ (that is, the other end of the path).

We know that, because $\hat{\pi}$ is Timeserving, that $k > |U|$. Consider instead taking the action $\publish(k-|U|,w)$. Observe first that this action is valid, because $\hat{\pi}$ is Orderly. Indeed, not only can $k-|U|$ of Miner $1$'s unpublished nodes be placed on top of $w$, but the exact same set of $k$ nodes which could be placed on top of $u$ can be placed on $w$ (if not, this would violate Orderly). Moreover, now that this action is valid, it is also Trimmed (because $w$ does not have an immediate descendant in $A(\chain)$ created by Miner $1$, by definition).

We just need to confirm that all future actions can be modified to be valid after this change, and that the reward to Miner $1$ is the same. Intuitively, this occurs because the new longest chain has \emph{exactly} the same height as the original, the owners of each block along the path are \emph{exactly} the same as the original (i.e .we replaced one sub-chain of length $|U|$, all created by Miner $1$ with $U$, another sub-chain of length $|U|$, all created by Miner $1$), and those new blocks were only created \emph{earlier} than the original blocks, making them easier to build upon.

Indeed, we claim that taking actions in all future rounds according to the LCM reduction of $\hat{\pi}$ results in feasible actions and identical payoffs. Indeed, the reward immediately after this action is swapped does not change. In all future rounds, because every block in $U$ was created before every block originally published, every future action is still valid. And moreover, the rewards of all future rounds are also identical. Because this holds pointwise, for all rounds, we can apply the same reduction ad infinitum to swap all non-Trimmed actions for Trimmed actions without changing the payoff at all.
\end{proof}

\section{Omitted Proofs from Section~\ref{sec:trimming-state}}\label{sec:trimming-state-appendix}
\subsection{Checkpoints}

\begin{proof}[Proof of Lemma~\ref{lemma:tree-growth}]
Observe a miner following the $\frontier$ strategy never publishes two blocks at the same height. This implies that $h(\chain(X_n))$ is at least the number of blocks mined by Miner 2 (note that these blocks are not necessarily themselves \emph{in} the longest path, but \emph{some} block of the same height must be). Thus $h(\chain(X_n)) \geq \sum_{i = 1}^n \ind{i \in T_2}$. From the Strong Law of Large Numbers, $\frac{1}{n}\sum_{i = 1}^n \ind{i \in T_2} \toas 1-\alpha$. This proves Lemma~\ref{lemma:tree-growth}.
\end{proof}
\begin{proof}[Proof of Corollary~\ref{cor:opt-revenue}]
For the lower bound, we use $\frontier$ to witness that $\rev(\pi) \geq \alpha$ because $\rev(\frontier) = \alpha$. If both miners follow the $\frontier$ strategy, all blocks Miner 1 and Miner 2 creates are part of the longest path. Thus $\rev_{\cstring}^{(n)}(\frontier) = \frac{\sum_{i = 1}^n \ind{i \in T_1}}{n}$. From the Strong Law of Large Numbers, $\frac{1}{n}\sum_{i = 1}^n \ind{i \in T_1} \toas \alpha$.

For the upper bound, we claim $\limsup_{n\to\infty}\rev_{\cstring}^{(n)}(\pi) \leq \frac{\alpha}{1-\alpha}$, for any strategy $\pi$, with probability 1. Observe the number of blocks Miner 1 owns in the longest path is at most the number of blocks Miner 1 creates. Moreover, Miner 2 never publishes two blocks at the same height implying the height of the longest chain is at least the number of blocks Miner 2 creates. Then
$$\rev_{\cstring}^{(n)}(\pi) \leq \frac{\frac{1}{n}\sum_{i = 1}^n \ind{i \in T_1}}{\frac{1}{n}\sum_{i = 1}^n \ind{i \in T_2}}.$$
From the Strong Law of Large Numbers, $\frac{\frac{1}{n}\sum_{i = 1}^n \ind{i \in T_1}}{\frac{1}{n}\sum_{i = 1}^n \ind{i \in T_2}} \toas \frac{\alpha}{1-\alpha}$ as desired. 
\end{proof}

\begin{proof}[Proof of Proposition~\ref{prop:checkpoint-inequality}]
\vspace{1mm}\noindent\textit{Proof of \ref{prop:checkpoint-inequality-1}.} The proof is clear from recursively applying the checkpoint definition.

\vspace{1mm}\noindent\textit{Proof of \ref{prop:checkpoint-inequality-2}.} Assume the converse. If $P_i$ is the highest checkpoint bellow $v$, $v$ is a checkpoint, a contradiction.

\vspace{1mm}\noindent\textit{Proof of \ref{prop:checkpoint-inequality-3}.} Let $P_k = \min\{P_j : P_j > v\}$. Then
\begin{align*}
|A(\chain) \cap (v, P_k]| &= |A(\chain) \cap (P_{k-1}, P_k]| - |A(\chain) \cap (P_{k-1}, v]|\\
&> |\unpublished \cap (P_{k-1}, P_k]| - |\unpublished \cap (P_{k-1}, v]|\\
&= |\unpublished \cap (v, P_k].
\end{align*}
The second line follows from (i) and (ii). The inequality above and (i) implies
\begin{align*}
|A(\chain) \cap (v, P_i] \cap T_1| &= |A(\chain) \cap (v, P_k] \cap T_1| + |A(\chain) \cap (P_k, P_i] \cap T_1|\\
&> |\unpublished \cap (v, P_k]|  + |\unpublished \cap (P_k, P_i]|\\
& =  |\unpublished \cap (v, P_i]|
\end{align*}
as desired.
\end{proof}

\subsection{Checkpoint Preserving Reduction}\label{sec:checkpoint-preserving-proof}

The next example highlights the main ideas behind the proof.
\begin{example}\label{example:checkpoint-preserving}
\begin{figure}[H]
\centering
\includegraphics[width=0.75\textwidth]{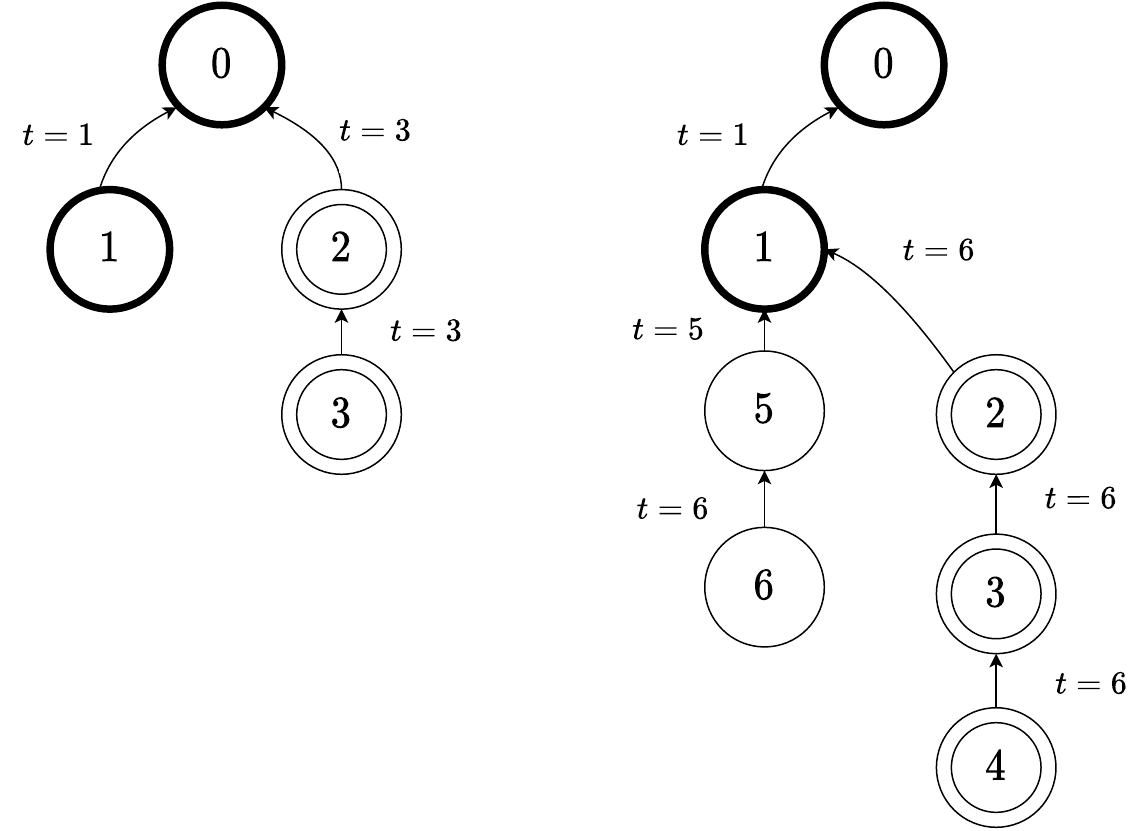}
\caption{The mining game from non-checkpoint preserving (left) and transformation to a checkpoint preserving strategy (right). Double circles denote blocks created by Miner 1 and single circles denote blocks created by Miner 2. Thicker blocks denote blocks that were checkpoints at some point in time.}
\label{fig:checkpoint-preserving}
\end{figure}
In Figure~\ref{fig:checkpoint-preserving}, consider a mining game where Miner 1 creates and withhold blocks 2 and 3, and Miner 2 creates block 1 and publishes $1 \to 0$. Then Miner 1 publishes $3 \to 2 \to 0$ during the third round forking the checkpoint $P_1 = 1$. To transform the original strategy into a checkpoint preserving strategy, observe that a new strategy could publishes $3 \to 2 \to 1$ instead of $3 \to 2 \to 0$. Note Miner 1 has an advantage of two blocks over Miner 2 so there is no risk into waiting to publishing $3 \to 2 \to 1$. Therefore, to further improve the new strategy, let the new strategy wait until the first time step where Miner 1's advantage reduces to a single block. For example, if Miner 1 creates and withhold block $4$ and Miner 2 creates blocks 5 and 6 and publishes $6 \to 5 \to 1$, Miner 1 can safely publish $4 \to 3 \to 2 \to 1$ forking blocks 5 and 6 during the sixth round.
\end{example}
\vspace{1mm}\noindent\textit{Remark.} Note it is not at all obvious the transformation in Example~\ref{example:checkpoint-preserving} guarantees the new strategy is at least as good as the original one. It is true the number of blocks Miner 1 owns in the longest path increases; however, the number of blocks Miner 2 owns in the longest path also increases. Thus the main challenge in proving Theorem~\ref{thm:checkpoint-preserving} is showing that on expectation, the payoff of the new strategy is at least as good as the payoff of the original one.

To show the transformed strategy in Example~\ref{example:checkpoint-preserving} is at least as good as the initial strategy, we will show that whenever Miner 1 takes $\publishpath(Q, v)$ (at time $t$) and forks a checkpoint $c \in A(\chain)$,
\begin{itemize}
\item $\publishpath(Q \cap (c, \infty), c)$ is a valid trimmed action Miner 1 can take instead.
\item Regardless if Miner 1 takes action $\publishpath(Q, v)$ or $\publishpath(Q \cap (c, \infty), c)$, $\max\ Q$ will becomes a checkpoint in the subsequent state.
\end{itemize}
By taking $\publishpath(Q \cap (c, \infty), c)$ instead of $\publishpath(Q, v)$, we show
\begin{itemize}
\item The number of blocks Miner 1 does not publish $|Q \cap (0, c)|$ is at most the number of blocks created by Miner 1 that Miner 1 does not fork $|A(\chain) \cap (v, c] \cap T_1|$. Thus Miner 1 receives at least the same reward by taking $\publishpath(Q \cap (c, \infty), c)$ instead of $\publishpath(Q, v)$.
\item If Miner 2's reward increases by $k = |A(\chain) \cap (v, c] \cap T_2|$ when Miner 1 does not fork $\Succ(v) \cap (v, c)$, then Miner 1 has a lead of $k+1$ blocks. That is
$$|Q \cap (c, \infty)| \geq |\Succ(c)| + k + 1$$
Thus instead of taking action $\publishpath(Q \cap (c, \infty), c)$ at time $t$, Miner 1 can safely wait until the first time step
$$\tau = \min\{t' \geq t: |T_2 \cap (t, t']| = |T_1 \cap (t, t']| + k\}$$
where Miner 2 creates $k$ more blocks than Miner 1 (since time $t+1$). At time $\tau$, we let Miner 1 take $\publishpath((Q \cap (c, \infty)) \cup (T_1 \cap (t, t']), c)$ forking $\Succ(c) \cup (T_2 \cap (t, t'])$. We show the expected number of blocks Miner 1 creates from time $t+1$ to $\tau$, outweighs the cost of allowing Miner 2 to walk away with $k$ additional blocks at time $t$.
\end{itemize}
\subsubsection{Preliminaries}\label{sec:checkpoint-preliminaries}

Next, we define a transformation of trimmed action $\publishpath(Q, v)$ that forks a checkpoint $c \in A(\chain)$ into another trimmed action $\publishpath(Q', v')$ that does not forks a checkpoint (with $Q' \subseteq Q$ and $v' > v$). Recall $\publishpath(Q, v)$ is trimmed if it is timeserving -- i.e., $Q$ becomes part of the longest path which requires $|Q| \geq |\Succ(v)| + 1$.
\begin{definition}[Lift]
For valid, timeserving $\publishpath(Q, v)$ action, let $Q' \subseteq Q$ and $v' \in \Succ(v)$. If $\publishpath(Q', v')$ is timeserving, then $\publishpath(Q', v')$ is a \emph{lift} of $\publishpath(Q, v)$.
\end{definition}
\begin{definition}[Safe Lift]\label{def:safe-lift}
$\publishpath(Q', v')$ is a \emph{safe lift} of $\publishpath(Q, v)$ if $\publishpath(Q', v')$ is a \emph{lift} of $\publishpath(Q, v)$ and taking action $\publishpath(Q', v')$ gives at least the same reward as taking action $\publishpath(Q, v)$. That is, $|(\Succ(v) \setminus \Succ(v')) \cap T_1| \geq |Q \setminus Q'|$.
\end{definition}
\begin{definition}[Checkpoint Lift]\label{def:checkpoint-lift}
Let $\publishpath(Q, v)$ be a trimmed action and assume $\Succ(v)$ contains a checkpoint. Let $c_v := \max\ \{P_i > v:\text{$P_i \in A(\chain)$ is a checkpoint}\}$ be the most recent checkpoint. Define $\publishpath(Q \cap (c_v, \infty), c_v)$ as the \emph{checkpoint lift} of $\publishpath(Q, v)$.
\end{definition}
\begin{lemma}[Checkpoint Lift Lemma]\label{lemma:checkpoint-lift}
Suppose Miner 1's strategy is trimmed. If $\publishpath(Q', v')$ is the \emph{checkpoint lift} of trimmed action $\publishpath(Q, v)$, then $\publishpath(Q', v')$ is trimmed and a \emph{safe lift} of $\publishpath(Q, v)$.
\end{lemma}
\begin{proof}
By assumption, $\Succ(v)$ contains a checkpoint and $v'$ be the most recent checkpoint in $\Succ(v)$. To show that $\publishpath(Q', v')$ is trimmed we need to show that
\begin{enumerate}
\item If $v'$ has any successors, $\min\ \Succ(v')$ was created by Miner 2.
\item $\publishpath(Q', v')$ is timeserving.
\end{enumerate}
For (1), suppose Miner 1 created $\min\ \Succ(v')$. Then $\min\ \Succ(v')$ is a checkpoint because Miner 1's strategy is orderly, a contradiction to $v'$ being the most recent checkpoint. This proves (1). For (2), suppose for contradiction $\publishpath(Q', v')$ is not timeserving. Then
\begin{align*}
|Q| &= |Q \cap (v, v']| + |Q \cap (v', \infty)|\\
&\leq |\unpublished \cap (v, v']| + |Q \cap (v', \infty)| &\quad \parbox{0.55\textwidth}{Because $\publishpath(Q, v)$ is orderly.}\\
&\leq |\unpublished \cap (v, v']| + |\Succ(v')| & \quad \parbox{0.55\textwidth}{From the assumption $\publishpath(Q', v')$ is not timeserving.}\\
&\leq |A(\chain) \cap (v, v'] | + |\Succ(v')| &\quad \parbox{0.55\textwidth}{From Proposition~\ref{prop:checkpoint-inequality} and the fact $v'$ is a checkpoint.}\\
& = |\Succ(v)| \Rightarrow \Leftarrow.
\end{align*}
The chain of inequalities contradicts the assumption that $\publishpath(Q, v)$ is timeserving. This proves (2) and proves $\publishpath(Q', v')$ is trimmed. Next, we show $\publishpath(Q', v')$ is a safe lift. Because $v'$ is a checkpoint,
\begin{align*}
|(\Succ(v) \setminus \Succ(v')) \cap T_1| &= |A(\chain) \cap (v, v'] \cap T_1|\\
&\geq |\unpublished \cap (v, v']| & \quad \text{From Proposition~\ref{prop:checkpoint-inequality}.}\\
&\geq |Q \cap (v, v']| & \quad \text{Because $\publishpath(Q, v)$ is orderly.}\\
&= |Q \setminus Q'| & \quad \text{Because $Q' = Q \cap (v', \infty)$}.
\end{align*}
The chain of inequalities witnesses $\publishpath(Q', v')$ is a safe lift of $\publishpath(Q, v)$.
\end{proof}
\begin{lemma}[Checkpoint Override Lemma]\label{lemma:checkpoint-override}
Let $\publishpath(Q, v)$ be a trimmed action. If $\{v\} \cup \Succ(v)$ contains a checkpoint, the new longest chain becomes a checkpoint after Miner 1 plays $\publishpath(Q, v)$.
\end{lemma}
\begin{proof}
We divide the proof into two cases:

\vspace{1mm} \noindent \textbf{Case 1.} If $v$ is a checkpoint, all blocks Miner 1 publishes become checkpoints because $\publishpath(Q, v)$ is orderly.

\vspace{1mm} \noindent \textbf{Case 2.} If $v$ is not a checkpoint, let $s = \max\{c \in A(v): \text{$c$ is a checkpoint}\}$ be the most recent checkpoint smaller than $v$ and let $c = \min\{c \in \Succ(v) : \text{$c$ is a checkpoint}\}$ be the oldest checkpoint bigger than $v$. Let $t \in Q$ be the block that would take the height of $c$ in the longest path once $Q \to v$ is published. It suffices to show that $t$ becomes a checkpoint because if $t$ becomes a checkpoint, $\max\ Q$ becomes a checkpoint because $\publishpath(Q, v)$ is orderly (i.e., $(\unpublished\setminus Q) \cap (t, \max\ Q) = \emptyset$).

Because $s < v$, $s$ will continue to be a checkpoint and to show that $t$ becomes a checkpoint, it suffices to show that $|(A(\chain) \cap (s, v] \cap T_1) \cup (Q \cap (v, t])| \geq |(\unpublished \setminus Q) \cap (s, t]|$. Indeed,
\begin{align*}
|(A(\chain) \cap (s, v] \cap T_1) \cup (Q \cap (v, t])| &\geq |A(\chain) \cap (s, c] \cap T_1|\\
&\geq |\unpublished \cap (s, c]| & \quad \parbox{0.4\textwidth}{Because $s$ and $c$ are checkpoints.}\\
&\geq |(\unpublished \setminus Q) \cap (s, c]|\\
&\geq |(\unpublished \setminus Q) \cap (s, t]|
\end{align*}
The first line observes $t$ will have the same height as $c$ and Miner 1 {\em will be the creator of all blocks in the path $v \to \ldots \to t$}. For the fourth line, the case where $t < c$ is clear. For the case where $t > c$, we claim $(\unpublished \setminus Q) \cap (c, t) = \emptyset$. If there exists $q \in (\unpublished \setminus Q) \cap (c, t)$, $q > v$ because $v$ is ancestor of $c$. Thus $q < t$ is a block Miner 1 could have published pointing to $v$ instead of $t$, a contradiction to $\publishpath(Q, v)$ being orderly. This proves $t$ (and $\max\ Q$) becomes a checkpoint as desired.
\end{proof}

\subsubsection{The Reduction}
In this section, we proof Theorem~\ref{thm:checkpoint-preserving}.

\begin{definition}[Potential reward]\label{def:potential-reward}
For a state $B$, let $B'$ be a {\em successor} of $B$ if there is a valid action Miner 1 can take at state $B$ such that the subsequent state is $B'$. Let $\Succs(B)$ be the set of all successors of $B$. The {\em potential reward} $\potr^k$ of Miner $k$ maps a state $B$ to the maximum absolute reward Miner $k$ can obtain from state $B$ to a successor of $B$. That is
$$\potr^k(B) = \max_{B' \in \Succs(B)} |\R^k(B, B')|.$$
\end{definition}
\begin{proposition}\label{prop:cp:base-case}
Let $\pi$ be a trimmed strategy. Let $B$ be a state satisfying the bullets:
\begin{itemize}
\item $B$ is reachable from a mining game starting at state $B_0$ with Miner 1 following strategy $\pi$.
\item At $B$, $\pi$ takes trimmed action $\publishpath(Q, v)$ forking the most recent checkpoint $P_i$.
\item Block $v$ reached finality with respect to $\pi$.
\end{itemize} 
Then there exists a valid, trimmed strategy $f(\pi)$ where $f(\pi)$ take the same actions as $\pi$ up to state $B$ and forks no checkpoint at state $B$. To define $f(\pi)$ at state $B$ onward, let $(X_t^{f(\pi)})_{t \geq 0}$ be a mining game starting at $X_0^{f(\pi)} = B$ and where Miner 1 follows strategy $f(\pi)$. Let $N = \max\ \vertex(B)$ and 
$$Q_t = |Q \cap (P_i, \infty)| + |T_1(X_t^{f(\pi)}) \cap (N, N + t]|,$$
$$Z_t = |\Succ^{B}(P_i)| + |T_2(X_t^{f(\pi)}) \cap (N, N + t]| \qquad \text{and} \qquad W_t = Q_t - Z_t - 1.$$
Let $\tau = \min\{t \geq 0 : W_t = 0\}$. At state $(X_t^\half)^{f(\pi)}$, for $t < \tau$, let $f(\pi)$ wait. At state $(X_\tau^\half)^{f(\pi)}$, let $f(\pi)$ take action
$$\publishpath(Q \cap (P_i, \infty) \cup T_1(X_\tau^{f(\pi)}) \cap (N, N + \tau], P_i)$$
forking blocks $\Succ^B(P_i) \cup T_2(X_t^{f(\pi)}) \cap (N, N + \tau]$. To define $f(\pi)$'s actions in future states, let $B'$ be the subsequent state to $B$ after $\pi$ publishes blocks $Q$. Let $(X_t)_{t \geq \tau}$ be a mining game starting at $X_\tau = B'$ and where Miner 1 follows strategy $\pi$. Complete the sequence $(X_t)_{t \geq 0}$ with 
$$X_0 = B \qquad \text{and} \qquad X_1 = X_2 = \ldots = X_{\tau - 1} = B'.$$
Let $(X_t, X_t^{f(\pi)})_{t \geq \tau}$ be a coupling where Miner $k$ creates block $t \geq \tau + 1$ in both games. For all $t \geq \tau + 1$, at state $(X_t^\half)^{f(\pi)}$, let $f(\pi)$ take the same actions $\pi$ takes at state $X_t^\half$.

Given the coupling $(X_t, X_t^{f(\pi)})_{t \geq 0}$, for all $t \geq 0$,
$$\rev_\cstring^{(t)}(f(\pi)) \geq \frac{|A(\chain(X_t)) \cap T_1| + \ \ind{t \geq \tau} \cdot \sum_{i = 1}^{W_0} Z_i - \potr^1(X_t) \cdot \ind{1 \leq t < \tau}}{h(\chain(X_t)) + \ind{t \geq \tau} \cdot \left(W_0 + \sum_{i = 1}^{W_0} Z_i \right)+ \potr^1(X_t) \cdot \ind{1 \leq t < \tau}}.$$
where $W_0 \geq 1$ and $(Z_i)_{i = 1}^{W_0}$ are i.i.d. random variables with expected value $\frac{\alpha}{1-2\alpha}$.

\end{proposition}
\begin{proof}
First, we show $W_0 \geq 1$.
\begin{claim}\label{claim:cp:w-0}
$W_0 \geq |\Succ^B(v) \setminus \Succ^B(P_i)| - |Q \cap (v, P_i]| \geq 1$.
\end{claim}
\begin{proof}
Because $\publishpath(Q, v)$ is timeserving, $Q$ successfully becomes part of the longest path once $\pi$ publishes $Q$ which implies $|Q| \geq |\Succ^B(v)| + 1$. Then
\begin{align*}
W_0 &= |Q \cap (P_i, \infty)| - |\Succ^B(P_i)| - 1\\
&= |Q| - |Q \cap (v, P_i]| - |\Succ^B(P_i)| - 1\\
&\geq |\Succ^B(v)| - |\Succ^B(P_i)| - |Q \cap (v, P_i]| &  \parbox{0.4\textwidth}{Because $\publishpath(Q, v)$ is timeserving.}\\
&= |\Succ^B(v)\setminus \Succ^B(P_i)| - |Q \cap (v, P_i]|.
\end{align*}
This proves the first inequality. To show the second inequality, recall $\publishpath(Q \cap (P_i, \infty), P_i)$ is the checkpoint lift of $\publishpath(Q, v)$ (and thus a safe lift according to Lemma~\ref{lemma:checkpoint-lift}), then, at state $B$, the number of blocks Miner 1 would add to the longest path by taking action $\publishpath(Q \cap (P_i, \infty), P_i)$ is at least the number of blocks Miner 1 would add to the longest path by taking action $\publishpath(Q, v)$. That is
\begin{equation}\label{eq:cp:safe-lift}
(\Succ^B(v) \setminus \Succ^B(P_i)) \cap T_1| \geq |Q \cap (v, P_i]|.
\end{equation}
Note $\Succ^B(v)$ is non-empty (since it contains a checkpoint). The fact $\pi$ is trimmed implies the immediate successor of block $v$ was created by Miner 2. Therefore,
$$|(\Succ^B(v) \setminus \Succ^B(P_i)) \cap T_2| \geq 1.$$
Combining the work above
\begin{align*}
|\Succ^B(v)\setminus \Succ^B(P_i)| - |Q \cap (v, P_i]|  &\geq |\Succ^B(v) \setminus \Succ^B(P_i)|\\
&\qquad - |(\Succ^B(v) \setminus \Succ^B(P_i)) \cap T_1| \\
&= |(\Succ^B(v) \setminus \Succ^B(P_i)) \cap T_2|\\
&\geq 1 
\end{align*}
The chain of inequalities witnesses $W_0 \geq 1$ as desired.
\end{proof}
Observe $\pi$ never forks $\chain(X_\tau)$. Suppose the converse. By assumption, $v$ reached finality with respect to $\pi$. Thus if $\pi$ forks $\chain(X_\tau)$, $\pi$ takes some action $\publishpath(Q', v')$ where $v'$ is in the path from $v$ to $\chain(X_\tau)$; however, the path from $v$ to $\chain(X_\tau)$ only contain blocks created by Miner 1 which implies the immediate successor of block $v'$ was created by Miner 1, a contradiction to $\pi$ being trimmed. This proves $\pi$ never forks $\chain(X_\tau)$ as desired. Next, observe
$$|\unpublished(X_\tau) \cap (\chain(X_\tau), \infty)| = |\unpublished(X_\tau^{f(\pi)}) \cap (\chain(X_\tau^{f(\pi)}), \infty)|.$$
Therefore, up to relabeling of blocks, states $X_\tau$ and $X_\tau^{f(\pi)}$ are equivalent with respect to trimmed strategy $\pi$. Thus it is valid for $f(\pi)$ to take the same actions as $\pi$ after time $\tau$. 

Let's check $f(\pi)$ is valid and trimmed. By inspection, $f(\pi)$ is valid because $\pi$ is valid. Up to state $B$, $f(\pi)$ is trimmed since $f(\pi)$ is equals to $\pi$ up to state $B$. From first to $(\tau-1)$-th step, $f(\pi)$ publishes no blocks and at step $\tau$, $f(\pi)$ publishes blocks pointing to $P_i$. If $\Succ(P_i)$ is the empty-set, $f(\pi)$'s action is clearly trimmed. If $\Succ(P_i)$ is non-empty, $\min\ \Succ(P_i)$ is a block created by Miner 2. Suppose the contrary, then $\min\ \Succ(P_i)$ existed at state $B$ (since this is the first time $f(\pi)$ publishes blocks since state $B$). But $B$ was reachable from a trimmed (and orderly) strategy which implies $\min\ \Succ(P_i)$ is also a checkpoint, a contradiction to $P_i$ being the most recent checkpoint at state $B$. Thus $f(\pi)$ is trimmed up to time $\tau$. After time $\tau$, $f(\pi)$ takes the same actions as $\pi$ which, by assumption, is a trimmed strategy. This proves $f(\pi)$ is a trimmed strategy as desired.

Let's now compare the revenue of $f(\pi)$ with $\pi$. From the first to $(\tau-1)$-th step, $\pi$ publishes $Q$ pointing to $v$, but $f(\pi)$ publishes no blocks. During the same time interval, in $(X_t^{f(\pi)})_{t \geq 0}$, Miner 2 publishes all blocks created from time $1$ to time $\tau-1$ while in $(X_t)_{t \geq 0}$, Miner 2 creates no blocks from time $1$ to $\tau$. Since $\potr^1(X_t^{f(\pi)})$ denotes the maximum reward Miner 1 can obtain from state $X_t$,
$$|A(\chain(X_t^{f(\pi)})) \cap T_1| + \potr^1(X_t^{f(\pi)}) \geq |A(\chain(X_t)) \cap T_1|.$$
$$h(\chain(X_t^{f(\pi)})) - \potr^1(X_t^{f(\pi)}) \leq h(\chain(X_t)).$$
At the $\tau$-th step, $f(\pi)$ publishes $Q \cap (P_i, \infty) \cup T_1(X_\tau^{f(\pi)}) \cap (N, N + \tau]$ pointing to $P_i$ and forking $\Succ^B(P_i) \cup T_2(X_\tau^{f(\pi)}) \cap (N, N + \tau]$. At the same time, $\pi$ publishes no blocks. Thus
\begin{align*}
|A(\chain(X_t^{f(\pi)})) \cap T_1| &= |A(\chain(X_t)) \cap T_1| - |Q \cap (v, P_i]| + |\Succ^B(v) \setminus \Succ^B(P_i) \cap T_1|\\
&\qquad + |T_1(X_t^{f(\pi)}) \cap (N, N + \tau]|\\
&\geq |A(\chain(X_t)) \cap T_1| + |T_1(X_t^{f(\pi)}) \cap (N, N + \tau]| \quad \text{From Equation~\ref{eq:cp:safe-lift}}.
\end{align*}
$$|A(\chain(X_t^{f(\pi)})) \cap T_2| = |A(\chain(X_t)) \cap T_2| + |(\Succ^B(v) \setminus \Succ^B(P_i)) \cap T_2|.$$
Adding $|A(\chain(X_t^{f(\pi)})) \cap T_1|$ and $|A(\chain(X_t^{f(\pi)})) \cap T_2|$, the height of the longest path is
\begin{align*}
h(\chain(X_t^{f(\pi)})) &= |A(\chain(X_t^{f(\pi)})) \cap T_1| + |A(\chain(X_t^{f(\pi)})) \cap T_2|\\
&= h(\chain(X_t)) + |\Succ^B(v) \setminus \Succ^B(P_i)| + |Q \cap (v, P_i]|\\
&\qquad + |T_1(X_t^{f(\pi)}) \cap (N, N + \tau]|\\
&\leq h(\chain(X_t)) + W_0 + |T_1(X_t^{f(\pi)}) \cap (N, N + \tau]|\qquad \text{From Claim~\ref{claim:cp:w-0}}.
\end{align*}
We will now derive a closed form for $|T_1(X_t^{f(\pi)}) \cap (N, N + \tau]|$.
\begin{claim}\label{claim:cp:random-walk}
$|T_1(X_t^{f(\pi)}) \cap (N, N + \tau]| = \sum_{i = 1}^{W_0} Z_i$ where $(Z_i)_{i = 1}^{W_0}$ are i.i.d. random variables with expected $\frac{\alpha}{1-2\alpha}$.
\end{claim}
\begin{proof}
Observe $(W_t)_{t \geq 0}$ is a one-dimensional biased random walk since $W_t$ increments whenever Miner 1 creates a block (with probability $\alpha$) and decrements whenever Miner 2 creates a block (with probability $1-\alpha$). That is,
$$W_t = \begin{cases}
W_{t-1} + 1 & \text{with probability $\alpha$,}\\
W_{t-1} - 1 & \text{with probability $1-\alpha$.}
\end{cases}$$
For $i \geq 0$, let $\tau_i = \min\{t \geq 0: W_t = W_0 - i\}$, then $\tau_{i+1} - \tau_i$ denotes the number of time steps the random walk takes to first reach state $W_0 - i$ since the first time step it reached stated $W_0 - i + 1$. Let $Z_{i+1} = |T_1(X_t^{f(\pi)}) \cap (N + \tau_i, N + \tau_{i+1}]|$ and note $Z_1, Z_2, \ldots$ are i.i.d. with
$$|T_1(X_t^{f(\pi)}) \cap (N, \tau_{W_0}]| = \sum_{i = 1}^{W_0} |T_1(X_t^{f(\pi)}) \cap (N + \tau_{i-1}, N + \tau_i]| = \sum_{i = 1}^{W_0} Z_i.$$
From Lemma~\ref{lemma:random-walk}, $\e{|T_1(X_t^{f(\pi)}) \cap (N + \tau_i, N + \tau_{i+1}]} = \frac{\alpha}{1-2\alpha}$ as desired.
\end{proof}
Combining the inequalities above, we obtain that for all $t \geq 0$,
\begin{align*}
\rev_\cstring^{(t)}(f(\pi)) &= \frac{|A(\chain(X_t^{f(\pi)})) \cap T_1|}{h(\chain(X_t^{f(\pi)}))}\\
&\geq \frac{|A(\chain(X_t)) \cap T_1| - \ind{t < \tau} \cdot \potr^1(X_t^{f(\pi)}) + \ind{t \geq \tau} \cdot \sum_{i = 1}^{W_0} Z_i}{h(\chain(X_t)) + \ind{t < \tau} \cdot \potr^1(X_t^{f(\pi)}) + \ind{t \geq \tau} \cdot \left( W_0 + \sum_{i = 1}^{W_0} Z_i \right)}
\end{align*}
as desired.
\end{proof}
To simplify the expression for the revenue of $f(\pi)$, we first show $\potr^1(X_t)$ is a negligible term by deriving that $\frac{\potr^1(X_t)}{t} \toas 0$.
\begin{lemma}[Asymptotic Reward Lemma]\label{lemma:reward-bound}
Let $(X_t)_{t \geq 0}$ be a mining game starting at state $X_0 = B_0$. For any $\epsilon > 0$, $\pr{\cup_{i = n}^\infty\{\potr^1(X_i) > i\epsilon\}} \leq e^{-\Omega(n)}$. Moreover,
\begin{itemize}
\item $\frac{\potr^1(X_n)}{n} \overset{a.s.}{\to} 0$.
\item $\frac{r^1(X_{n-1}, X_n)}{n} \overset{a.s.}{\to} 0$.
\end{itemize} 
\end{lemma}
\begin{proof}
Event $\potr^1(X_i) > i \epsilon$ implies there is a timeserving action $\publishpath(Q, v)$ Miner 1 can take at time step $i$ with $|Q| > i \epsilon$. The fact $\publishpath(Q, v)$ is timeserving implies Miner 1 creates more blocks than Miner 2 from time $v$ to $i$. To derive an upper bound on $v$, observe at most $i$ blocks were created by time $i$. Therefore, $|Q| + v \leq i$ which implies $v < (1-\epsilon)i$ since $|Q| > i\epsilon$. Thus
\begin{align*}
&\pr{\cup_{i = n}^\infty \left\{\potr^1(X_i) > i\epsilon \right\}} \leq \sum_{i = n}^\infty \pr{\potr^1(X_i) > i \epsilon} & \text{From union bound.}\\
&\qquad\leq \sum_{i = n}^\infty \pr{\exists v < (1-\epsilon)i, |T_1 \cap (v, i] > |T_2 \cap (v, i]|}\\
&\qquad\leq \sum_{i = n}^\infty \sum_{v = 0}^{(1-\epsilon)i} \pr{|T_1 \cap (v, i] > |T_2 \cap (v, i]|} & \text{From union bound.}
\end{align*}
Observe $|T_k \cap (v, i]| = \sum_{j = v+1}^i \ind{j \in T_k}$ is the sum of Bernoulli random variables where $\pr{j \in T_1} = 1 - \pr{j \in T_2} = \alpha$. Thus event $|T_1 \cap (v, i]| > |T_2 \cap (v, i]|$ is equivalent to $\sum_{j = v+1}^i \ind{j \in T_1} > \frac{i-v}{2}$. To bound $\pr{\sum_{j = v+1}^i \ind{j \in T_1} > \frac{i-v}{2}}$, we will use the Chernoff bound and the fact $\alpha < 1/2$.
\begin{theorem}[Chernoff Bound]\label{thm:chernoff-bound}
If $X_1, X_2, \ldots, X_n$ are independent indicator random variables where $\e{X_i} = \mu$, then for any $\delta > 0$,
$$\pr{\sum_{i = 1}^n X_i \geq (1+\delta)\mu} \leq e^{-\frac{\delta^2 \mu}{2+\delta}}.$$
\end{theorem}
\begin{claim}\label{claim:reward-bound}
Let $X_1, X_2, \ldots, X_n$ be i.i.d. copies of Bernoulli random variable $X$ where $\mathbb E[X] = \alpha < 1/2$. Then $Pr[\sum_{i = 1}^{n} X_i > n/2] \leq e^{-\frac{(1-2\alpha)n}{4}}$
\end{claim}
\begin{proof}
Let $\mu = E[\sum_{i = 1}^n X_i] = \alpha n$ and $\delta = (1-2\alpha)/(2\alpha)$. From Chernoff Bound and the fact $\alpha < 1/2$,
$$\pr{\sum_{i = 1}^n X_i > n/2} = \pr{\sum_{i = 1}^n X_i > \mu(1+\delta)}\leq e^{-\frac{\delta^2 \mu}{2+\delta}} < e^{-\frac{(1-2\alpha)n}{4}}.$$
\end{proof}
Therefore, $\pr{\sum_{j = v+1}^i \ind{j \in T_1} > \frac{i-v}{2}} \leq e^{-(1-2\alpha)(i - v)/2} < e^{-(1-2\alpha)\epsilon i}$ where the last inequality is the fact $v < (1-\epsilon)i$. We conclude
\begin{align*}
\pr{\cup_{i = n}^\infty \left\{\potr^1(X_i) > i\epsilon \right\}} &\leq \sum_{i = n}^\infty \sum_{v = 0}^{(1-\epsilon)i} e^{-(1-2\alpha)\epsilon i} \leq \sum_{i = n}^\infty e^{-\Omega(i)} = e^{-\Omega(n)}
\end{align*}
as desired. For the ''Moreover'' part, we just observe $\frac{\potr^1(X_n)}{n} \toas 0$ is equivalent to
$$\lim_{n \to \infty} \pr{\cup_{i = n}^\infty \left\{\potr^1(X_i) > i \epsilon \right\}} = 0 \qquad \text{for all $\epsilon > 0$.}$$
This is clear since $e^{-\Omega(n)} \to 0$. For the second part, note from state $X_{n-1}$ to $X_n$, a single block is created; therefore, $r^1(X_{n-1}, X_n) \leq \potr^1(X_n) + 1$. Thus
$$\frac{r^1(X_{n-1}, X_n)}{n} \leq \frac{\potr^1(X_n) + 1}{n} \toas 0$$
as desired.
\end{proof}
\begin{proof}[Proof of Theorem~\ref{thm:checkpoint-preserving}]
Let $\pi$ be a trimmed strategy. We claim there exists a strategy $f(\pi)$ that is checkpoint preserving. We will interactively transform $\pi$ into $f(\pi)$. Initialize $f(\pi)$ to be equal to $\pi$.

\textbf{Step 1.} Whenever $f(\pi)$ reaches a state $B$ where $f(\pi)$ is about to take action $\publishpath(Q, v)$ where $\Succ(v)$ contains a checkpoint, we will transform $f(\pi)$ so that it is does not fork a checkpoint. We consider two cases:

\textbf{Case 1.} Consider the case block $v$ reached finality with respect to $f(\pi)$. Then $f(\pi)$ and $B$ satisfy the conditions for Proposition~\ref{prop:cp:base-case}. Thus there is a trimmed strategy $\pi'$ where $\pi'$ is equals to $f(\pi)$ up to state $B$ and at state $B$, $\pi'$ follows the algorithm in Proposition~\ref{prop:cp:base-case}. Redefine $f(\pi)$ to be equal to $\pi'$ and repeat until $f(\pi)$ reaches another state where $f(\pi)$ is about to fork a checkpoint. If no such state exists, $f(\pi)$ is checkpoint preserving.

\textbf{Case 2.} Consider the case block $v$ does not reach finality with respect to $f(\pi)$. From Lemma~\ref{lemma:checkpoint-override}, $\max\ Q$ becomes a checkpoint after $f(\pi)$ publishes $Q$ pointing to $v$. Observe that in the future, $f(\pi)$ will take action $\publishpath(Q', v')$ with $h(v') < h(v)$ at most $h(v)$ times; otherwise, there is a state where $f(\pi)$ takes action $\publishpath(Q', v')$ and the immediate successor of block $v'$ was created by Miner 1, a contradiction to $f(\pi)$ being trimmed. Let $B'$ be the last state reachable from $B$ where $f(\pi)$ takes action $\publishpath(Q', v')$ with $h(v') < h(v)$. Clearly $v'$ reaches finality with respect to $f(\pi)$. Note $f(\pi)$ forks a checkpoint by publishing $Q'$ pointing to $v'$ since $\max\ Q$ became a checkpoint and any action thereafter that forks $\max\ Q$ induces the longest chain to become a checkpoint. Thus state $B'$ and strategy $f(\pi)$ satisfy the conditions for Proposition~\ref{prop:cp:base-case}. Thus there is a trimmed strategy $\pi'$ that is equals to $f(\pi)$ up to state $B'$ and at state $B'$, $\pi'$ follows the algorithm in Proposition~\ref{prop:cp:base-case}. Redefine $f(\pi)$ to be equal to $\pi'$ and return to Step 1 with strategy $f(\pi)$ and state $B$.

Since we visit Case 2 with state $B$ at most $h(v)$ times, eventually, block $v$ reaches finality with respect to $f(\pi)$ and we execute Case 1. Once Case 1 is executed with strategy $f(\pi)$ and state $B$, $f(\pi)$ becomes checkpoint preserving on a larger set of states. Ad infinitum $f(\pi)$ becomes checkpoint preserving as desired.

This proves there is a strategy $f(\pi)$ that is valid, trimmed and checkpoint preserving as desired. Next, we check $\rev(f(\pi)) \geq \rev(\pi)$. Let $(X_t^{f(\pi)})_{t \geq 0}$ be a mining game starting at $X_0^{f(\pi)} = B_0$ and where Miner 1 follows $f(\pi)$. Let $t_i$ be the $i$-th time step where $f(\pi)$ would take action $\publishpath(Q, v)$ and fork a checkpoint $P_i \in \Succ(v)$, but, instead, we execute the algorithm in Proposition~\ref{prop:cp:base-case}. That is, $f(\pi)$ waits until time $\tau_i \geq t_i$ where $f(\pi)$ publishes $(Q \cap (P_i, \infty]) \cup (T_1(X_{\tau_i}^{f(\pi)}) \cap (t_i, \tau_i])$ pointing to $P_i$. Let $\tau_0 = 0$, $N_t = \sum_{i = 1}^\infty \ind{\tau_i \leq t}$ and observe $\tau_{N_t} \leq t < \tau_{N_t+1}$.
\begin{corollary}\label{cor:cp-revenue}
For any strategy $\pi$, there is a trimmed, checkpoint preserving strategy $f(\pi)$ where
$$\rev_\cstring^{(t)}(f(\pi)) \geq \frac{|A(\chain(X_t)) \cap T_1| + \sum_{j = 1}^{N_t}\sum_{i = 1}^{W_0^j} Z_i^j - \potr^1(X_t)}{h(\chain(X_t)) + \sum_{j = 1}^{N_t} W_0^j + \sum_{j = 1}^{N_t} \sum_{i = 1}^{W_0^j} Z_i^j + \potr^1(X_t)}.$$
where $(X_t)_{t \geq 0}$ is a mining game with $X_0 = B_0$ where Miner 1 follows $\pi$. For all $j$, $W_0^j \geq 1$ and $(Z_i^j)_{i, j}$ are i.i.d random variables with expected value $\frac{\alpha}{1-2\alpha}$.
\end{corollary}
\begin{proof}
The work above transform a strategy $\pi$ into a trimmed, checkpoint preserving strategy $f(\pi)$. The revenue of $f(\pi)$ with respect to the initial strategy $\pi$ follows from applying Proposition~\ref{prop:cp:base-case} each time Case 1 is executed.
\end{proof}
From supperaddivity of $\liminf$, 
\begin{align*}
\liminf_{t \to \infty} \rev_\cstring^{(t)}(f(\pi)) &\geq \liminf_{t\to\infty} \rev_\cstring^{(t)}(\pi) + \liminf_{t \to \infty} \left(\rev_\cstring^{(t)}(f(\pi)) - \rev_\cstring^{(t)}(\pi)\right)
\end{align*}
It suffices to show $\liminf_{t \to \infty} \left(\rev_\cstring^{(t)}(f(\pi)) - \rev_\cstring^{(t)}(\pi)\right) \geq 0$. Note Corollary~\ref{cor:cp-revenue} provides a lower bound on $\rev_\cstring^{(t)}(f(\pi))$.
\begin{proposition}\label{prop:cp-revenue}
$\liminf_{t \to \infty} \left(\frac{|A(\chain(X_t)) \cap T_1| + \sum_{j = 1}^{N_t}\sum_{i = 1}^{W_0^j} Z_i^j - \potr^1(X_t)}{h(\chain(X_t)) + \sum_{j = 1}^{N_t} W_0^j + \sum_{j = 1}^{N_t} \sum_{i = 1}^{W_0^j} Z_i^j + \potr^1(X_t)} - \rev_\cstring^{(t)}(\pi)\right) \geq 0$.
\end{proposition}
\begin{proof}
Recall Miner 2 never publishes two block at the same height; therefore, $h(\chain(X_t)) \geq |T_2(X_t)|$. Since $|T_2(X_t)|$ is the sum of $t$ i.i.d. random variables with expected value $1-\alpha$, from the strong law of large numbers
$$\frac{|T_2(X_t)|}{t} \toas 1-\alpha \qquad \text{and} \qquad \frac{|T_1(X_t)|}{t} = \frac{t - |T_2(X_t)|}{t} \toas \alpha$$
From Lemma~\ref{lemma:reward-bound}, $\frac{\potr^1(X_t)}{h(\chain(X_t))} \toas 0$ and
\begin{align*}
&\liminf_{t \to \infty} \left(\frac{|A(\chain(X_t)) \cap T_1| + \sum_{j = 1}^{N_t}\sum_{i = 1}^{W_0^j} Z_i^j - \potr^1(X_t)}{h(\chain(X_t)) + \sum_{j = 1}^{N_t} W_0^j + \sum_{j = 1}^{N_t} \sum_{i = 1}^{W_0^j} Z_i^j + \potr^1(X_t)} - \rev_\cstring^{(t)}(\pi)\right)\\
&= \liminf_{t \to \infty}\left( \frac{|A(\chain(X_t)) \cap T_1| + \sum_{j = 1}^{N_t}\sum_{i = 1}^{W_0^j} Z_i^j}{h(\chain(X_t)) + \sum_{j = 1}^{N_t} W_0^j + \sum_{j = 1}^{N_t} \sum_{i = 1}^{W_0^j} Z_i^j} - \frac{|A(\chain(X_t)) \cap T_1|}{h(\chain(X_t))}\right)\\
& =\liminf_{t \to \infty} \frac{t \cdot \sum_{j= 1}^{N_t} W_0^j}{t \cdot \sum_{j = 1}^{N_t} W_0^j}\\
&\qquad \left( \frac{h(\chain(X_t))\sum_{j = 1}^{N_t}\sum_{i = 1}^{W_0^j} Z_i^j - |A(\chain(X_t)) \cap T_1|\left(\sum_{j = 1}^{N_t} W_0^j + \sum_{j = 1}^{N_t} \sum_{i = 1}^{W_0^j} Z_i^j\right)}{\left(h(\chain(X_t)) + \sum_{j = 1}^{N_t} W_0^j + \sum_{j = 1}^{N_t} \sum_{i = 1}^{W_0^j} Z_i^j\right) h(\chain(X_t))} \right)\\
& = \liminf_{t \to \infty} a_t.
\end{align*}
First, consider the case $\lim_{t \to \infty} N_t < \infty$ which implies 
$$\limsup_{t \to \infty} \sum_{j = 1}^{N_t} \sum_{i = 1}^{W_0^j} Z_i^j \quad \text{and} \quad \limsup_{t \to \infty} \sum_{j = 1}^{N_t} W_0^j < \infty.$$
Thus
$$\frac{\sum_{j = 1}^{N_t} \sum_{i = 1}^{W_0^j} Z_i^j}{h(\chain(X_t))} \toas 0 \qquad \frac{\sum_{j = 1}^{N_t} W_0^j }{h(\chain(X_t))} \toas 0.$$
Therefore, $\liminf_{t \to \infty} a_t = 0$ as desired. Next, we consider the case $\lim_{t \to \infty} N_t = \infty$. From supermultiplicativity of $\liminf$,
\begin{align*}
\liminf_{t \to \infty} a_t &\geq \frac{\liminf_{t \to \infty} \frac{h(\chain(X_t))\sum_{j = 1}^{N_t}\sum_{i = 1}^{W_0^j} Z_i^j - |A(\chain(X_t)) \cap T_1|\left(\sum_{j = 1}^{N_t} W_0^j + \sum_{j = 1}^{N_t} \sum_{i = 1}^{W_0^j} Z_i^j\right)}{t \cdot \sum_{j= 1}^{N_t} W_0^j}}{\limsup_{t \to \infty} \frac{\left(h(\chain(X_t)) + \sum_{j = 1}^{N_t} W_0^j + \sum_{j = 1}^{N_t} \sum_{i = 1}^{W_0^j} Z_i^j\right) h(\chain(X_t))}{t \cdot \sum_{j= 1}^{N_t} W_0^j}}\\
&= \frac{\liminf_{t \to \infty} b_t}{\limsup_{t \to \infty} c_t}.
\end{align*}
Next, we check $\frac{\liminf_{t \to \infty} b_t}{\limsup_{t \to \infty} c_t}$ is well-defined. For that, we will show $\liminf_{t \to \infty} b_t$ is lower bounded by 0 and $\limsup_{t \to \infty} c_t$ is bounded away from 0. Recall $\liminf_{t \to \infty} \frac{h(\chain(X_t))}{t} \geq 1-\alpha$ almost surely. Thus
$$\limsup_{t \to \infty} c_t = \limsup_{n \to \infty} \frac{h(\chain(X_t) + \sum_{j = 1}^{N_t} W_0^j + \sum_{j = 1}^{N_t} \sum_{i = 1}^{W_0^j} Z_i^j}{\sum_{j = 1}^{N_t} W_0^j}\frac{h(\chain(X_t))}{t} \geq 1-\alpha > 0.$$
From Proposition~\ref{prop:cp:base-case}, $\sum_{j = 1}^{N_t} \sum_{i = 1}^{W_0^j} Z_i^j$ is the sum of $\sum_{j = 1}^{N_t} W_0^j$ i.i.d. random variables with expected value $\frac{\alpha}{1-2\alpha}$. From the strong law of large numbers,
$$\frac{h(\chain(X_t))}{t}\frac{\sum_{j = 1}^{N_t}\sum_{i = 1}^{W_0^j} Z_i^j}{\sum_{j = 1}^{N_t} W_0^j} \geq \frac{|T_2(X_t)|}{\sum_{j = 1}^{N_t}\sum_{i = 1}^{W_0^j} Z_i^j } \toas (1-\alpha)\frac{\alpha}{1-2\alpha}.$$
Observe $A(\chain(X_t) \cap T_1 \subseteq T_1(X_t)$. Then
\begin{align*}
\frac{|A(\chain(X_t)) \cap T_1|}{t}\frac{\sum_{j = 1}^{N_t} W_0^j + \sum_{j = 1}^{N_t} \sum_{i = 1}^{W_0^j} Z_i^j}{\sum_{j = 1}^{N_t} W_0^j} &\geq \frac{|T_1(X_t)|}{t} \left(1 + \frac{\sum_{j = 1}^{N_t} \sum_{i = 1}^{W_0^j} Z_i^j}{\sum_{j = 1}^{N_t} W_0^j}\right)\\
&\toas \alpha \left(1 + \frac{\alpha}{1-2\alpha}\right)
\end{align*}
The work above proves
\begin{align*}
\liminf_{t \to \infty} b_t &= \liminf_{t \to \infty} \frac{h(\chain(X_t))\sum_{j = 1}^{N_t}\sum_{i = 1}^{W_0^j} Z_i^j - |A(\chain(X_t)) \cap T_1|\left(\sum_{j = 1}^{N_t} W_0^j + \sum_{j = 1}^{N_t} \sum_{i = 1}^{W_0^j} Z_i^j\right)}{t \cdot \sum_{j= 1}^{N_t} W_0^j}\\
&\geq \frac{\alpha(1-\alpha)}{1-2\alpha} - \frac{\alpha(1-\alpha)}{1-2\alpha} = 0
\end{align*}
Thus $\liminf_{t \to \infty} a_t \geq 0$ as desired.
\end{proof}
From Proposition~\ref{prop:cp-revenue}, $\liminf_{t \to \infty} \rev_\cstring^{(t)}(f(\pi)) \geq \liminf_{t \to \infty} \rev_\cstring^{(t)}(\pi)$. Taking the expected value proves $\rev(f(\pi)) \geq \rev(\pi)$ as desired.
\end{proof}

\subsection{Opportunistic Reduction}\label{sec:opportunistic-proof}

First, we give an example that highlights the main ideas of the proof.
\begin{example}\label{example:opportunistic}
\begin{figure}[H]
\centering
\includegraphics[width=0.75\textwidth]{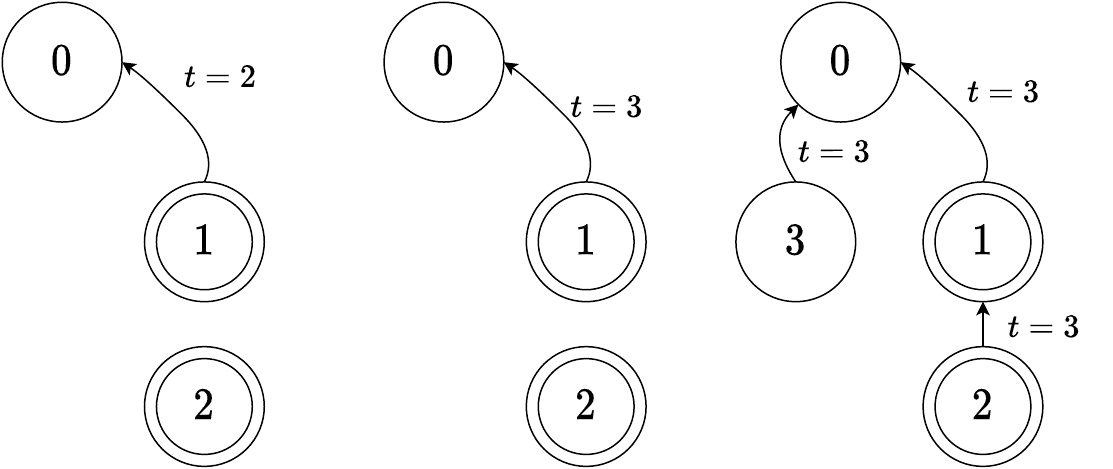}
\caption{In the left, we have a checkpoint preserving strategy $\pi$ that is not $1$-Opportunistic. In the center and right, we have transform $\pi$ into a $1$-Opportunistic strategy $\hat \pi$. In the center, we consider the case where Miner 1 creates block $3$ and in the right, we consider the case where Miner 2 creates block $3$.}
\label{fig:opportunistic}
\end{figure}
In the left of Figure~\ref{fig:opportunistic}, consider a state where Miner 1 withholds blocks 1 and 2 and publishes only $1 \to 0$ with block 1 reaching finality -- that is, Miner 1 is following a strategy that will never fork block 1. Thus Miner 1's strategy is not opportunistic because block 2 remains unpublished. To transform Miner 1's strategy into an opportunistic one, Miner 1 will wait during the second round. For future rounds, we consider two cases depending on whom creates the third block.

\noindent\vspace{1mm}\textbf{Case 1.} Consider the case Miner 1 creates block 3 (center of Figure~\ref{fig:opportunistic}). Let Miner 1 publish $1 \to 0$ and copy all the future actions of the original strategy as if nothing happened.

\noindent\vspace{1mm}\textbf{Case 2.} Consider the case Miner 2 creates block 3 (right of Figure~\ref{fig:opportunistic}). Let the new strategy publish $2 \to 1 \to 0$. In future rounds, the new strategy will publishes all edges $u \to s$ the original strategy publishes except when $u = 2$ or $s = 3$. If $u = 2$, the new strategy ignores edge $u \to s$ because $u \to s$ was already published. That is, the original strategy must be publishing edge $2 \to 1$ (because block 1 reached finality), but the new strategy already published $2 \to 1$ during the second round. If $s = 3$, the new strategy publishes $u \to 2$ instead. The only difference between the game trees is that block 3 is replaced by block 2 (which was created by Miner 1).
\end{example}
\begin{proposition}\label{prop:opportunistic}
Let $\pi$ be a trimmed strategy. Let $B$ be a state where:
\begin{itemize}
\item State $B$ is reachable from a mining game starting at state $B_0$ with Miner 1 following strategy $\pi$.
\item $\pi$ takes action $\publishpath(Q, v)$.
\item $v$ reached finality with respect to $\pi$.
\item $Q \subset \unpublished(B) \cap (v, \infty)$.
\end{itemize}
Then there is a trimmed strategy $f(\pi)$ that is equals to $\pi$ up to state $B$. At state $B$, $f(\pi)$ waits. Let $N = \max\ V(B) + 1$. To define $f(\pi)$ at future states, let $(X_t^\pi)_{t \geq 0}$ and $(X_t^{f(\pi)})_{t \geq 0}$ be mining games where Miner 1 follows strategy $\pi$ and $f(\pi)$ respectively with $X_0^\pi = X_0^{f(\pi)} = B$. Define the coupling $(X_t^\pi, X_t^{f(\pi)})_{t \geq 0}$ where Miner $k$ creates block $t \geq N$ in both games. Let $q = \min\ (\unpublished(B) \cap (v, \infty)) \setminus Q$. If Miner 1 creates block $N$, at state $(X_1^\half)^{f(\pi)}$, let $f(\pi)$ publish $Q$ pointing to $v$. Else, let $f(\pi)$ publish $Q \cup \{q\}$ pointing to $v$. Moreover, at all time steps $t \geq 1$, whenever $\pi$ publishes $u \to s$, we consider the cases:
\begin{itemize}
\item If $u = q$, $f(\pi)$ does not publish.
\item If $s = N$, $f(\pi)$ publishes $u \to q$ instead.
\item If $u \neq q$ and $s \neq N$, $f(\pi)$ also publishes $u \to s$.
\end{itemize}
Then, for all time steps $t \geq 0$,
$$\rev_\cstring^{(t)}(f(\pi)) \geq \frac{|A(\chain(X_t^\pi)) \cap T_1| - \ind{t = N-1} \cdot \potr^1(B)}{h(\chain(X_t^\pi))}.$$
\end{proposition}
\begin{proof}
Let's check $f(\pi)$ is a trimmed and valid strategy. Up to state $B$, $f(\pi)$ is equals to $\pi$ which, by assumption, is trimmed and valid. At state $B$, we consider the case where Miner 1 or Miner 2 creates block $N$ separately. For the case Miner 1 creates block $N$, $f(\pi)$ is equals to $\pi$ except $f(\pi)$ delays one round to publish $Q$ pointing to $v$. Then $f(\pi)$ copy all future actions of $\pi$. Thus for the case Miner 1 creates block $N$, $f(\pi)$ is trimmed and valid since $\pi$ is trimmed and valid.

For the case Miner 2 creates block $N$, $f(\pi)$ first publishes $Q \cup \{q\}$ pointing to $v$ at step $N$. Let's check this action is valid and trimmed. Note $q = (\unpublished(B) \cap (v, \infty)) \setminus Q$ is well-defined since, by assumption, $Q$ is a strict subset of $\unpublished(B) \cap (v, \infty)$. Moreover, the immediate successor of block $v$ must be a block created by Miner 2; otherwise, $\publishpath(Q, v)$ would not be a trimmed action at state $B$. Thus, at step $N$, publishing $Q \cup \{q\}$ pointing to $v$ is valid and trimmed action.

Next, still considering the case Miner 2 creates block $N$, we proof the simulations of $\pi$ and $f(\pi)$ satisfy the invariant that the longest paths $A(\chain(X_t^\pi))$ and $A(\chain(X_t^{f(\pi)}))$ are identical except for the edge $z \to \max\ Q$. In $\pi$'s longest path, $z$ is either equals to $q$ or $N$, while in $f(\pi)$'s longest path, $z$ is always equals to $q$. The invariant is clearly satisfied after $\pi$ publishes $Q \cup \{q\}$ pointing to $v$ so we need to check that, in the future, the invariant is preserved whenever $\pi$ or $f(\pi)$ publishes blocks.

First, observe all blocks $Q$ reached finality with respect to $\pi$. To check, by assumption, $v$ reached finality with respect to $\pi$ and the fact $\pi$ is trimmed implies $\pi$ would never take an action $\publishpath(Q', v')$ where the immediate successor of block $v'$ is in $Q$. Since $v' \geq v$ (since $v$ reached finality), we must have $v' \geq \max\ Q$. This proves $\max\ Q$ (in fact all blocks in $Q$) reached finality with respect to $\pi$. Next, we prove the invariant and $f(\pi)$'s actions are valid and trimmed. Consider the event $\pi$ publishes $u \to s$. We divide the proof into three cases:
\begin{itemize}
\item $u = q$. For this case, we claim $s = \max\ Q$. The fact $\max\ Q$ reached finality with respect to $\pi$ implies $s \geq \max\ Q$. Because $q$ was an unpublished block at state $B$ an $\pi$ is about to publish $q \to s$, we have $\max\ Q \leq s < q < N$. Thus $s$ is block created by Miner 1 since any blocks Miner 2 publishes pointing to $\max\ Q$ are bigger or equal than $N$. Note $\publishpath(Q \cup \{q\}, v)$ was an orderly action at state $B$ since $\publishpath(Q, v)$ was an orderly action at state $B$ and $q = \min\ \unpublished(B) \cap (v, \infty) \setminus Q$. Thus $\pi$ has no unpublished blocks between $\max\ Q$ and $q$ which implies $s = \max\ Q$. This proves $\pi$ is publishing edge $q \to \max\ Q$ as desired. Observe $f(\pi)$ already published this edge at step $N$. Because $\pi$ is timeserving, $q \to \max\ Q$ is about to become an edge in the longest path (and $N \to \max\ Q$ is about to become an orphaned edge) as desired. Once $q$ becomes part of the longest path, $q$ reaches finality with respect to $\pi$ since $\max\ Q$ reached finality with respect to $\pi$. To check, observe the fact $\pi$ is trimmed implies $\pi$ would never take an action $\publishpath(Q', \max\ Q)$ since the immediate successor of $\max\ Q$ would be block $q$ (a block created by Miner 1).

\item $s = N$. Recall $\max\ Q$ must still be a block in the longest path since $\max\ Q$ reached finality with respect to $\pi$. Thus $N \to \max\ Q$ is still an edge in the longest path while, by the inductive hypothesis, $q \to \max\ Q$ is the equivalent edge in the longest path of $f(\pi)$'s simulation. By the inductive hypothesis, the immediate successor of block $N$ and $q$ are identical. Moreover, the immediate successor of block $N$ is a block created by Miner 2 (since $\pi$ is trimmed and $\pi$ is about to publish edge $u \to N$). Thus publishing $u \to q$ is a trimmed action for $f(\pi)$. Moreover, publishing edge $u \to q$ instead of $u \to N$ preserves the invariant as desired.

\item $u \neq q$ and $s \neq N$. First, we claim $s > \max\ Q$. Recall block $\max\ Q$ reached finality with respect to $\pi$. Then $s \geq \max\ Q$. If $s = \max\ Q$, then $u = q$ because $\pi$ is trimmed (and orderly) and $q$ was one of $\pi$'s unpublished blocks at state $B$. This contradicts $u \neq q$ and proves $s > \max\ Q$ as desired. From the inductive hypothesis, $s$ is also a block in the longest path of $f(\pi)$'s simulation. Therefore, publishing edge $u \to s$ is a valid and trimmed action for $f(\pi)$ that preserves the invariant.
\end{itemize}
This proves $f(\pi)$ is valid and trimmed and proves the invariant that the longest path in $\pi$'s and $f(\pi)$'s simulation are identical except when edge $N \to \max\ Q$, in $\pi$'s simulation, is replaced by edge $q \to \max\ Q$, in $f(\pi)$'s simulation. Let's now compare the revenue of $f(\pi)$ and $\pi$. Clearly, the height of the longest path in $f(\pi)$ is at most the height of the longest path in $\pi$. In fact, the heights are identical until the $(N-2)$-th step. During the $(N-1)$-th step, the height can be strictly lower for $f(\pi)$ when $f(\pi)$ waits but $\pi$ publishes $Q$ pointing to $v$. After the $N$-th step, the longest paths have the same height.

Let's now compare how many blocks $\pi$ and $f(\pi)$ have in the longest path. Until the $(N-2)$-th step, both $\pi$ and $f(\pi)$ have the same number of blocks in the longest path. During the $(N-1)$-th step, $f(\pi)$ can have less blocks in the longest path when $f(\pi)$ waits, but $\pi$ publishes $Q$ pointing to $v$; however, the difference is bounded by $\potr^1(B)$, Definition~\ref{def:potential-reward}, the maximum reward $f(\pi)$ can obtain from state $B$. After the $N$-th step, $f(\pi)$ have at least the same number of blocks in the longest path as $\pi$. In fact, $f(\pi)$ can have one more block when Miner 2 creates block $N$ and edge $N \to \max\ Q$ is replaced by edge $q \to \max\ Q$. Thus for all $t$,
$\rev_\cstring^{(t)}(f(\pi)) \geq \frac{|A(\chain(X_t^\pi)) \cap T_1| - \ind{t = N-1} \cdot \potr^1(B)}{h(\chain(X_t^\pi))}$
as desired.
\end{proof}
\begin{proof}[Proof of Theorem~\ref{thm:opportunistic}]
Without loss of generality, let $\pi$ be a trimmed, checkpoint preserving strategy. Otherwise, from Theorem~\ref{thm:checkpoint-preserving}, there is a valid, trimmed strategy with the same revenue as $\pi$. Initialize $f(\pi)$ to be equal to $\pi$. To transform $f(\pi)$ into a trimmed, checkpoint preserving and opportunistic strategy execute the following procedure.

\textbf{Step 1.} While $f(\pi)$ is not opportunistic, there is a state $B$ where $f(\pi)$ takes action $\publishpath(Q, v)$ where $v$ reaches finality with respect to $\pi$, but $Q \subset \unpublished(B) \cap (v, \infty)$. Let $B$ be the first of such states that $f(\pi)$ encounters during a mining game starting at state $B_0$. From Proposition~\ref{prop:opportunistic}, there exists a strategy that is equals to $f(\pi)$ up to state $B$ and at state $B$ onward, such strategy executes the algorithm described in Proposition~\ref{prop:opportunistic}. Update $f(\pi)$ to be such strategy. Note $f(\pi)$ is trimmed and both opportunistic and checkpoint preserving up to state $B$, but might not be opportunistic nor checkpoint preserving on subsequent states. From Proposition~\ref{prop:cp:base-case}, there is a strategy that is equals to $f(\pi)$ up to state $B$ and is checkpoint preserving on all subsequent states. Let $f(\pi)$ be such strategy. At this point $f(\pi)$ is trimmed, checkpoint preserving, and opportunistic up to state $B$, then return to Step 1.

Each time we execute Step 1, $f(\pi)$ is trimmed, checkpoint preserving, and opportunistic on a larger set of states. Therefore, ad infinitum $f(\pi)$ becomes an opportunistic strategy. Observe the end result of executing Step 1 ad infinitum is that whenever Miner 1 was about to take a non-opportunistic action with a lead of $k$ blocks, Miner 1 waits until the lead reduces to a single block, then Miner 1 publishes. This sequence of actions is equivalent to the selfish mining strategy at state $B_{2, 0}$.

Next, we compare the revenue of $f(\pi)$ and $\pi$. Let $(X_t)_{t \geq 0}$ be a mining game with initial state $X_0 = B_0$ where Miner 1 follows strategy $\pi$. From Proposition~\ref{prop:opportunistic} and Corollary~\ref{cor:cp-revenue}, the revenue of $f(\pi)$ with respect to $\pi$ is given by
$$\rev_\cstring^{(t)}(f(\pi)) \geq \frac{|A(\chain(X_t)) \cap T_1| + \sum_{j = 1}^{N_t}\sum_{i = 1}^{W_0^j} Z_i^j - \potr^1(X_t)}{h(\chain(X_t)) + \sum_{j = 1}^{N_t} W_0^j + \sum_{j = 1}^{N_t} \sum_{i = 1}^{W_0^j} Z_i^j + \potr^1(X_t)}.$$
where $N_t$ denotes the number of times $f(\pi)$ was transformed by Proposition~\ref{prop:cp:base-case}, $W_0^j \geq 1$ and $(Z_i^j)_{i, j}$ are i.i.d random variables with expected value $\frac{\alpha}{1-2\alpha}$. From superadditivity of $\liminf$ 
$$\liminf_{t \to \infty} \rev_\cstring^{(t)}(f(\pi)) \geq \liminf_{t \to \infty} \rev_\cstring^{(t)}(\pi) + \liminf_{t \to \infty} \left(\rev_\cstring^{(t)}(f(\pi))  - \rev_\cstring^{(t)}(\pi)\right).$$
From Proposition~\ref{prop:cp-revenue} and our lower bound on $\rev_\cstring^{(t)}(f(\pi))$,
$$\liminf_{t \to \infty} \left(\rev_\cstring^{(t)}(f(\pi))  - \rev_\cstring^{(t)}(\pi)\right) \geq 0.$$
Taking the expected value proves $\rev(f(\pi)) \geq \rev(\pi)$ as desired.
\end{proof}

\subsection{The Strong Recurrence Theorem}\label{sec:strong-recurrence-proof}

For a mining game $(X_t)_{t \geq 0}$ starting at state $X_0 = B_0$ where Miner 1 follows optimal checkpoint recurrent strategy $\pi$, define $t_0 = 0$ and let $t_1, t_2, \ldots$ be the sequence of time steps where Miner 1 capitulates to $B_0$. Let $\tau_i = \tau_i - t_{i-1}$ and recall $(\tau_i)_{i \geq 1}$ is an i.i.d. sequence of positive random variables. Thus $N_n = \sum_{i  = 1}^\infty \ind{t_i \leq n}$ (with $t_0 = 0$) is a counting process (Definition~\ref{def:counting-process}) and satisfy the following zero-one law.
\begin{claim}[Zero-One Law]\label{claim:zero-one-law}
One of the following holds,
\begin{itemize}
    \item If Miner 1's strategy is transient, $\pr{\lim_{n \to \infty} N_n < \infty} = 1$.
    \item If Miner 1's strategy is recurrent, $\pr{\lim_{n \to \infty} N_n = \infty} = 1$.
\end{itemize}
\end{claim}
\begin{proof}
Without loss of generality, let $X_0 = B_0$. Let $p$ be the probability that Miner 1 capitulates to state $B_0$ at time $\tau \geq 1$. If Miner 1's strategy is transient, $p < 1$ and we get
\begin{align*}
\pr{\lim_{n \to \infty} N_n < \infty} &= \sum_{i = 0}^\infty \pr{\lim_{n \to \infty} N_n = i} = \sum_{i = 1}^\infty p^i (1-p) =(1-p)\frac{1}{1-p} = 1.
\end{align*}
If $p = 1$, $\pr{\lim_{n \to \infty} N_n < \infty} = 0$ which implies $\lim_{n \to \infty} N_n = \infty$ almost surely.
\end{proof}

For the case where Miner 1's strategy is transient, Claim~\ref{claim:zero-one-law} implies that with probability 1, the game will reach a time step $t_i$ where Miner 1 never defines another checkpoint. From the definition of checkpoints, this implies Miner 1 never publishes more than half of all the blocks they creates after time step $t_i$. This simple observation suffices to witness that $\frontier$ has at least the same payoff as $\pi$.

\begin{claim}\label{claim:transient}
If $\pi$ is checkpoint recurrent and transient, $\rev(\pi) \leq \alpha$.
\end{claim}
\begin{proof}
Recall $t_{N_n} \leq n < t_{N_n+1}$ (because $N_n$ is a counting process). From Claim~\ref{claim:zero-one-law}, $t_{N_n} < \infty$ almost surely as $n \to \infty$ (because $N_n < \infty$ almost surely as $n \to \infty$). Thus
\begin{align*}
\rev^{(n)}_{\cstring}(\pi) &= \frac{|A(\chain(X_{t_{N_n}})) \cap T_1| + \sum_{t = t_{N_n} + 1}^n r^1(X_{t-1}, X_t)}{h(\chain(X_n))}\\
&\leq \frac{t_{N_n} + \sum_{t = t_{N_n} + 1}^n r^1(X_{t-1}, X_t)}{h(\chain(X_n))}.
\end{align*}
During time step $t_{N_n} \leq t < t_{N_n+1}$, Miner 1 has not yet reached a new checkpoint. Thus Miner 1 publishes at most half of all blocks created from time $t_{N_n}+1$ to $t$:
$$\sum_{ t = t_{N_n} + 1}^n r^1(X_{t-1}, X_t) \leq \frac{1}{2}\sum_{i = 1}^n \ind{i \in T_1}$$
where the inequality is Corollary~\ref{cor:checkpoint-reward-bound}. Recall the height of the longest path $h(\chain(X_t))$ at time $t$ is at least $\sum_{i = 1}^n \ind{i \in T_2}$ -- the number of blocks Miner 2 creates up to time $t$. Thus
\begin{align*}
\frac{t_{N_n} + \sum_{t = t_{N_n + 1}}^n r^1(X_{t-1}, X_t)}{h(\chain(X_n))} &\leq \frac{t_{N_n} + \frac{1}{2}\sum_{i = 1}^n \ind{i \in T_1}}{h(\chain(X_n))}\\
&\leq \frac{t_{N_n} + \frac{1}{2}\sum_{i = 1}^n \ind{i \in T_1}}{\sum_{i = 1}^n \mathbbm 1_{i \in T_2}}\\
& = \frac{\frac{1}{n}t_{N_n} + \frac{1}{2n}\sum_{i = 1}^n \ind{i \in T_1}}{\frac{1}{n}\sum_{i = 1}^n \ind{i \in T_2}} \toas \frac{\alpha/2}{1-\alpha}
\end{align*}
The last step observes $\frac{t_{N_n}}{n} \toas 0$, and uses the strong law of large numbers and the fact $\pr{i \in T_1} = \alpha$, and $\pr{i \in T_2} = 1-\alpha$. Since $\alpha < 1/2$, we get $\rev(\pi) \leq \alpha$.
\end{proof}
Next, we consider the case where $\pi$ is checkpoint recurrent and null recurrent which implies $\e{\tau_i} = \infty$ for $i \geq 1$. To see that that $\rev(\pi) \leq \rev(\frontier)$, let 
$$A_{S} = \frac{\epsilon \sum_{i \in S} \tau_i}{|S|}$$
for any $\epsilon > 0$. Then partition $\{\tau_i\}_{1 \leq i \leq N}$ into $D_N = \{1 \leq i \leq N : \tau_i > A_{[N]}\}$ and $F_N = [N] \setminus D_N$. From the strong law of large numbers, we obtain $A_{[N]} \toas \infty$ because $\pi$ is null recurrent. Interestingly, this implies $\min_{i \in D_N} \tau_i \toas \infty$ since $\min_{i \in D_N} \tau_i \geq A_{[N]}$.

Recall Miner 1 capitulates to state $B_0$ at time steps $t_i$ and $t_{i+1}$ (whenever Miner 1 defines a new checkpoint). Let $k_i$ be the number of blocks Miner 1 creates from time $t_i+1$ to $t_{i+1}$. By definition, Miner 1 does not define a new checkpoint from time $t_i + 1$ to $t_{i+1}-1$; otherwise, Miner 1 would have capitulated to state $B_0$ between time $t_i+1$ and $t_{i+1} - 1$. Thus up to time step $t_{i+1} - 1$, Miner 1 publishes less than $k_i/2$ blocks (Corollary~\ref{cor:checkpoint-reward-bound}). If we hope for Miner 1's strategy to be better than $\frontier$, Miner 1's strategy must publish a significant fraction $\epsilon k_i$, for some $\epsilon > 0$, of their blocks at time step $t_{i+1}$. We expect the probability of Miner 1 publishing $\epsilon k_i$ blocks diminishes exponentially in $k_i$ because Miner 1 must create more blocks than Miner 2 (but Miner 2 has probability $1-\alpha > 1/2$ of creating each block). Lemma~\ref{lemma:reward-bound} formalizes this intuition.
\begin{claim}\label{claim:null-recurrent}
If $\pi$ is checkpoint and null recurrent, $\rev(\pi) \leq \alpha$.
\end{claim}
\begin{proof}\def\currentprefix{claim:null-recurrent}
Recall $t_{N_n} \leq n < t_{N_n + 1}$ (because $N_n$ is a counting process). Thus
\begin{align}\locallabel{eq:1}
\ &|A(\chain(X_n)) \cap T_1|= \sum_{t = 1}^n r^1(X_{t-1}, X_t)\\
    &\qquad = \sum_{j = 0}^{N_n-1} \bigg( r^1(X_{t_{j+1}-1}, X_{t_{j+1}}) + \sum_{t = t_j + 1}^{t_{j+1}-1} r^1(X_{t-1}, X_t)\bigg) + \sum_{t = t_{N_n}}^n r^1(X_{t-1}, X_t).
\end{align}
From time step $t_j+1$ to $t_{j+1}$, for $j \geq 0$, Miner 1 publishes only blocks created from time step $t_j+1$ to $t_{j+1}$ (because Miner 1 capitulates to state $B_0$ at time steps $t_j$ and $t_{j+1}$). Observe Miner 1 did not reached a new checkpoint by time step $t_{j+1}-1$ (because the strategy is checkpoint recurrent). This implies Miner 1 publishes less than half of all the blocks created from time step $t_j + 1$ to $t_{j+1}-1$ by time step $t_{j+1}-1$. From Corollary~\ref{cor:checkpoint-reward-bound}, we get
\begin{equation}\locallabel{eq:2}
\sum_{t = t_j + 1}^{t_{j+1}-1} r^1(X_{t-1}, X_t) \leq \frac{1}{2}\sum_{i = t_j + 1}^{t_{j+1}} \ind{i \in T_1}.
\end{equation}
From (\localref{eq:1}) and (\localref{eq:2}),
$$|A(\chain(X_n)) \cap T_1| \leq \frac{1}{2} \sum_{i = 1}^n \ind{i \in T_1} + \sum_{i = 1}^{N_n} r^1(X_{t_i-1}, X_{t_i}).$$
Recall the height of the longest chain is at least the number of blocks Miner 2 creates. That is $h(\chain(X_n)) \geq \sum_{t = 1}^n \ind{i \in T_2}$. Thus
\begin{align*}
\rev(\pi) &= \e{\liminf_{n \to \infty} \rev_{\cstring}^{(n)}(\pi)} \leq \e{\liminf_{n \to \infty} \frac{\frac{1}{2n} \sum_{i = 1}^n \ind{i \in T_1} + \frac{1}{n}\sum_{j = 1}^{N_n} r^1(X_{t_i-1}, X_{t_i})}{\frac{1}{n} \sum_{i = 1}^n \ind{i \in T_2}}}\\
&= \e{\liminf_{n \to \infty} \frac{\frac{\alpha}{2} +  \frac{1}{n}\sum_{j = 1}^{N_n} r^1(X_{t_i-1}, X_{t_i})}{1-\alpha}}\quad\text{From the strong law of large numbers.}\\
&\leq \alpha + \e{\liminf_{n\to\infty}\frac{1}{n}\sum_{i = 1}^{N_n} r^1(X_{t_i-1}, X_{t_i})} \quad \text{Because $\alpha < 1/2$.}
\end{align*}
From the Zero-One Law there is there is a sequence of time steps $t_1, t_2, \ldots$ where Miner 1 capitulates to state $B_0$ (and $t_0 = 0$). Fix any $\epsilon > 0$. Let $\tau_i = t_i - t_{i-1}$ for $i \geq 1$. For a subset of $t_1, t_2, \ldots$, let $S \subset \mathbb N$, $C_{S} = \sum_{i \in S} \tau_i$ and $A_{S} = \frac{\epsilon C_S}{|S|}$ and observe $C_{[N]} = \tau_N$ for all $N \in \mathbb N$. Note $n \geq t_{N_n}$ (because $N_n$ is a counting process). Then
\begin{align*}
\frac{1}{n}\sum_{i = 1}^{N_n} r^1(X_{t_i-1}, X_{t_i}) &\leq \frac{1}{t_{N_n}} \sum_{i = 1}^{N_n}  r^1(X_{t_i - 1}, X_{t_i}) = \frac{1}{C_{[N_n]}} \sum_{i = 1}^{N_n}  r^1(X_{t_i - 1}, X_{t_i})\\
&= \frac{1}{C_{[N_n]}} \sum_{i = 1}^{N_n} \left(r^1(X_{t_i - 1}, X_{t_i}) \cdot  \ind{\tau_i < A_{[N_n]}} +  r^1(X_{t_i - 1}, X_{t_i}) \cdot \ind{\tau_i \geq A_{[N_n]}}\right).
\end{align*}
To bound the first term, note from time $t_{i-1} + 1$ to $t_i$, Miner 1 only publishes blocks created from time $t_{i-1}+1$ to $t_i$ (and at most $\tau_i$ blocks were created in this time interval). Thus
$$\frac{1}{C_{[N]}} \sum_{i = 1}^N  r^1(X_{t_i - 1}, X_{t_i}) \cdot \ind{\tau_i < A_{[N]}} \leq \frac{1}{C_{[N]}} \sum_{i = 1}^N \tau_i \cdot \ind{\tau_i < A_{[N]}} < \frac{N A_{[N]}}{C_{[N]}} = \epsilon.$$
To bound the second term,
\begin{align*}
&\frac{1}{C_{[N]}} \sum_{i = 1}^N  r^1(X_{t_i - 1}, X_{t_i}) \cdot \ind{\tau_i \geq A_{[N]}} = \frac{1}{C_{[N]}} \sum_{i = 1}^N \tau_i \frac{ r^1(X_{t_i - 1}, X_{t_i}) \cdot \ind{\tau_i \geq A_{[N]}}}{\tau_i}\\
&\qquad\leq \frac{\sum_{i = 1}^N \tau_i}{C_{[N]}} \max_{i \in [N]} \frac{ r^1(X_{t_i - 1}, X_{t_i})}{\tau_i} \cdot \ind{\tau_i \geq A_{[N]}}\\
&\qquad= \max_{i \in [N]} \frac{r^1(X_{t_i - 1}, X_{t_i})}{\tau_i} \cdot \ind{\tau_i \geq A_{[N]}} \quad \text{By definition $C_{[N]} = \sum_{i = 1}^N \tau_i$}
\end{align*}
Next, we claim $\max_{i \in [N]} \frac{r^1(X_{t_i - 1}, X_{t_i})}{\tau_i} \cdot \ind{\tau_i \geq A_{[N]}} \toas 0$. Recall $A_{[N]}$ is the average of i.i.d. random variables $\tau_i$ and the fact Miner 1's strategy is null recurrent implies $\e{\tau_i} = \infty$. Thus from the strong law of large numbers, $A_{[N]} \toas \infty$. Therefore, the event
$$\limsup_{N \to \infty} \left\{\max_{i \in [N]} \frac{r^1(X_{t_i - 1}, X_{t_i})}{\tau_i} \cdot \ind{\tau_i \geq A_{[N]}} > 0\right\}$$
implies there is a mining game $(Y_t)_{t \geq 0}$ where $\limsup_{t \to \infty} \frac{r^1(Y_{t - 1}, Y_t)}{t} > 0$; however, from Lemma~\ref{lemma:reward-bound}, $\frac{r^1(Y_{t - 1}, Y_{t})}{t} \toas 0$. Summing up,
$$\rev(\pi) \leq \alpha + \epsilon + \e{\liminf_{N \to \infty} \max_{i \in [N]} \frac{r^1(X_{t_i-1}, X_{t_i})}{\tau_i} \cdot \ind{\tau_i \geq A_{[N]}}}= \alpha + \epsilon$$
as desired.
\end{proof}
\begin{proof}[Proof of Strong Recurrence Theorem~\ref{thm:strong-recurrence}]
From the Weak Recurrence Theorem~\ref{thm:weak-recurrence} there exists an optimal strategy $\pi$ that is checkpoint recurrent. Recall $\frontier$ is a potential candidate since $\frontier$ is checkpoint and positive recurrent (Observation~\ref{obs:frontier-positive-recurrent}), and $\rev(\frontier) = \alpha$ (Corollary~\ref{cor:opt-revenue}). From Claim~\ref{claim:transient} and Claim~\ref{claim:null-recurrent}, $\rev(\pi) \leq \alpha$ unless $\pi$ is positive recurrent.  Thus there exists an optimal strategy $\pi$ that is checkpoint and positive recurrent. Suppose not, then $\rev(\pi) \leq \alpha = \rev(\frontier)$ which witnesses $\frontier$ is optimal.
\end{proof}

\section{Omitted Proofs From Section~\ref{sec:nash-equilibrium}} \label{sec:nash-equilibrium-proof}

The main observation to show $\frontier$ is optimal (for sufficiently small $\alpha$) is to note that any mining game starting at state $B_0$ will transition to either state $B_{0, 1}$ or state $B_{1, 0}$ (see Figure~\ref{fig:states}). Recall the value faction $\va$ (Definition~\ref{def:value-function}) evaluated at $B_0$ is $\va(B_0) = 0$ (Lemma~\ref{lemma:b-0-optimization}). From the recursive definition of the value function,
\begin{align*}
0 = \va(B_0) &= \e{\R_{\lambda^*}(X_0, X_\tau)} = \alpha \va(B_{1, 0}) + (1-\alpha)(\va(B_{0, 1}) - \lambda^*)
\end{align*}
where $\lambda^* = \max_\pi \rev(\pi)$ and $\tau \geq 1$ is the first time step where Miner 1 capitulates to state $B_0$. Note block 1 is a checkpoint in state $B_{0, 1}$. From Theorem~\ref{thm:strong-recurrence}, Miner 1 has an optimal strategy that never forks a checkpoint (or block 1 in state $B_{0, 1}$). Therefore, there is an optimal strategy that capitulates to state $B_0$ from state $B_{0, 1}$. The following results will give us a closed form for the value function $\va$ evaluated at state $B_{0, 1}$.
\begin{proposition}\label{prop:value-lower-bound}
For any state $B$, $\va(B) \geq 0$.
\end{proposition}
\begin{proof}
From Bellman's principle of optimality (Lemma~\ref{lemma:bellman}),
$$\va(B) \geq \va_\pi^{\lambda^*}(B)$$
for any positive recurrent strategy $\pi$. Let $\pi$ be a strategy that capitulates to state $B_0$ from state $B$. From state $B$, Miner 1 can follow any optimal strategy $\pi^*$ as if it was at state $B_0$ by treating the longest chain $\chain(B)$ as block 0. Thus $\va(B) \geq \va(B_0) = 0$.
\end{proof}
\begin{proposition}\label{prop:value-capitulation}
If there is an optimal positive recurrent strategy that capitulates to state $B_0$ from state $B$, $\va(B) = 0$.
\end{proposition}
\begin{proof}
Let $\pi^*$ be an optimal strategy that capitulates to state $B_0$ from state $B$.  From state $B_0$, Miner 1 can follow strategy $\pi^*$ as if it was at state $B$ by treating block 0 as the longest chain $\chain(B)$. Then $0 = \va(B_0) \geq \va(B)$. From Proposition~\ref{prop:value-lower-bound}, $\va(B) \geq 0$. Thus $\va(B) = 0$ as desired.
\end{proof}
\begin{corollary}\label{cor:b_0_1}
Once at state $B_{0, 1}$, the optimal action for Miner 1 is to capitulate to state $B_0$. Thus $\va(B_{0, 1}) = 0$.
\end{corollary}
\begin{proof}
From the strong recurrence theorem, it is optimal for Miner 1 to capitulate to state $B_0$ from state $B_{0, 1}$. Thus state $B_{0, 1}$ satisfy the conditions for Proposition~\ref{prop:value-capitulation} and $\va(B_{0, 1}) = 0$.
\end{proof}
To show $\frontier$ is optimal (for sufficiently small $\alpha$), it suffices to show that $\va(B_{1, 0})$ is maximized when Miner 1 publishes $1 \to 0$ and capitulates to state $B_0$. Then there is an optimal strategy taking the same actions as $\frontier$ and capitulates to state $B_0$ every round as desired.

Formally, consider a strategy that, at state $B_{1, 0}$, publishes $1 \to 0$ and capitulates to state $B_0$. The game reward (Definition~\ref{def:game-reward}) is $1-\lambda^*$ since Miner 1 adds one block to the longest path.  From Bellman's principle of optimality,
$$\va(B_{1, 0}) \geq 1-\lambda^*.$$
Next, suppose strategy $\pi$ does not publish $1 \to 0$ at state $B_{1, 0}$. By showing $\va_\pi^{\lambda^*}(B_{1, 0}) \leq 1-\lambda^* \leq \va(B_{1, 0})$ (when $\alpha$ is sufficiently small), we will derive a certificate that there is an optimal strategy that publishes $1 \to 0$ at state $B_{1, 0}$. For that, we observe that starting from state $B_{1, 0}$, and under the assumption Miner 1 does not publishes any blocks until at least the next round, the subsequent state is either $B_{2, 0} = ((\{0\}, \emptyset), \{1, 2\}, \{1, 2\})$ when Miner 1 creates and withhold block 2, or $B_{1, 1} = ((\{0, 2\}, \{2 \to 0\}), \{1\}, \{1\})$ when Miner 2 creates block 2 and publishes $2 \to 0$. Then
$$\va_\pi^{\lambda^*}(B_{1, 0}) = \alpha \va_\pi^{\lambda^*}(B_{2, 0}) + (1-\alpha)(\va_\pi^{\lambda^*}(B_{1, 1}) - \lambda^*) \leq \alpha \va(B_{2, 0}) + (1-\alpha)(\va(B_{1, 1}) - \lambda^*)$$
where the inequality is Bellman's principle of optimality.

As a warmup, we will show how we can derive the upper bound of $\va(B_{1, 1}) \leq \frac{\alpha}{1-\alpha}$. As a thought experiment, image Miner 1 never forks block 2. Then block 1 is ``useless'' since it cannot point to any block $\geq 2$. We would like to formalize what it means for Miner 1 to ``forget'' blocks 1 and 2 from state $B_{1, 1}$. Once we forget blocks 1 and 2 from  state $B_{1, 1}$, state $B_{1, 1}$ is equivalent to $B_0$ (or Miner 1 capitulates to state $B_0$).

At state $B$, we say a block $q \in \va(B) \cup \unpublished(B)$ can {\em reach height} $\ell$ (from state $B$) if one of the two holds:
\begin{itemize}
\item If $q \in \va(B)$ was already published, then $h(q) \geq \ell$.
\item If $q \in \unpublished(B)$ is unpublished, then there is an action that Miner 1 can take from state $B$ such that $h(q) \geq \ell$ in the subsequent state.
\end{itemize}
Recall any blocks created in the future can point to any block $\leq q$ but any block $\leq q$ cannot point to any block $> q$. Thus if block $q$ cannot reach height $\ell$ from state $B$, then $q$ cannot reach height $\ell$ from any state reachable from $B$.
\begin{example}
At state $B_{2, 0} = ((\{0\}, \emptyset), \{1, 2\}, \{1, 2\})$, block 2 can reach heights 1 and 2 because Miner 1 can publish $2 \to 1 \to 0$, but block 2 cannot reach height $\geq 3$.
\end{example}
\begin{definition}[Induced Subgraph]
Let $G = (V, E)$ be a graph and let $S \subseteq V$ be any subset of the vertices of $G$. The {\em induced subgraph} $G[S]$ is the graph whose vertex set is $S$ and whose edge set consists of all edges in $E$ with both end points in $S$.
\end{definition}
\begin{definition}[State Capitulation]
Let $B = (\tree, \unpublished, T_1)$ be a state and let $c \leq h(\chain(B))$. Define $D \subseteq A(\chain) \cup \unpublished \setminus \{0\}$ as the set of blocks that cannot reach height $\geq c + 1$ from state $B$. Define the $c$-Capitulation of $B$ as the state
$$B[V \setminus D] := (\tree[V \setminus D], \unpublished \setminus D, T_1\setminus D)$$
where $\tree[V \setminus D]$ is an induced subgraph of $\tree$ obtained by deleting blocks $D$.
\end{definition}
\begin{example}
For state $B_{2, 2} = ((\{0, 2, 3\}, \{3 \to 2 \to 0\}), \{1, 4\}, \{1, 4\})$, its $1$-Capitulation deletes blocks 1 and 2, but not blocks 3 (since it is already at height 2) and 4 (since it can reach height 3 if Miner 1 publishes $4 \to 3$ or height 2 if Miner 1 publishes $4 \to 2$). Thus the $1$-Capitulation of state $B_{2, 2}$ is the state
$$B_{2, 2}[\{3, 4\}] = ((\{0, 3\}, \{3\to0\}), \{4\}, \{4\}).$$
Additionally, we could capitulate at height 2 to get state
$$B_{2, 2}[\{4\}] = ((\{0\}, \emptyset), \{4\}, \{4\})$$
since block 4 can reach height 3 if Miner 1 publishes $4 \to 3$, but all other blocks can only reach up to height 3.
\end{example}
Recall $H_i(B)$ denotes the block at height $i$ in state $B$ -- i.e., $h(H_i(B)) = i$. Next, we show how to upper bound $\va(B)$ in terms of $\va(B')$ where $B'$ is the $c$-Capitulation of $B$.
\begin{lemma}\label{lemma:truncation}
Let $B$ be a state, $c \leq h(\chain(B))$, and let $B'$ be its $c$-Capitulation. Then
\begin{equation}\label{eq:truncation}
    \va(B) \leq \va(B') + \R_{\lambda^*}(B_0, B') - \R_{\lambda^*}(B_0, B) + \sum_{i = 1}^c \left(\pr{H_i(X_\tau) \in T_1 | X_0 = B}- \lambda^*\right)
\end{equation}
where $\lambda^* = \max_\pi \rev(\pi)$ is the optimal revenue, and $\tau$ is the first time step Miner 1 capitulates to state $B_0$ in mining game $(X_t)_{t \geq 0}$ starting from state $X_0 = B$ where Miner 1 follows an optimal strategy.
\end{lemma}
For intuition behind Lemma~\ref{lemma:truncation}, observe a possible strategy at state $B'$ is to copy the optimal strategy at state $B$ by ignoring everything that happens at height $\leq c$. From Bellman's principle of optimality, $\va(B')$ is lower bounded by the reward obtained by such strategy. Second, observe $\va(B) + r_{\lambda^*}(B_0, B)$ is equals to the expected number of blocks Miner 1 has in the longest path at $X_\tau$ minus the length of the longest path times $\lambda^*$. Finally, Lemma~\ref{lemma:truncation} breaks down $\va(B)$ into four pieces: the first and second terms count the contributions from height $> c$ which is at most $\va(B') +  r_{\lambda^*}(B_0, B')$; the fourth term, counts the contributions from heights $\leq c$ that our strategy for $B'$ ignores.
\begin{example}
Consider $B$ where Miner 1 has on unpublished block 1, and Miner 2 has published $4 \to 3 \to 2 \to 0$. Let $c = 3$, then the $c$-Capitulation $B'$ of $B$ is $B_0$. Note $\va(B')$ is $0$, $\R_{\lambda^*}(B_0, B')$ is $0$ and $\R_{\lambda^*}(B_0, B)$ is $-3\lambda^*$. Thus Lemma~\ref{lemma:truncation} implies
$$\va(B) \leq 0 + 0 + 3\lambda^* + \sum_{i = 1}^3 (\pr{H_i(X_\tau) \in T_1 | X_0 = B} - \lambda^*) \leq 3.$$
\end{example}
\begin{proof}[Proof of Lemma~\ref{lemma:truncation}]\def\currentprefix{lemma:truncation}
Let $T = \max\{T_1(B), T_2(B)\}$ be the last block created at $B$. Define the mining game $(X_t)_{t \geq 0}$ starting at state $X_0 = B$ with Miner 1 following any optimal strategy $\pi$. Without loss of generality, $\pi$ is a trimmed, positive recurrent strategy $\pi$ (Theorem~\ref{thm:strong-recurrence}). Define the mining game $(X_t')_{t \geq 0}$ starting at state $X_0' = B'$ with Miner 1 following a strategy $\pi'$ (that will depend on $\pi$). Define a coupling between each game where at time $t \geq 1$, Miner 1 creates block $t + T$ with probability $\alpha$ in both games; otherwise, Miner 2 creates block $t + T$ in both games. Define $\pi'$ as follows: 
\begin{itemize}
    \item If $\pi$ plays $\wait$ in state $X_t^\half$, $\pi'$ plays $\wait$ in state $X_t'^\half$.
    \item If $\pi$ capitulates to state $B_0$ from state $X_t$, $\pi'$ capitulates to state $B_0$ from state $X_t'$.
    \item If $\pi$ plays $\publishpath(Q, v)$ at state $X_t^\half$, we will consider two distinct cases. Let
    $$Q' = \{q \in Q : \text{$h(q) \geq c + 1$ after $\pi$ plays $\publishpath(Q, v)$}\}$$
    Then, at state $X_t'^\half$,
    \begin{itemize}
        \item If $h(v) > c$, $\pi'$ plays $\publishpath(Q', v)$.
        \item If $h(v) \leq c$, $\pi'$ plays $\publishpath(Q', 0)$.
    \end{itemize}
\end{itemize}
For the graph $\tree(B) = (\vertex(B), \edge(B))$, $\vertex(B)$ denote the set of blocks that were already published at state $B$. Let $S(X_t)$ be the blocks $q \in \vertex(X_t)$ with height $h(q) \geq c+1$ and let $S(X_t') = \vertex(X_t') \setminus \{0\}$. Then for all $t$, the coupling between $(X_t)_{t \geq 0}$ and $(X_t')_{t \geq 0}$ induces
\begin{equation}\locallabel{eq:1}
\tree(X_t')[S(X_t')] = \tree(X_t)[S(X_t)].
\end{equation}
Let $\tau \geq 1$ be the time step Miner 1 capitulates to state $B_0$. Recall that for a state $B''$ reachable from state $B$, $r_\lambda(B, B'')$ denotes the game reward obtained from state $B$ to $B''$. If $B'$ is an intermediate state between $B$ and $B''$, we naturally have $r_\lambda(B, B') + r_\lambda(B', B'') = r_\lambda(B, B'')$.
\begin{observation}\label{obs:path-independence}
For any $\lambda \in \mathbb R$ and states $B$, $B'$, and $B''$,
$$r_\lambda(B, B') + r_\lambda(B', B'') = r_\lambda(B, B'').$$
\end{observation}
From definition of $\va(B)$ and Observation~\ref{obs:path-independence},
\begin{equation}\locallabel{eq:3}
\R_{\lambda^*}(B_0, B) + \va(B) = \e{\R_{\lambda^*}(B_0, B) + \R_{\lambda^*}(X_0, X_\tau)} = \e{\R_{\lambda^*}(B_0, X_\tau)}.
\end{equation}
\begin{observation}\label{obs:game-reward}
For any states $B$ and $B'$,
$$r_\lambda(B, B') = |A(\chain(B')) \cap T_1| - |A(\chain(B)) \cap T_1| - \lambda \left(h(\chain(B')) - h(\chain(B))\right)$$
\end{observation}
From Observation~\ref{obs:game-reward},
$$\e{\R_{\lambda^*}(B_0, X_\tau)} = \e{\sum_{i = 1}^{h(\chain(X_\tau))} \ind{H_i(X_\tau) \in T_1} - \lambda^* h(\chain(X_\tau))}$$
\begin{align*}
    &= \e{\sum_{i = 1}^c \ind{H_i(X_\tau) \in T_1} - \lambda^* c + \sum_{i = c + 1}^{h(\chain(X_\tau))} \ind{H_i(X_\tau) \in T_1} - \lambda^* (h(\chain(X_\tau)) - c)}\\
    &= \sum_{i = 1}^c (\pr{H_i(X_\tau) \in T_1} - \lambda^*) + \e{\sum_{i = 1}^{h(\chain(X_\tau'))} \ind{H_i(X_\tau') \in T_1} - \lambda^* h(\chain(X_\tau'))} \quad \text{From (\localref{eq:1}).}
\end{align*}
From Observation~\ref{obs:game-reward}, 
\begin{align*}
 &\e{\sum_{i = 1}^{h(\chain(X_\tau'))} \ind{H_i(X_\tau') \in T_1} - \lambda^* h(\chain(X_\tau'))} = \e{\R_{\lambda^*}(B_0, X_\tau')} &\\
	&= \e{\R_{\lambda^*}(B_0, X_0') + \R_{\lambda^*}(X_0', X_\tau')} & \text{From Observation~\ref{obs:path-independence},}\\
    &= \R_{\lambda^*}(B_0, B') + \va_{\pi'}^{\lambda^*}(B')\\
    &\leq \R_{\lambda^*}(B_0, B') + \va(B') & \text{From Lemma~\ref{lemma:bellman}}.
\end{align*}
This proves
$$\va(B) \leq \va(B') + \R_{\lambda^*}(B_0, B') - \R_{\lambda^*}(B_0, B) + \sum_{i = 1}^c \left(\pr{H_i(X_\tau) \in T_1} - \lambda^*\right)$$
as desired.
\end{proof}
We are ready to show $\va(B_{1, 1}) \leq \frac{\alpha}{1-\alpha}$. First, we will observe that the probability Miner 1 adds block 1 to the longest path once the game reaches state $B_{1, 1}$ is at most $\frac{\alpha}{1-\alpha}$.
\begin{lemma}\label{lemma:tie-breaking}
Suppose at state $B$ Miner 1 needs at least $0 \leq \ell \leq 2$ blocks to fork $\chain(B)$---i.e., for all $v \in A(\chain(B))$, $h(\chain(B)) \geq h(v) + |\unpublished(B)\cap (v, \infty)| + \ell - 1$. Then for a mining game starting a state $X_0 = B$, the probability Miner 1 removes $\chain(B)$ from the longest path is at most $\big(\frac{\alpha}{1-\alpha}\big)^{\ell}$.
\end{lemma}
For the proof, we will use a well known fact about one-dimensional random walks.
\begin{lemma}\label{lemma:winning-probability}
Let $(M_t)_{t \geq 0}$ be a biased one-dimensional random walk with initial state $M_0 = i$ and for $t \geq 1$, $M_t = M_{t-1} - 1$ with probability $\alpha < \frac{1}{2}$ and $M_t = M_{t-1} + 1$ otherwise. Then the probability $M_\tau = 0$ for some $\tau \geq 0$ is $\big(\frac{\alpha}{1-\alpha}\big)^{i}$.
\end{lemma}
\begin{proof}
Let $E_n = \cup_{t = 0}^n \{M_t = 0\}$ be the event $M_\tau = 0$ for some $0 \leq \tau \leq n$. Note $\lim_{n \to \infty} E_n$ denotes the event where $M_\tau = 0$ for some $\tau \geq 0$. Since $E_n \subseteq E_{n+1}$ for all $n \geq 0$, the continuity theorem for probabilities implies
$$\pr{\lim_{n \to \infty} E_n | M_0 = i} = \lim_{n \to \infty} \pr{E_n | M_0 = i}.$$
Let $p_i = \pr{E_n | M_0 = i}$. Thus it suffices to compute $p_i$ and take the limit of $n \to \infty$.

Clearly $p_0 = 1$ and $p_i = 0$ for $i \geq n + 1$ (since $M_n \geq i - n \geq 1$). For $1 \leq i \leq n$,
$$p_i = \alpha p_{i-1} + (1-\alpha) p_{i+1}.$$
Writing $p_i = \alpha p_i + (1-\alpha) p_i$, we get
$$p_i - p_{i+1} = \frac{\alpha}{1- \alpha}(p_{i-1} - p_i).$$
\begin{claim}\label{claim:winning-prob-1}
For $1 \leq i \leq n$, $p_i - p_{i+1} = \left(\frac{\alpha}{1-\alpha}\right)^{i}(1-p_1)$.
\end{claim}
\begin{proof}
The proof is by induction. For the base, $i = 1$, we have $p_1 - p_2 = (1-p_1)\alpha/(1-\alpha)$. For $i \geq 2$, the inductive hypothesis gives $p_{i-1} - p_i = \left(\frac{\alpha}{1-\alpha}\right)^{i-1}(1-p_1)$. Then
\begin{align*}
p_i - p_{i+1} &= \left(\frac{\alpha}{1-\alpha}\right)(p_{i-1} - p_i)\\
&= \left(\frac{\alpha}{1-\alpha}\right)^i(1-p_1).
\end{align*}
\end{proof}
Note $\sum_{i = 1}^n (p_i - p_{i+1}) = p_1 - p_{n+1} = p_1$. Thus
\begin{align*}
    p_1 = \sum_{i = 1}^n (p_i - p_{i+1}) &= (1-p_1)\sum_{i = 1}^n \left(\frac{\alpha}{1-\alpha}\right)^i \\
    &= (1-p_1) \frac{\alpha - (1-\alpha)\left(\frac{\alpha}{1-\alpha}\right)^{n+1}}{1-2\alpha}.
\end{align*}
Because $\alpha < \frac{1}{2}$, $\left(\frac{\alpha}{1-\alpha}\right)^n \to 0$. Rearranging and taking the limit of $n \to \infty$, proves $p_1 = \frac{\alpha}{1-\alpha}$. Next, we claim that $p_{i-1} = \left(\frac{\alpha}{1-\alpha}\right)^{i-1}$ for $i \geq 2$ as $n \to \infty$. As inductive hypothesis, assume the statement holds for $i - 1$ (where the base case is implied from $p_1 = \frac{\alpha}{1-\alpha}$). From Claim~\ref{claim:winning-prob-1},
$$p_i = p_{i-1} -\left(\frac{\alpha}{1-\alpha}\right)^{i-1} (1-p_1) = \left(\frac{\alpha}{1-\alpha}\right)^{i-1} - \left(\frac{\alpha}{1-\alpha}\right)^{i-1}\left(1- \left(\frac{\alpha}{1-\alpha}\right)\right) = \left(\frac{\alpha}{1-\alpha}\right)^i.$$
This proves $p_i = \left(\frac{\alpha}{1-\alpha}\right)^i$ as desired.
\end{proof}

\begin{proof}[Proof of Lemma~\ref{lemma:tie-breaking}]
Consider a mining game starting at state $X_0 = B$. Let $(Y_t)_{t \geq 0}$ be biased one-dimensional random walk with $Y_0 = \ell$. For $t \geq 1$, let $Y_t = Y_{t-1} - 1$ if Miner 1 creates block $t$ and $Y_t = Y_{t-1} + 1$ if Miner 2 creates block $t$. Miner 1 can only remove block $\chain(B)$ from the longest path if Miner 1 creates $\ell$ more blocks than Miner 2 since the beginning of the game. In other words, we must have $Y_t = 0$ for some $t \geq 0$. Thus, the probability Miner 1 forks block $\chain(B)$ is at most the probability $Y_t = 0$ for some $t \geq 0$. From Lemma~\ref{lemma:winning-probability},
$$\pr{\cup_{t = 1}^\infty \{Y_t = 0\}} = \left(\frac{\alpha}{1-\alpha}\right)^{\ell}.$$
\end{proof}
\begin{proposition}\label{prop:b_1_1}
$\va(B_{1, 1}) \leq \frac{\alpha}{1-\alpha}.$
\end{proposition}
\begin{proof}
The $1$-Capitulation of state $B_{1, 1}$ is the state $B_0$ because blocks 1 and 2 can only be at height 1. From the Lemma~\ref{lemma:truncation},
\begin{align*}
\va(B_{1, 1}) &\leq \R_{\lambda^*}(B_0, B_0) + \va(B_0) + \lim_{t \to \infty} \pr{H_1(X_t) \in T_1 | X_0 = B_{1, 1}} - \lambda^* - \R_{\lambda^*}(B_0, B_{1, 1})\\
&= \lim_{t \to \infty} \pr{H_1(X_t) \in T_1 | X_0 = B_{1, 1}}
\end{align*}
The equality observes $\R_{\lambda^*}(B_0, B_0) = 0$, $\va(B_0) = 0$ (Lemma~\ref{lemma:b-0-optimization}) and $\R_{\lambda^*}(B_0, B_{1, 1}) = -\lambda^*$. At state $B_{1, 1}$, Miner 2 owns block 2 at height 1 and Miner 1 can only own the block at height 1 if Miner 1 removes block 2 from the longest path. For that, we must have a time $t \geq 1$ where Miner 1 creates one more block than Miner 2 from time 1 to $t$. From Lemma~\ref{lemma:tie-breaking}, the probability of this event is at most $\frac{\alpha}{1-\alpha}$ which proves $\va(B_{1, 1}) \leq \frac{\alpha}{1-\alpha}$ as desired.
\end{proof}
Next, we will show that it is optimal for Miner 1 to wait at state $B_{2, 0}$. More generally, consider $B_{k, 0} := \{(\{0\}, \emptyset), [k], [k]\}$ (Equation~\ref{eq:b_k_0}) as the state where Miner 1 creates and withholds blocks $1, 2, \ldots, k$. Intuitively, there is no advantage for Miner 1 to publish at $B_{k, 0}$, for $k \geq 2$, because Miner 1 can safely wait to publish in the subsequent state. That is, even if Miner 2 publishes the subsequent block, Miner 1 would still have sufficient blocks to fork the block Miner 2 just published. We can further extend this intuition for a larger class of states where Miner 2 has published blocks in the longest path. Formally, we define $\ca(B)$ as the collection of states $B'$ where 
\begin{itemize}
\item Miner 1 has no blocks in the longest path.
\item Miner 1 has $h(\chain(B'))$ hidden blocks that cannot reach height $> h(\chain(B'))$.
\item Capitulating $B'$ at the longest chain result at state $B$. That is, if Miner 1 forgets all blocks that cannot reach height $> h(\chain(B'))$, we obtain state $B$.
\end{itemize}
In other words,
\begin{align*}
\ca(B) := \{\text{$B'$ is a state} : \qquad &A(\chain(B')) \cap T_1 = \emptyset,\\
& |T_1(B')| - |T_1(B)| = h(\chain(B')),\\
& \text{$h(\chain(B'))$-Capitulation of $B'$ is state $B$}\}.
\end{align*}
As an example, consider a state $B \in \ca(B_{2, 0})$ where Miner 1 creates blocks 1, 2, 3 and Miner 2 publishes block $4 \to 0$. Observe if $B \in \ca(B_{k, 0})$, then the subsequent state is $B' \in \ca(B_{k+1, 0})$ when Miner 1 creates and withholds the next block or $B' \in \ca(B_{{k-1}, 0})$ when Miner 2 creates and publishes the next blocks. When $k \geq 2$, we will show there is an optimal strategy that waits at state $B \in \ca(B_{k, 0})$ until the game reaches a state $B' \in \ca(B_{1, 0})$. Once at state $B'$, the optimal strategy is to publish all blocks and fork the whole longest path when $\alpha$ is sufficiently small.

Let's check why the three conditions for a state $B$ to belong to $\ca(B_{k, 0})$ might be necessary for this result. The condition that Miner 1 has no blocks in the longest path hints that forking all the blocks in the longest path is optimal since Miner 1 has no blocks in there. The condition that Miner 1 could fork all blocks in the longest path at $B$ will imply it is optimal for Miner 1 to wait at state $B$. For example, consider the state where Miner 1 withholds blocks 1, 5, 6 and Miner 2 published $3 \to 2 \to 0$. This state satisfy the first and third conditions, but not the second. As a result, it is not clear how to argue that publishing $6 \to 5 \to 2$ is not optimal. The condition $B_{k, 0}$ is the capitulation of $B$ gives the property that the subsequent state is either in $\ca(B_{k-1, 0})$ or $\ca(B_{k+1, 0})$.
\begin{proposition}\label{prop:b_k_0}
Let $\alpha \leq \frac{1}{2}(3 - \sqrt 5)$. At state $B \in \ca(B_{k, 0})$, for $k \geq 2$, it is optimal to play $\wait$.
\end{proposition}
\begin{proof}
Define a mining game $(X_t')_{t \geq 0}$ starting at state $X_0' = B$ where Miner 1 follows an optimal positive recurrent strategy $\pi^*$. Define a mining game $(X_t)_{t \geq 0}$ starting at state $X_0 = B$ where Miner 1 follows a positive recurrent strategy $\pi$ (to be defined later). Couple the mining games $(X_t)_{t \geq 0}$ and $(X_t')_{t \geq 0}$ so that Miner $k$ creates block $n \geq 1 + \max \vertex(B)$ in the first game if and only if Miner $k$ creates block $n$ in the second game. Let $\tau$ (respectively $\tau'$) be the first time step Miner 1 capitulates to state $B_0$ in game $(X_t)_{t \geq 0}$ (respectively $(X_t')_{t \geq 0}$). Without loss of generality, assume $\tau' \geq \tau$. Let $T$ be the first time step $t \geq 1$ where $X_t^\half \in \ca(B_{1, 0})$. Define $\pi$ as follows:
\begin{itemize}
\item For all $t \leq T - 1$, note $X_t^\half \in \ca(B_{k, 0})$ with $k \geq 2$. Then $\pi$ plays $\wait$ at state $X_t^\half$.
\item If $X_T^\half \neq X_T'^\half$, $\pi$ plays $\publishpath(T_1(X_T), 0)$ at state $X_T^\half$. Then $\pi$ capitulates to state $B_0$.
\item If $X_T^\half = X_T'^\half$, $\pi$ plays the same action as $\pi^*$ at state $X_t^\half$ for all $t \geq T$.
\end{itemize}
To show it is optimal for Miner 1 to play $\wait$ at state $B \in \ca(B_{k, 0})$, it suffices to show $\pi$ is an optimal strategy. It is without loss of generality to assume $B \in \ca(B_{2, 0})$ because if $\pi$ is optimal for a game starting at $B \in \ca(B_{2, 0})$, $\pi$ is also optimal for a game starting at state $B \in \ca(B_{k, 0})$, for $k \geq 2$.

Let $E$ be the event $X_T'^\half = X_T^\half$ and let $E^c$ be the complement of $E$. If $X_T'^\half = X_T^\half$, then $X_t' = X_t$ for all $t \geq 0$ since $\pi^*$ played the same actions as $\pi$ up to time $T-1$ and $\pi$ plays the same actions as $\pi^*$ after time $T$ (inclusive). Then
$$\e{\R_{\lambda^*}(X_0', X_{\tau'}') | E} = \e{\R_{\lambda^*}(X_0, X_\tau) | E}.$$
If $X_T'^\half \neq X_T^\half$, Miner 1 owns at least one block in $A(\chain(X_T'^\half))$ (because they published at least one block up to time $T-1$). Next, we consider separately the case where Miner 1 has at most one block in the longest path $A(\chain(X_T'^\half))$ and the case where Miner 1 has at least two blocks in the longest path $A(\chain(X_T'^\half))$.

\noindent\textbf{Case 1:} Consider the case Miner 1 has at most one block in the longest path $A(\chain(X_T'^\half))$. For this case, $h(\chain(X_T'^\half)) = |T_1(X_T)|$. Using the fact $B_0$ is the $h(\chain(X_T'^\half))$-Capitulation of state $X_T'^\half$ in Lemma~\ref{lemma:truncation} gives
\begin{align*}
\R_{\lambda^*}(X_0', X_T'^\half) + \va(X_T'^\half) &\leq \R_{\lambda^*}(B_0, B_0) + \va(B_0)\\
&\qquad + \sum_{i = 1}^{h(\chain(X_T'^\half))} \left(Pr[H_i(X_{\tau'}') \in T_1] - \lambda^*\right) + \lambda^* h(\chain(B))\\
&\leq |T_1(X_T)|(1-\lambda^*) + \lambda^* h(\chain(B)).
\end{align*}
The second line observes $\va(B_0) = 0$ (Lemma~\ref{lemma:b-0-optimization}) and $\R_{\lambda^*}(B_0, B_0) = 0$. 

\noindent\textbf{Case 2:} Consider the case Miner 1 has at least two blocks in the longest path $A(\chain(X_T'^\half))$. For $k = 1, 2$, let $M_k = \{i \leq h(\chain(X_T'^\half)) : H_i(X_T'^\half) \in T_k\}$ be the heights Miner $k$ owns blocks in the longest path $A(\chain(X_T'^\half))$. Using the fact $B_0$ is the $h(\chain(X_T'^\half))$-Capitulation of state $X_T'^\half$ in Lemma~\ref{lemma:truncation} gives
\begin{align*}
\R_{\lambda^*}(X_0', X_T'^\half) + \va(X_T'^\half) &\leq \R_{\lambda^*}(B_0, B_0) + \va(B_0)\\
&\qquad + \sum_{i \in M_1 \cup M_2} (\pr{H_i(X_{\tau'}') \in T_1} - \lambda^*) + \lambda^* h(\chain(B))\\
&= |M_1|(1-\lambda^*) + \sum_{i \in M_2} (\pr{H_i(X_{\tau'}') \in T_1} - \lambda^*) + \lambda^* h(\chain(B))\\
&\leq |T_1(X_T)|(1-\lambda^*) + \lambda^* h(\chain(B))  + \sum_{i \in M_2} (\pr{H_i(X_{\tau'}') \in T_1} - \alpha)\\
&\leq |T_1(X_T)|(1-\lambda^*) + \lambda^* h(\chain(B))  + \sum_{i \in M_2}\left(\left(\frac{\alpha}{1-\alpha}\right)^2 - \alpha\right)\\
&\leq |T_1(X_T)|(1-\lambda^*) + \lambda^* h(\chain(B))
\end{align*}
The second line observes $|M_1|$ is at most the number of blocks Miner 1 creates up to time $\tau$ and $\lambda^* = \max_\pi \rev(\pi) \geq \alpha$ because $\rev(\frontier) = \alpha$. For the third line, note the event $H_i(X_{\tau'}') \in T_1$ for $i \in M_2$ implies Miner 1 forks the longest chain $\chain(X_T'^\half)$ at some point in the future. Because Miner 1 publishes at least two blocks in the longest path $A(\chain(X_T'^\half))$ and $X_T^\half \in \ca(B_{1, 0})$, Miner 1 needs at least two blocks to fork the longest chain $\chain(X_T'^\half)$. Thus from Lemma~\ref{lemma:tie-breaking}, the probability of $H_i(X_{\tau'}') \in T_1$ for $i \in M_2$ is at most $\left(\frac{\alpha}{1-\alpha}\right)^2$. The last line observes $\left(\frac{\alpha}{1-\alpha}\right)^2 - \alpha \leq 0$ for $\alpha \leq \frac{1}{2}(3 - \sqrt 5)$. This concludes the proof of the second case.

Case 1 and 2 proves
\begin{align*}
\e{\R_{\lambda^*}(X_0', X_{\tau'}') | E^c} &= \e{\R_{\lambda^*}(X_0', X_\tau'^\half) + \R_{\lambda^*}(X_\tau'^\half, X_{\tau'}') | E^c}\\
&= \e{\R_{\lambda^*}(X_0', X_\tau'^\half) + \va(X_\tau'^\half) | E^c}\\
&\leq \e{|T_1(X_T)|(1-\lambda^*) + \lambda^* h(\chain(B)) | E^c}
\end{align*}
as desired. We claim
$$\e{\R_{\lambda^*}(X_0, X_\tau) | E^c} = \e{|T_1(X_T)|(1-\lambda^*) + \lambda^* h(\chain(B)) | E^c}.$$
Recall, at state $X_T^\half$, $\pi$ takes action $\publishpath(T_1(X_T), 0)$ and capitulates to state $B_0$ whenever $X_T^\half \neq X_T'^\half$. The game reward from state $X_0$ to $X_\tau$ is $|T_1(X_T)|(1-\lambda^*)$, because Miner 1 adds $|T_1(X_T)|$ blocks in the longest path, plus $\lambda^*h(\chain(B))$, because Miner 2 has $h(\chain(B))$ blocks removed from the longest path. Thus
$$\e{\R_{\lambda^*}(X_0, X_\tau) | E^c} = \e{|T_1(X_T)|(1-\lambda^*) + \lambda^* h(\chain(B)) | E^c} \geq \e{\R_{\lambda^*}(X_0', X_{\tau'}') | E^c}.$$
We conclude $\va_\pi^{\lambda^*}(B) \geq \va(B)$. From Bellman's principle of optimality, $\pi$ is optimal.
\end{proof}
Because $B_{2, 0} \in \ca(B_{2, 0})$, we proved that once the mining game reaches state $B_{2, 0}$, Miner 1 will prefer to wait until it reaches a state in $B \in \ca(B_{1, 0})$. Next, we prove that at state $B \in \ca(B_{1, 0})$, it is optimal for Miner 1 to publish all their hidden blocks (when $\alpha$ is sufficiently small).
\begin{proposition}\label{prop:b_1_0-1}
Let $\alpha \leq \frac{1}{2}(3-\sqrt 5)$. At state $B \in \ca(B_{1, 0})$, it is either optimal to take action $\wait$ or $\publishpath(T_1(B), 0)$.
\end{proposition}
\begin{proof}
If waiting at state $B$ is not optimal, then one of the following is optimal:
\begin{enumerate}
\item Miner 1 takes action $\publishpath(T_1(B), 0)$.
\item Miner 1 takes action $\publishpath(Q, v)$ for some $Q \subset T_1(B)$.
\end{enumerate}
We will show the action in the first case is strictly better than the action in the second case.

In the first case, let $B'$ be the subsequent state after Miner 1 plays $\publishpath(T_1(B), 0)$. From Proposition~\ref{prop:value-capitulation}, $\va(B') \geq 0$ since Miner 1 can capitulate to state $B_0$ from state $B'$. The game reward from state $B$ to $B'$ is $\lambda^* h(\chain(B))$, because Miner 1 removes $h(\chain(B))$ of Miner 2's blocks in the longest path, plus $(1-\lambda^*)(h(\chain(B)) + 1)$, because Miner 1 adds $h(\chain(B)) + 1$ blocks in the longest path. Thus the game reward from state $B$ until Miner 1 capitulates to state $B_0$ is
$$\va(B) \geq h(\chain(B)) + (1-\lambda^*) + \va(B') \geq h(\chain(B)) + (1-\lambda^*).$$
In the second case, let $B'$ be the subsequent state after Miner 1 takes action $\publishpath(Q, v)$. Let $M_k = \{i \leq h(\chain(B')) : H_i(B') \in T_k\}$ be the heights of the blocks Miner $k$ owns in the longest path $A(\chain(B'))$. The $h(\chain(B'))$-Capitulation of $B'$ is the state $B_0$. Thus from Lemma~\ref{lemma:truncation},
\begin{align*}
\va(B) &= \R_{\lambda^*}(B, B') + \va(B') = \R_{\lambda^*}(B_0, B') + \va(B') - \R_{\lambda^*}(B_0, B)\\
&\leq \R_{\lambda^*}(B_0, B_0) + \va(B_0) + \sum_{i \in M_1 \cup M_2} \left(\pr{H_i(X_\tau) \in T_1| X_0 = B'} - \lambda^*\right) - \R_{\lambda^*}(B_0, B)\\
&= |M_1|(1-\lambda^*) + \sum_{i \in M_2} \left(\pr{H_i(X_\tau) \in T_1 | X_0 = B'} - \lambda^*\right) + \lambda^* h(\chain(B))\\
&\leq |M_1|(1-\lambda^*) + \sum_{i \in M_2} \left(\left(\frac{\alpha}{1-\alpha}\right)^2 - \alpha\right) + \lambda^* h(\chain(B))\\
&\leq |M_1|(1-\lambda^*) + \lambda^* h(\chain(B))\\
&= h(\chain(B)) + (1-\lambda^*)
\end{align*}
The third line observes $\va(B_0) = 0$ (Lemma~\ref{lemma:b-0-optimization}) and $\R_{\lambda^*}(B_0, B_0) = 0$. The fourth line observes $\lambda^* = \max_\pi \rev(\pi) \geq \alpha$ and Miner 1 needs at least 2 blocks to fork $\chain(B')$ from the longest path. Thus from Lemma~\ref{lemma:truncation}, the probability Miner 1 will own the block at height $i \in M_2$ is at most $\left(\frac{\alpha}{1-\alpha}\right)^2$. The fifth line observes $\left(\frac{\alpha}{1-\alpha}\right)^2 \leq \alpha$ for $\alpha \leq \frac{1}{2}(3 -\sqrt 5)$. The sixth line observes Miner 1 owns at most $|T_1(B)| = h(\chain(B)) + 1$ blocks in the longest path. The chain of inequalities witnesses taking action $\publishpath(T_1(B), 0)$ is strictly better than taking action $\publishpath(Q, v)$ for some $Q \subset T_1(B)$.
\end{proof}
We are ready to show that for any state $B \in \ca(B_{1, 0})$, it is optimal for Miner 1 to play $\publishpath(T_1(B), 0)$ (when $\alpha$ is sufficiently small).
\begin{proposition}\label{prop:b_1_0}
Let $\frac{\alpha(1-\alpha)^2}{(1-2\alpha)^2} \leq c + 1$ where $c = h(\chain(B))$. At state $B \in \ca(B_{1, 0})$, it is optimal for Miner 1 to take action $\publishpath(T_1(B), 0)$.
\end{proposition}
\begin{proof}
Let $\lambda^* = \max_\pi \rev(\pi)$. The proof will be by induction on $c$. For the base case, we will proof the statement holds for $c \geq C$ where $C$ is an absolute constant. Then we will prove the statement when $c = C-1, C-2, \ldots, 0$. If Miner 1 takes action $\publishpath(T_1(B), 0)$, the game reward is $(1-\lambda^*)(c + 1)$, because Miner 1 adds $|T_1(B)|= c + 1$ blocks in the longest path, plus $\lambda^* c$, because Miner 2 has $c$ blocks removed from the longest path. Therefore,
$$\va(B) \geq c + (1-\lambda^*) + \va(B') \geq c + (1-\lambda^*)$$
where $B'$ is the subsequent state. The inequality observes $\va(B') \geq 0$ because Miner 1 can capitulate to $B_0$ from state $B'$ (Proposition~\ref{prop:value-capitulation}). By assumption $\alpha(1-\alpha)^2 \leq (1-2\alpha)^2$ since $c \geq 0$. Solving the inequality gives $\alpha \leq \frac{1}{2}(3-\sqrt{5})$. From Proposition~\ref{prop:b_1_0-1}, the only alternative to playing $\publishpath(T_1(B), 0)$ is to play $\wait$. Let $\pi$ be any positive recurrent strategy that plays $\wait$ at state $B$. Let $Z_1 \in \ca(B_{2, 0})$ be the subsequent state if Miner 1 creates and withholds the next block and let $Z_2 \in \ca(B_0)$ be the subsequent state when Miner 1 creates and publishes the next block. Then
$$\va_\pi^{\lambda^*}(B) = \alpha \va_\pi^{\lambda^*}(Z_1) + (1-\alpha)(\va(Z_2) - \lambda^*) \leq \alpha \va(Z_1) + (1-\alpha) (\va(Z_2) - \lambda^*)$$
where the inequality observes $\va_\pi^{\lambda^*}(B') \leq \va(B')$ for any states $B'$ and strategies $\pi$ (Lemma~\ref{lemma:bellman}). Observe the $(c+1)$-Capitulation of state $Z_2$ is state $B_0$. Then from Lemma~\ref{lemma:truncation},
\begin{align*}
\va(Z_2) &\leq \R_{\lambda^*}(B_0, B_0) + \va(B_0) + \sum_{i = 1}^{c+1} \left(\pr{H_i(X_\tau) \in T_1 | X_0 = Z_2} - \lambda^*\right) -\R_{\lambda^*}(B_0, Z_2)\\
&= \sum_{i = 1}^{c+1} \left(\pr{H_i(X_\tau) \in T_1 | X_0 = Z_2} - \lambda^*\right) + \lambda^* (c+1)\\
&\leq (c+1) \frac{\alpha}{1-\alpha}.
\end{align*}
The first line is Lemma~\ref{lemma:truncation} applied to states $Z_2$ and $B_0$. The second line observes Miner 2 owns $c+1$ blocks in the longest path. Thus $\R_{\lambda^*}(B_0, Z_2) = -\lambda^* (c + 1)$. The third line observes Miner 1 will only own the block at height $i \leq c+1$ if Miner 1 can fork the longest chain $\chain(Z_2)$. But Miner 1 needs at least one block to fork the longest chain $\chain(Z_2)$. From Lemma~\ref{lemma:tie-breaking}, the probability Miner 1 forks $\chain(Z_2)$ is at most $\frac{\alpha}{1-\alpha}$.

Next, we upper bound $\va(Z_1)$. Observe the $c$-Capitulation of state $Z_1$ is state $B_{2, 0}$. From Lemma~\ref{lemma:truncation},
\begin{align*}
\va(Z_1) \leq \R_{\lambda^*}(B_0, B_{2, 0}) + \va(B_{2, 0}) - \R_{\lambda^*}(B_0, Z_1) = \va(B_{2, 0}) + c(1-\lambda^*) + \lambda^* c = \va(B_{2, 0}) + c
\end{align*}
Recall the revenue $\lambda^* = \max_\pi \rev(\pi)$ of the optimal strategy is at least $\alpha$ and at most $\frac{\alpha}{1-\alpha}$ (Corollary~\ref{cor:opt-revenue}). Then
\begin{equation}\label{eq:opt-lambda}
\alpha \leq \lambda^* \leq \frac{\alpha}{1-\alpha}
\end{equation}
The bounds on $\lambda^*$, allow us to derive an upper bound on $\va(B_{2, 0})$. 
\begin{claim}\label{claim:b_1_0}
$\va(B_{1, 0}) \leq 1$.
\end{claim}
\begin{proof}
At state $B_0$, the subsequent state is either $B_{1, 0}$ or $B_{0, 1}$. Then
$$0 = \va(B_0) = \alpha \va(B_{1, 0}) + (1-\alpha)(\va(B_{0, 1}) - \lambda^*).$$
From Proposition~\ref{prop:value-lower-bound}, $\va(B_{0, 1}) \geq 0$. Thus
$$\va(B_{1, 0}) = \lambda^* \frac{1-\alpha}{\alpha} \leq 1.$$
\end{proof}
\begin{claim}\label{claim:b_2_0}
$\va(B_{2, 0}) \leq \frac{1}{\alpha} + 1$.
\end{claim}
\begin{proof}
At state $B_{1, 0}$, if Miner 1 takes action $\wait$, the subsequent state is either $B_{2, 0}$ or $B_{1, 1}$. Then
$$1 \geq \va(B_{1, 0}) \geq \alpha \va(B_{2, 0}) + (1-\alpha)(\va(B_{1, 1}) - \lambda^*) \geq \alpha \va(B_{2, 0}) - (1-\alpha)\lambda^*$$
The first inequality is Claim~\ref{claim:b_1_0}. The second inequality is Bellman's principle of optimality (Lemma~\ref{lemma:bellman}). The third inequality uses the fact $\va(B_{1, 1}) \geq 0$ (Proposition~\ref{prop:value-lower-bound}). Solving for $\va(B_{2, 0})$ gives
$$\va(B_{2, 0}) \leq \frac{1}{\alpha} + \frac{1-\alpha}{\alpha}\lambda^* \leq \frac{1}{\alpha} + 1$$
\end{proof}
From Claim~\ref{claim:b_2_0}, we obtain
$$\va(Z_1) \leq \frac{1}{\alpha} + 1 + c$$
Combining the upper bounds for $\va(Z_1)$ and $\va(Z_2)$, we obtain
$$\va_\pi^{\lambda^*}(B) \leq 1 + \alpha + \alpha c + \alpha(c + 1) - \lambda^* (1-\alpha) \leq c  + (1-\lambda^*) + 2\alpha + \frac{\alpha^2}{1-\alpha} - c(1-2\alpha).$$
We would like to conclude by saying $V_\pi^{\lambda^*}(B) < \va(B)$. Unfortunately, the above upper bound on $\va_\pi^{\lambda^*}(B)$ is not strong enough for all $c$. Fortunately, for sufficiently large $c$, we obtain
$$\va_\pi^{\lambda^*}(B) < \va(B).$$
as desired. By Bellman's principle of optimality, we have a certificate that it is sub-optimal to wait at state $B$. The intuition is that Miner 1 {\em risks} not publishing blocks $T_1(B)$ by waiting at state $B$ (because if Miner 2 creates and publish the next block, the probability Miner 1 can publish $T_1(B)$ is at most $\frac{\alpha}{1-\alpha}$). Thus when $c = |T_1(B)| - 1$ is large, the cost of not publishing $T_1(B)$ outweighs the expected gain from not publishing $T_1(B)$ at state $B$.

To extend the proof to smaller $c$, we will do induction on $c$. The work above proves the base case for any $c > \frac{\alpha(2-\alpha)}{(1-\alpha)(1-2\alpha)}$. As inductive hypothesis, assume for all states $B' \in \ca(B_{1, 0})$ with $h(\chain(B')) \geq c + 1$, it is optimal for Miner 1 to take action $\publishpath(T_1(B'), 0)$. Note $\max T_1(B')$ becomes a checkpoint. From Theorem~\ref{thm:strong-recurrence}, there is an optimal checkpoint recurrent strategy which implies Miner 1 capitulates to state $B_0$ once a new checkpoint is defined. Thus
$$\va(B') = h(\chain(B')) + (1-\lambda^*).$$
As before, let $B \in \ca(B_{1, 0})$ and $c = h(\chain(B))$. To improve our upper bound on $\va_\pi^{\lambda^*}(B)$, we will improve our upper bound on $\va(Z_1)$. In fact, our inductive hypothesis {\em will allow us to derive a closed form on $\va(Z_1)$}. Recall $Z_1 \in \ca(B_{2, 0})$. Let $(X_t)_{t \geq 0}$ be a mining game starting at state $X_0 = Z_1$. From Proposition~\ref{prop:b_k_0}, Miner 1 will wait until the first time step $\tau$ where $X_\tau^\half \in \ca(B_{1, 0})$. Note $h(\chain(X_\tau^\half)) \geq c + 1$ because Miner 2 published at least one block in the longest path since state $X_0$. From the inductive hypothesis, Miner 1 will take action $\publishpath(T_1(X_\tau), 0)$ at state $X_\tau^\half$ and
$$\va(X_\tau^\half) = h(\chain(X_\tau^\half)) + (1-\lambda^*).$$
From state $Z_1$ to $X_\tau^\half$, Miner 2 publishes $h(\chain(X_\tau)) - h(\chain(Z_1))$ blocks. Thus the game reward from state $Z_1$ to $X_\tau^\half$ is
$$\R_{\lambda^*}(Z_1, X_\tau^\half) = -\lambda^*(h(\chain(X_\tau^\half)) - h(\chain(Z_1))).$$
The random variable $h(\chain(X_\tau^\half)) - h(\chain(Z_1))$ denotes the number of blocks Miner 2 creates from state $Z_1 \in \ca(B_{2, 0})$ until state $X_\tau^\half \in \ca(B_{1, 0})$. From Lemma~\ref{lemma:selfish-mining-reward}, we can construct a coupling $h(\chain(X_\tau^\half)) - h(\chain(Z_1))$ with a random walk starting at state $Y_0 = 1$ and for $t \geq 1$, $Y_t = Y_{t-1} + 1$ whenever Miner 1 creates the block at time $t$ (with probability $\alpha$); otherwise, $Y_t = Y_{t-1} - 1$. Then $h(\chain(X_\tau^\half)) - h(\chain(Z_1))$ counts the number of time steps $t \leq \tau$ where $Y_t = Y_{t-1} - 1$ (i.e., the number of blocks Miner 2 creates). Thus
$$\e{h(\chain(X_\tau^\half)) - h(\chain(Z_1))} = \frac{1-\alpha}{1-2\alpha}.$$
Note $h(\chain(Z_1)) = h(\chain(B)) = c$ since no blocks are published from state $B$ to state $Z_1$. Combining the work above, we derive
\begin{align*}
\va(Z_1) &= \e{\R_{\lambda^*}(Z_1, X_\tau^\half) + \va(X_\tau^\half)}\\
&= \e{h(\chain(Z_1)) + (1-\lambda^*)(h(\chain(X_\tau^\half)) - h(\chain(Z_1))) + (1-\lambda^*)}\\
&= c + (1-\lambda^*)\frac{1-\alpha}{1-2\alpha}  + (1-\lambda^*)
\end{align*}
Substituting $\va(Z_1)$ and our upper bound for $\va(Z_2)$ in our upper bound for $\va_\pi^{\lambda^*}(B)$,
\begin{align*}
\va_\pi^{\lambda^*}(B) &\leq \alpha \va(Z_1) + (1-\alpha)(\va(Z_2) - \lambda^*) \\
&\leq \alpha\left(c + (1-\lambda^*)\frac{1-\alpha}{1-2\alpha} + 1-\lambda^*\right) + (1-\alpha) \left((c+1)\frac{\alpha}{1-\alpha} + 1 -\lambda^* - 1\right)\\
&= \alpha c + (1-\alpha)c - (1-\alpha)c + (1-\lambda^*) + \alpha(1-\lambda^*)\frac{1-\alpha}{1-2\alpha} + \alpha(c+1) - (1-\alpha) \\
&\leq c + (1-\lambda^*) + \alpha\frac{(1-\alpha)^2}{1-2\alpha} - c(1-2\alpha) - (1-2\alpha) \quad \text{Because $\lambda^* \geq \alpha$}\\
&\leq \va(B) + \alpha\frac{(1-\alpha)^2}{1-2\alpha} - (c + 1)(1-2\alpha) \quad \text{Because $\va(B) \geq c + (1-\lambda^*)$}.
\end{align*}
By assumption, $\frac{\alpha(1-\alpha)^2}{(1-2\alpha)^2} \leq c + 1$. Then
$$\va_\pi^{\lambda^*}(B) \leq \va(B)$$
as desired. The inequality witnesses that it is optimal for Miner 1 to take action $\publishpath(T_1(B), 0)$ at state $B \in \ca(B_{1, 0})$.
\end{proof}
We can now conclude that for $\alpha \leq 0.307979$, $\frontier$ is an optimal strategy.
\begin{proof}[Proof of Theorem~\ref{thm:nash-equilibrium}]
Assume Miner 1 follows an optimal strategy. From Theorem~\ref{thm:strong-recurrence}, it is without loss of generality to assume such strategy is checkpoint recurrent and positive recurrent. Starting from state $B_0$, the subsequent state is either $B_{0, 1}$ when Miner 2 creates block 1 and publishes $1 \to 0$, or $B_{1, 0}$ when Miner 1 creates block 1, but yet did not decide between publishing $1 \to 0$ or waiting for the next round. From Corollary~\ref{cor:b_0_1}, at state $B_{0, 1}$, it is optimal for Miner 1 to capitulate to state $B_0$ (for any $\alpha$). At Proposition~\ref{prop:b_1_0}, let $B = B_{1, 0}$. Then $c = 0$ and as long as $\alpha \leq 0.307979$, we satisfy the condition $\frac{\alpha(1-\alpha)^2}{(1-2\alpha)^2} \leq 1$. Thus it is optimal for Miner 1 to publish $1 \to 0$ at $B_{1, 0}$. Once Miner 1 take this action, from Definition~\ref{def:checkpoint}, block 1 becomes a checkpoint. Since Miner 1's strategy is checkpoint recurrent, Miner 1 capitulates to state $B_0$ once block 1 becomes a checkpoint. Thus Miner 1 capitulates to state $B_0$ by the end of round 1 with probability 1. Additionally, up to round 1, Miner 1 is taking the same actions as $\frontier$. This witnesses $\frontier$ is an optimal strategy as desired.
\end{proof}

\newpage 
\section{Table of Notation}\label{sec:table}

\begin{longtable}[c]{|| c | c | L{0.6\textwidth} ||}
\caption{Notation.\label{tab:notation}}\\
\hline
Symbol           & Domain                            & Usage\\[0.5ex]

\hline
\hline
\endfirsthead
\hline
Symbol           & Domain                            & Usage\\[0.5ex]

\hline
\hline

\endhead

$n$              & $\mathbb{N}_+$                    & Round, or block created during a round\\

\hline

$\gamma_n$       & $\{1,2\}$                         & The miner during round $n$\\

\hline
$T_i$            & $2^{\mathbb{N}_+}$                    & The set of rounds where Miner $i$ mines\\

\hline
$\tree$          & Directed trees & Set of all published blocks, evolves over time\\
& with a single sink &\\

\hline
$V$              & $\mathbb{N}_+$                   & Nodes in $\tree$\\

\hline
$E$              & $\mathbb{N}_+$ & Edges in $\tree$\\

\hline

$\unpublished_i$ & $\mathbb{N}_+$                    & Blocks created by Miner $i$, but not yet published, evolves over time\\

\hline

$A(b)$ & $2^{\mathbb{N}}$ & Ancestors of block $b \in V$\\ 

\hline
$\Succ(b)$ & $2^{\mathbb{N}_+}$ & Successors of block $b \in V$\\

\hline

$h(b)$ & $\mathbb{N}_+$ &$:=|A(b)|-1$, height of block $b \in V$\\

\hline

$\chain$ & $\mathbb{N}_+$ & The longest chain and the block $v \in V$ with highest height\\

\hline

$H_i$ & $\mathbb N_+$ & Block $v \in A(\chain)$ with height $i = h(v)$.\\

\hline

$\text{$(B)$}$         & N/A                               & Modifies $\{\tree, V, E, \unpublished_i, \chain, H\}$ to denote its state at $B$.\\

\hline

$B^\half$      & N/A                               & Denote prior state prior to state $B$ after the most recent block was created, Miner 2 has acted, but Miner 1 has not.\\

\hline

$r^k(B, B')$ & $\mathbb N$ & $:= |A(\chain(B')) \cap T_k| - |A(\chain(B)) \cap T_k|$, Miner $k$ reward from state $B$ to $B'$\\

\hline

$\rew(B, B')$ & $\mathbb R$ & $:= (1-\lambda)r^1(B, B') - \lambda r^2(B, B')$, game reward from state $B$ to $B'$\\
 
\hline

$\rev(\pi)$ & $\mathbb R_+$ & $:= \e{\liminf_{n \to \infty}\frac{ r^1(X_0, X_n)}{r^1(X_0, X_n) + r^2(X_0, X_n) }| X_0 = B_0}$, revenue of strategy $\pi$\\

\hline

$\Succs(B)$ & Set of states & $:= \{B' : (B')^\half = B\}$, all states $B'$ reachable from state $B$ when Miner 1 takes a single action at $B$.\\

\hline

$\potr^k(B)$ & $\mathbb N$ & $:= \max_{B' \in \Succs(B)} |r^k(B, B')|$, Miner $k$ potential reward at state $B$.\\

\hline

$\va_\pi^\lambda(B)$ & $\mathbb R$ & $:= \e{r_{\lambda}(X_0, X_\tau) \bar X_0 = B}$ where $\tau$ is the first time step Miner 1 capitulates to state $B_0$\\

\hline

$\va(B)$ & $\mathbb R$ & $:= V_{\pi^*}^{\rev(\pi^*)}(B)$ where $\pi^* = \arg\max_\pi \rev(\pi)$\\

\hline
\end{longtable}

\end{document}